%% file: main.tex
\documentclass[
12pt, 
english, 
singlespacing, 
headsepline, 
]{MastersDoctoralThesis} 

\usepackage[utf8]{inputenc} 
\usepackage[T1]{fontenc} 

\usepackage{mathpazo} 
\usepackage{amsmath}
\usepackage{amsthm}
\usepackage{siunitx}
\usepackage{enumitem}
\usepackage{caption}
\usepackage{subcaption}
\usepackage{glossaries}
\usepackage{breakcites}
\usepackage{listings}
\usepackage{threeparttable}
\usepackage{multirow}
\usepackage{microtype}
\usepackage{xcolor}

\definecolor{codegreen}{rgb}{0,0.6,0}
\definecolor{codegray}{rgb}{0.5,0.5,0.5}
\definecolor{codepurple}{rgb}{0.58,0,0.82}
\definecolor{backcolour}{rgb}{0.95,0.95,0.92}

\lstdefinestyle{mystyle}{
    backgroundcolor=\color{backcolour},   
    commentstyle=\color{codegreen},
    keywordstyle=\color{magenta},
    numberstyle=\tiny\color{codegray},
    stringstyle=\color{codepurple},
    basicstyle=\ttfamily\footnotesize,
    breakatwhitespace=false,         
    breaklines=true,                 
    captionpos=b,                    
    keepspaces=true,                 
    numbers=left,                    
    numbersep=5pt,                  
    showspaces=false,                
    showstringspaces=false,
    showtabs=false,                  
    tabsize=2
}

\usepackage{float}
\usepackage[useregional,showdow=false]{datetime2}
\usepackage{array,tabularx}
\usepackage{titlesec}

\usepackage[backend=bibtex,citestyle=authoryear,natbib=true]{biblatex} 

\renewcommand\cite{\citep}

\usepackage{tocloft}

\usepackage[autostyle=true]{csquotes} 
\thesistitle{\textit{Predictive simulation of single-leg landing scenarios for ACL injury risk factors evaluation}} 
\supervisor{Moustakas Konstantinos} 
\examiner{} 
\degree{Master of Science} 
\author{Evgenia Moustridi \textsc{}} 
\authorid{1079079}
\addresses{} 

\subject{} 
\university{Univeristy of Patras} 
\department{\href{}{Department of Electrical and Computer Engineering}} 
\master{Biomedical Engineering} 
\group{\href{}{}} 
\faculty{\href{}{}} 

\AtBeginDocument{
\hypersetup{pdftitle=\ttitle} 
\hypersetup{pdfauthor=\authorname} 
\hypersetup{pdfkeywords=\keywordnames} 
}

\newacronym{fem}{FEM}{Finite Element Method}
\newacronym{fea}{FEA}{Finite Element Analysis}
\newacronym{fe}{FE}{Finite Element}
\newacronym{fd}{FD}{Forward Kinematics}
\newacronym{dofs}{DoFs}{Degrees of Freedom}
\newacronym{dof}{DoF}{Degree of Freedom}
\newacronym{ik}{IK}{Inverse Kinematics}
\newacronym{rms}{RMS}{Root Mean Square}
\newacronym{id}{ID}{Inverse Dynamics}
\newacronym{so}{SO}{Static Optimization}
\newacronym{rra}{RRA}{Residual Reduction Algorithm}
\newacronym{jra}{JRA}{Joint Reaction Analysis}
\newacronym{jrf}{JRF}{Joint Reaction Forces}
\newacronym{jrm}{JRM}{Joint Reaction Moments}
\newacronym{kjrf}{kJRF}{Knee Joint Reaction Forces}
\newacronym{kjrm}{kJRM}{Knee Joint Reaction Moments}
\newacronym{acl}{ACL}{Anterior Cruciate Ligament}
\newacronym{pcl}{PCL}{Posterior Cruciate Ligament}
\newacronym{mcl}{MCL}{Medial Collateral Ligament}
\newacronym{lcl}{LCL}{Lateral Collateral Ligament}
\newacronym{grf}{GRF}{Ground Reaction Forces}
\newacronym{grm}{GRM}{Ground Reaction Moments}
\newacronym{vgrf}{vGRF}{vertical Ground Reaction Force}
\newacronym{pgrf}{pGRF}{posterior Ground Reaction Force}
\newacronym{3D}{3D}{3 - Dimensional}
\newacronym{2D}{2D}{2 - Dimensional}
\newacronym{bw}{BW}{Body Weight}
\newacronym{q/h}{Q/H force ratio}{Quadricep / Hamstring force ratio}
\newacronym{gta}{GTA force ratio}{Gastrocnemius / Tibialis Anterior force ratio}
\newacronym{emg}{EMG}{Electromyography}
\newacronym{com}{COM}{Center of Mass}
\newacronym{cop}{CoP}{Center of Pressure}
\newacronym{igc}{IGC}{Initial Ground Contact}
\newacronym{af}{AF}{Anterior force}
\newacronym{cf}{CF}{Compressive force}
\newacronym{mf}{MF}{Medial force}
\newacronym{fm}{FM}{Flexion moment}
\newacronym{em}{EM}{Extension moment}
\newacronym{irm}{IRM}{Internal rotation moment}
\newacronym{erm}{ERM}{External rotation moment}
\newacronym{abdm}{AbdM}{Abduction moment}
\newacronym{addm}{AddM}{Adduction moment}
\newacronym{mocap}{MoCap}{Motion Capture}

\makeglossaries{}

\begin{document}
\frontmatter 

\pagestyle{plain} 

{\let\cleardoublepage\clearpage


 

 

 

 

\begin{titlepage}
    \begin{center}
    
    \vspace*{.01\textheight}
    \bigskip
    \begin{figure}
    \begin{subfigure}{.5\linewidth}
        \centering
        \includegraphics[width = \textwidth]{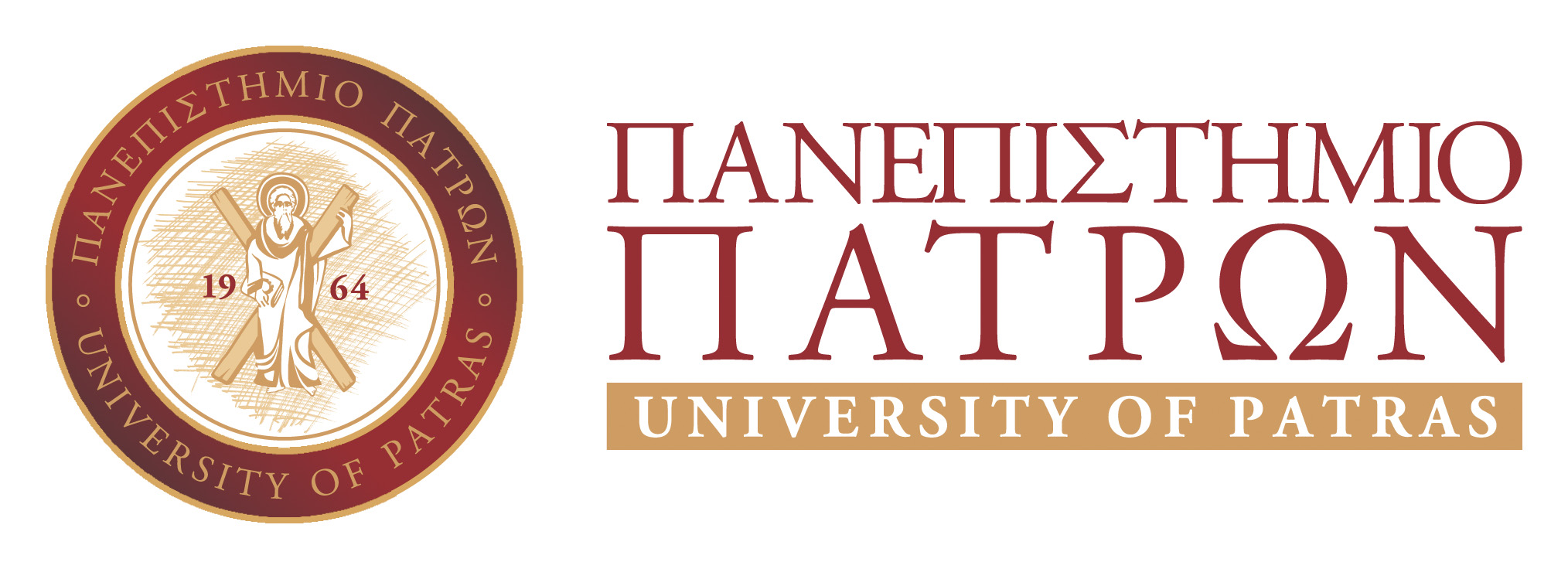} 
    \end{subfigure}
    \hfill
    \begin{subfigure}{.4\linewidth}
        \centering
        \includegraphics[width = \textwidth]{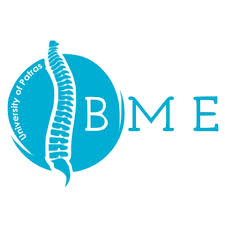}
    \end{subfigure}
    \end{figure}

\Large MSc Program \\ 
    \bigskip
    "\mastername" \\[0.5cm] 
    \bigskip
    
    \Large MSc Thesis \\ 
    of  \authorname \\
    Dipl. Electrical Engineer and Computer Engineer \\ 
    A.M.: \idnumber \\[0.5cm]

    \HRule \\[0.4cm] 
    {\Large \bfseries \ttitle\par}\vspace{0.4cm} 
    \HRule \\[1.5cm] 

    \emph{Supervisor:} \\
    Prof. \supname 
        
    
    \textsc{\Large Thesis number: } 
    

    \vfill
    
    {\large 08/02/2022}\\[4cm] 
        
    \end{center}
\end{titlepage}


\begin{declaration}
\vfill

\noindent\textit{University of Patras, Electrical and Computer Engineering Department} \\
\authorname \\
\copyright 2022 - All rights reserved

\bigskip

\noindent The thesis reports original material developed by \authorname and
does not violate Intellectual Property Rights in any way.
Reuse of existing material is properly referenced, while potential use of figures,
graphs, etc is being reproduced after receiving the respective license from the IPR
holder.

\rule[0.5em]{25em}{0.5pt} 
\end{declaration}






\begin{abstract}
\addchaptertocentry{\abstractname} 

The \acrlong{acl} rupture is a very common knee injury during sport activities. Landing after a jump is one of the most prominent human body movement scenarios that can lead to such an injury. The landing - related \gls{acl} injury risk factors have been in the spotlight of research interest. Over the years, researchers and clinicians acquire knowledge about human movement during daily - life activities by organizing complex in \textit{vivo} studies using \acrlong{mocap} equipment. The recorded motion data are further analyzed to estimate joint angles and forces and provide insight to the functionality and response of internal structures such as soft tissues, ligaments and muscles. Nevertheless, these experiments feature high complexity, costs and technical and most importantly physical challenges. Computational modeling and simulation overcomes all these limitations and allows for studying musculoskeletal systems. Specifically, predictive simulation approaches offer researchers the opportunity to predict and study new biological motions without the demands of acquiring experimental data. In this thesis, we present a pipeline that aims to predict and identify key parameters of interest that are related to \gls{acl} injury during single - leg landings. We examined the following conditions of single - leg landing: a) initial landing height, b) hip internal and external rotation, c) lumbar forward - backward leaning, d) lumbar medial - lateral bending, e) lumbar internal - external rotation, f) muscle forces permutations and g) effort goal weight. Identified on related research studies, we evaluated the following risk factors: \acrlong{vgrf}, knee joint \acrlong{af}, knee joint \acrlong{abdm}, and \acrlong{q/h}. Our study clearly demonstrated that \gls{acl} injury is a rather complicated mechanism with many associated risk factors which are evidently correlated. Nevertheless, our results were mostly in agreement with other research studies regarding the \gls{acl} risk factors. Despite the limitations regarding the adopted modeling assumptions, our pipeline clearly showcased promising potential of predictive simulations to evaluate different aspects of complicated phenomena, such as the \gls{acl} injury. Therefore, further improvements can lead to the development of a workflow that can be used by physiotherapists and clinicians on adjusting rehabilitation and training plans based on subject - specific characteristics.  

\vspace*{0.05\textheight}


\end{abstract}

\renewcommand{\abstractname}{\emph{Περίληψη}}
\begin{abstract}

Η ρήξη του Πρόσθιου Χιαστού Συνδέσμου (ΠΧΣ) είναι πολύ συχνός τραυματισμός της άρθρωσης του γονάτου κατά την διάρκεια αθλητικών δραστηριοτήτων. Η προσγείωση μετά από ένα άλμα είναι μια κίνηση του ανθρώπινου σώματος που μπορεί να οδηγήσει σε τέτοιου είδους τραυματισμό. Οι παράμετροι επικινδυνότητας ρήξης του ΠΧΣ που σχετίζονται με άλματα είναι στο επίκεντρο του ερευνητικού ενδιαφέροντος. Όλα αυτά τα χρόνια, οι ερευνητές και οι κλινικοί γιατροί αποκτούν γνώσεις σχετικά με τις καθημερινές κινήσεις του ανθρώπινου σώματος οργανώνοντας σύνθετες "in vivo" μελέτες με χρήση εξοπλισμού καταγραφής κίνησης. Τα δεδομένα κίνησης αναλύονται περαιτέρω ώστε να εκτιμηθούν οι γωνίες και οι δυνάμεις στις αρθρώσεις παρέχοντας με αυτό τον τρόπο μια εικόνα στη λειτουργία και την απόκριση των εσωτερικών δομών, όπως οι μαλακοί ιστοί, οι σύνδεσμοι και οι μύες. Ωστόσο, αυτά τα πειράματα χαρακτηρίζονται από υψηλή πολυπλοκότητα, και παρουσιάζουν μεγάλο κόστος, σύνθετες τεχνικές και κυρίως φυσικές προκλήσεις. Η υπολογιστική μοντελοποίηση και προσομοίωση επιτρέπει την μελέτη μυοσκελετικών συστημάτων ξεπερνώντας αυτούς τους περιορισμούς. Πιο συγκεκριμέ\-να, οι μέθοδοι προσομοιώσεων με τεχνικές πρόβλεψης προσφέρουν στους ερευνητές την ευκαιρία να προβλέπουν και να μελετούν νέες βιολογικές κινήσεις χωρίς την απαίτηση καταγραφής πειραματικών δεδομένων. Σε αυτή τη διπλωματική παρου\-σιάζουμε μια μεθοδολογία που στοχεύει στην πρόβλεψη και αναγνώριση βασικών παραμέτρων που σχετίζονται με τραυματισμούς του ΠΧΣ κατά την εκτέλεση προσγειώσεων με ένα πόδι μετά από άλμα. Οι περιπτώσεις προσγείωσης με ένα πόδι που εξετάστηκαν ήταν οι ακόλουθες: α) αρχικό ύψος προσγείωσης, β) εσωτερική και εξωτερική περιστροφή του ισχίου, γ) οσφυϊκή κλίση προς τα εμπρός και προς τα πίσω, δ) οσφυϊκή κλίση προς τα δεξιά και προς τα αριστερά, ε) οσφυϊκή εσωτερική και εξωτερική περιστροφή, στ) μεταβολές μυϊκών δυνάμεων και τέλος ζ) επίπεδο ενεργοποίησης μυών. Βασιζόμενοι σε σχετικές ερευνητικές μελέτες αξιολογήσαμε τις ακόλουθες παραμέτρους κινδύνου: κατακόρυφη δύναμη αντίδρασης εδάφους, πρόσθια δύναμη στην άρθρωση του γονάτου, ροπή απαγωγής της άρθρωσης του γονάτου και λόγος δύναμης τετρακέφαλου μυός / οπίσθιου μηριαίου μυός. Η μελέτη μας έδειξε ξεκάθαρα την περιπλοκότητα του τραυματισμού του ΠΧΣ και την συσχέ\-τιση του με πολλούς παράγοντες κινδύνου τραυματισμού. Παρ' όλα αυτά, τα αποτε\-λέσματα μας ήταν ως επί το πλείστον σε συμφωνία με σχετικές ερευνητικές μελέτες όσον αφορά τους παράγοντες τραυματισμού του ΠΧΣ. Παρά τους περιορισμούς σχετικά με την υιοθετούμενη μοντελοποίηση, η μελέτη μας παρουσίασε υποσχόμενες δυνατότητες προσομοίωσης πρόβλεψης και αξιολόγησης διαφορετικών πτυχών περίπλοκων φαινομένων, όπως αυτό της ρήξης του ΠΧΣ. Περαιτέρω βελτιώσεις μπορούν να οδηγήσουν στην ανάπτυξη ενός εργαλείου που θα μπορεί να χρησιμοποιηθεί από φυσικοθεραπευτές και κλινικούς γιατρούς για την προσαρμογή σχεδίων αποκατάστασης και εκπαίδευσης με βάση συγκεκριμένα χαρακτηριστικά του κάθε υποκειμένου.

\textbf{Λέξεις κλειδιά:} Πρόσθιος Χιαστός, ΠΧΣ, ρήξη ΠΧΣ, προσγέιωση, προσομοίωση πρόβλεψης, Δύναμη Αντίδρασης εδάφους, Δυνάμεις Αντίδρασεις άρθρωσης
\end{abstract}

\begin{acknowledgements}
\addchaptertocentry{\acknowledgementname} 

First of all, I would like to express my sincere gratitude to my thesis supervisor Prof. Konstantinos Moustakas for his guidance, support and trust for letting me be part of the Visualization and Virtual Reality Group.

Furthermore, I would like to express my gratitude and appreciation to phd candidate Konstantinos Risvas whose guidance, support and encouragement has been invaluable throughout this study.

In addition, I would like to express my acknowledgements to my sister, Maria and friends, for their support and continuous encouragement  throughout the years of study and through the process of researching and writing this thesis. 

Finally, I am grateful for my parents whose constant love and support keep me motivated and encourage me throughout the years of my studies. This accomplishment would not have been possible without them. 

\end{acknowledgements}

\renewcommand{\contentsname}{\hfill\bfseries Table of Contents\hfill}
{
  \hypersetup{linkcolor=black}
  \tableofcontents
}

\listoffigures 

\listoftables 

\printglossary[type=\acronymtype,title=Acronyms]



\mainmatter 

\pagestyle{thesis} 


\titlespacing{\chapter}{0pt}{*0}{*4}
\titleformat{\chapter}[display]
{\bfseries\Large}
{\filleft \Huge \color{mdtRed} {\chaptertitlename} {\thechapter}}
{2ex}
{\titlerule
\vspace{2ex}%
\centering}
[\vspace{2ex}%
\titlerule]

\renewcommand*{\chaptermarkformat}{}

\include{Chapters/Chapter1}
\include{Chapters/Chapter2}

\include{Chapters/Chapter3}
\include{Chapters/Chapter4}

\include{Chapters/Chapter5}
\include{Chapters/Conclusions}


\appendix 




\printbibliography[heading=bibintoc]


\end{document}

%% file: Chapters/Chapter1.tex

\chapter{Introduction}\label{Introduction} 



\begin{figure}[H]
    \begin{subfigure}[t]{0.49\textwidth}
        \centering
        \includegraphics[width=0.75\linewidth]{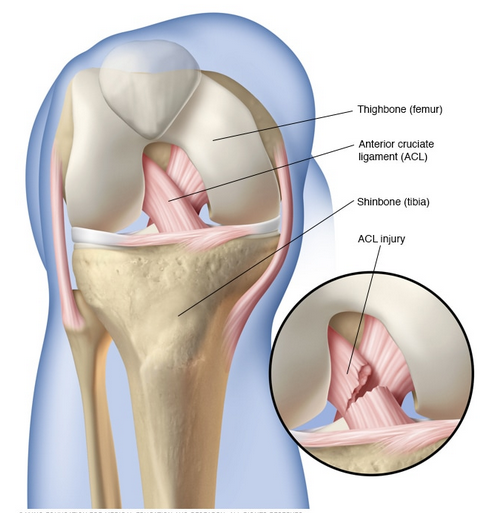}
        \caption{\gls{acl} rupture representation. Reproduced from \cite{MayoClinic}.}
        \label{fig:acl_tear}
    \end{subfigure}
     \begin{subfigure}[t]{0.49\textwidth}
        \centering
        \includegraphics[width=0.9\linewidth]{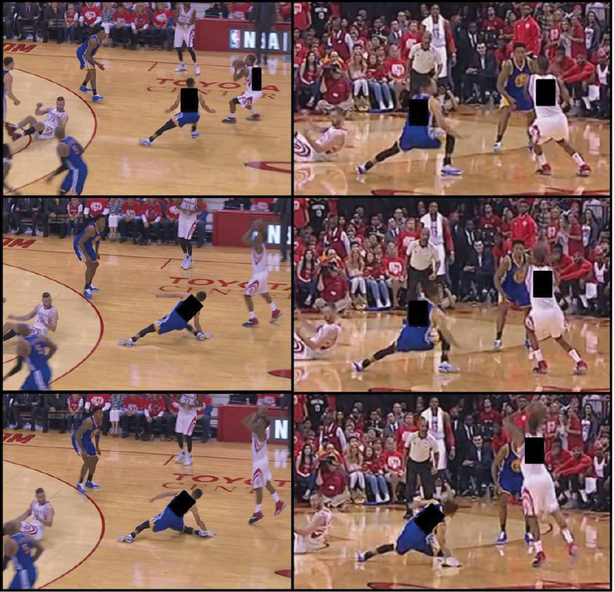}
        \caption{Mechanism of ACL rupture during change of direction. An increased injury is introduced with the foot planted, knee externally rotated or in valgus position and a varus or internal moment is applied. Reproduced from \cite{Brukner2016}.}
        \label{fig:injury}
    \end{subfigure}
    \caption{Left: \gls{acl} rupture. Right: ACL rupture during change of direction.}
\end{figure}

The knee joint anatomical structure is one of the most complex human body joints. Its stability and functionality during daily life activities are maintained mainly by the articulations, ligaments, menisci and muscle forces. Among the knee joint ligaments, \gls{acl} is of prominent importance as it acts as a stabilizer by restricting excessive posterior and anterior knee displacement during dynamic movements. This is the reason why \gls{acl} rupture is one of the most common knee injuries in competitive sports activities, like football, basketball, or skiing  (\autoref{fig:acl_tear}) \cite{Prodromos2008}. These sports involve movement patterns of high risk, such as sudden stops, abrupt changes of direction, landing after a jump and deceleration before changing direction \cite{Sugimoto2014,Boden2000}. These are labeled as non - contact \gls{acl} injury conditions (\autoref{fig:injury}). However, injuries of the \gls{acl} can also happen when direct contact with another athlete takes place, although only 1/3 of the total \gls{acl} injuries occur due to direct contact ~\cite{Mokhtarzadeh2017,Bulat2019}.

In general, \gls{acl} injury involves a complete or partial ligament tear. This condition as well as the desired post injury activity levels affect the chosen treatment procedure. Typically, there are two main approaches. The first option involves a basic rehabilitation plan including physiotherapy and bracing support. However, this option requires low levels of activity in the future. On the other hand, if high levels of activity are desired, surgery for \gls{acl} restoration is unavoidable. This procedure named as "\gls{acl} reconstruction" has emerged as the golden standard approach when it comes to \gls{acl} injuries \cite{Looney2020}. It requires replacement of the ligament by a tendon graft extracted by the same patient or a donor. Also, artificial grafts can be used. Post - surgery, rehabilitation strategies involve muscle strengthening and gradual restoration of knee motion range. A post - effect of a complete \gls{acl} tear is the higher risk of osteoarthritis development on the knee cartilage, especially when the meniscus is also damaged \cite{Monk2014}. \gls{acl} reconstruction is a complex surgery that depends on a plethora of parameters and requires long periods of rehabilitation, that can span a 6 - 12 month range \cite{Kvist2004,Goes2020}. Moreover, graft failures are common and full restoration of knee kinematics is not always achievable. Therefore, complete and accurate knowledge regarding knee joint biomechanics is crucial to develop training strategies that aim to limit \gls{acl} injury risk or assist surgery and rehabilitation plans post - injury.  


\begin{figure}[H]
    \begin{subfigure}[t]{0.49\textwidth}
        \centering
        \includegraphics[width=0.4\linewidth]{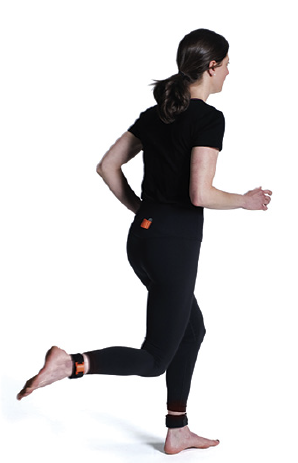}
        \caption{Inertial measurement units, the orange sensors  measure linear accelerations and angular velocities.\\ Reproduced from \cite{Uchida2020}}
        \label{kneemotions}
    \end{subfigure}
     \begin{subfigure}[t]{0.49\textwidth}
        \centering
        \includegraphics[width=0.9\linewidth]{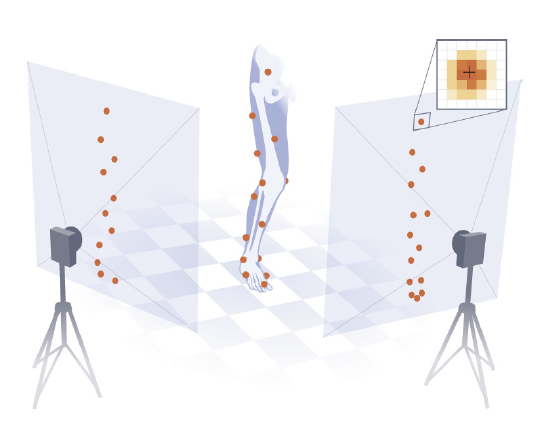}
        \caption{Optical motion capture (mocap) is a technique for estimating the motion of the underlying skeleton from the trajectories of points on the skin.\\ Reproduced from \cite{Uchida2020}}
        \label{anatomicalplanes}
    \end{subfigure}
    \caption{Left: Inertial Measurement Units (IMUs). Right: Optical Motion Capture (mocap).}
    \label{fig:motion_caption}
\end{figure}

Traditionally, researchers and clinicians acquire knowledge about human movement during daily - life activities by organizing complex \textit{in vivo} studies. However, these experiments feature high complexity, costs and technical and physical challenges. Particularly in potentially precarious dynamic movements, such as landing, ethical issues may arise that can restrain the availability of participants in these scenarios. These limitations and uncertainties gave prominence to computational modeling and simulation as a competent tool at the disposal of biomechanics professional and research community when studying musculoskeletal systems, especially with the rapidly increased computational power that significantly reduces time of numerical studies' execution.

\begin{figure}[H]
    \captionsetup{width=1\linewidth,justification=justified, singlelinecheck=false}
    \centering
    \includegraphics[width=0.6\linewidth, height=5cm]{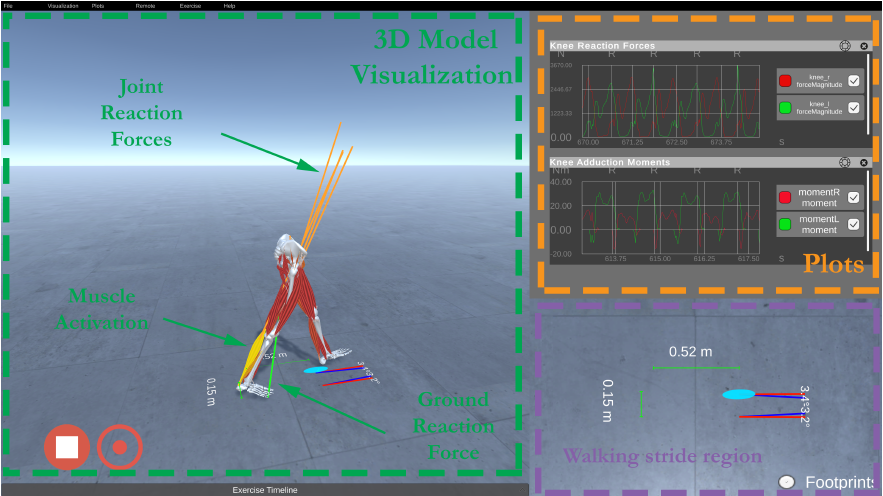}
    \caption{Visualization front-end composed of a musculoskeletal visualizer, real - time plotting, and footprint visualization. Reproduced from \cite{Stanev2021_realtime}.}
    \label{fig:real_time}
\end{figure}

In the most common approach, biomechanical studies usually require human motion data that are recorded using dedicated equipment. Examples of such devices are mobile sensors (Inertial Measurement Units or IMUs) and \gls{mocap} equipment using video technologies and retroreflective markers \autoref{fig:motion_caption}. The recorded data are further analyzed using software to estimate joint angles and forces. That way researchers gain insight into the functionality and response of internal structures such as soft tissues, ligaments and muscles. Estimation of these kinematic and kinetic parameters is also achievable with real - time biofeedback as can bee seen in \autoref{fig:real_time} \cite{Stanev2021_realtime}. 

An integral part of a standard biomechanics analysis is the development of computer models that mathematically describe all the aspects of the physical systems that they represent. The response of these models is numerically evaluated by applying rigid body dynamics, \gls{fem} or computational fluid dynamics (CFD) depending on the phenomena under consideration \cite{Engel2011}. 

Usually, rigid body dynamics are applied to study musculoskeletal system on a high - level through estimation of forces at a certain number of body points or joints. The musculoskeletal system is modeled as rigid bodies in series linked together by joints that represent real life articulation. In clinical practice, the main methods regarding multi-body systems are the \gls{id} and the \gls{fd} methods. \gls{id} analysis starts from the movement (effect) and computes the muscle excitation signals (cause), while in \gls{fd} analysis the muscles are triggered and the resultant movement is estimated \cite{Stanev2015_Proprioceptive}. On the other hand, when detailed response of anatomical structures is demanded, \gls{fe} models are more suitable, since they allow for body deformation evaluation and estimation of stress. \gls{fea} allows to capture the geometry and material effects on the response of the anatomical structures \cite{Benos2020,Naghibi2020}. Combining these two approaches is also a common route for researchers \cite{Kłodowski2015}. 

Examples of biomechanics application fields using mutli - body systems are neuromuscular pathologies, study of muscle coordination and surgery modeling and simulation. Regarding neuromuscular pathologies, multiple research studies examine the analysis of human motion dynamics, and especially gait, to obtain biometric features through which disturbances can be observed from normal and abnormal motion patterns \cite{Nandy2017,Akbas2018,Arnold2006}. Furthermore, muscle redundancy is a common biomechanics research topic where scientists attempt to investigate the muscle contribution in each movement considering that there exist infinitely many solutions for muscle forces resulting in identical movement \cite{Stanev2019_Stiffness}. Many research studies focus on modifying and improving the developed algorithms of efficient computation of these parameters \cite{Stanev2018_task,Stanev2019_null}. Finally, biomechanics are also applied to model and simulate surgery procedures aiming to develop assisting tools and treatment plans \cite{Stanev2015_aclreconstruction}. In general, all these studies require experimental data, such as \gls{mocap} and biosensor measurements (\gls{emg}, or sensor implants). Therefore the demands of "state of the art" technology facilities, equipment and complex experimental setups with multiple trials is unavoidable.  

In an attempt to overcome these challenges, predictive simulation approaches have emerged as a valuable counterpart to \gls{mocap} equipment that offer researchers the opportunity to predict new motions without the demands of acquiring experimental data. These modern simulation approaches as part of the trajectory optimization domain allow researchers to generate and study new biological motions. That way, predominant factors that can cause injuries can be identified and determined overcoming the necessity of real experiments. Predictive simulations can be utilized to provide answers in multiple and repeated what - if scenarios that would require laborious and demanding experimental setups. The outcomes of predictive subject - specific simulations can be used as the scaffold of personalized treatment plans to improve medical therapy and rehabilitation. Moreover, physiotherapists and clinicians can organize retraining procedures such as muscle strengthening and posture improvement to minimize injury risk factors. Finally, prototypes of assistive devices can be assessed to improve their functionality in minimizing injury risk factors.

In this thesis, we use biomechanics tailored trajectory prediction tools to generate a single - leg landing motion and study what - if scenarios that can potentially contribute to an \gls{acl} rupture. This is accomplished by evaluating permutations of body posture, individual muscle forces and simulations settings.

\section{Thesis contribution}

In this work, we propose a predictive simulation approach that allows prediction of a single - leg landing motion without using experimental data. The aim of the workflow is to perform multiple case studies of single - leg landings, compare them and identify factors that could contribute to an \gls{acl} injury.

In more detail, our analysis is divided in three main parts. First, we predict the motion of a single - leg landing using the SCONE software with a simplified OpenSim model. Then, we track that motion using a more complex OpenSim model and the Moco tool. Finally, we predict multiple single - leg landing motions using MOCO tool, a more complex model and the previously tracked motion as an initial guess. The examined case studies are divided in five main categories. First, we predict landings from six different initial heights. Then we predict single - leg landings with different hip rotation angles. Also, we predict drop - landings with deviations on the trunk orientation, and in particular the following cases: lumbar flexion - extension, lumbar medial - lateral bending and internal - external rotation. Furthermore, we predict single - leg landings with different combinations of knee joint agonist and antagonist muscle forces. Finally, we examine the effect of an effort goal on the results of the analysis since knee stability during single - leg landing can be affected by muscle forces and proprioception. 

The presented pipeline offered us an insight into the parameters associated with \gls{acl} injuries during single - leg landings which was the main objective. For all case studies we were able to compare numerous parameters that are related to increased \gls{acl} injury risk based on thorough literature review of similar research studies. Moreover, when comparing these risk factors we clearly witnessed and showcased their high correlation. 

The primary objective of the work presented in this thesis was to lay the technical framework that can allow the easy setup of multiple case studies for evaluating single - leg landings to answer what - if scenarios of our interest. Capitalizing on this, we consider that our work  offers multiple additional assets. First, changing the   data of the musculoskeletal model can be easily done through the OpenSim interface and allow for subject - specific simulations. This is crucial for identifying or simulating pathological conditions such as muscle weakening following an injury. Furthermore, modifications of the foot - contact modeling setup can support different settings such as inclined landing planes or material properties for footwear.    

We envision that the presented pipeline could be used by physiotherapists and clinicians on adjusting rehabilitation, training plans and an overall improve in medical therapy.


\section{Thesis structure}

The next chapters of the presented thesis are organized in the following way:

In \textbf{Chapter 2} we introduce the anatomy and biomechanics of the lower - limb, following by a description of  the muscle physiology, architecture and dynamics. Then, we present basic concepts of \gls{acl} injuries during dynamic movements like drop - landings and associated parameters. Finally, we make a high - level description of mathematical and physic concepts, such as multi - body dynamics and optimization strategies.

In \textbf{Chapter 3}, we describe the methods of this thesis. First, we present a description of the musculoskeletal models and the applied modifications. Then, we present the simulations run in SCONE in order to create an initial predicted motion of a single - leg landing. Finally, the simulations conducted using OpenSim MOCO are described. Information about all studied scenarios are thoroughly described. 

In \textbf{Chapter 4}, we present results for the investigated scenarios. For every case, we showcase the following parameters of evaluation: \gls{grf}, \gls{grm}, \gls{jrf}, \gls{jrm}, muscle forces and muscle force ratios. The results are analyzed and presented using intuitive plots for the entire motion and tables with values at time of peak \gls{vgrf}. 

In \textbf{Chapter 5}, we discuss the findings of our study and compare them by referencing the \gls{acl} risk factors pointed out by similar research studies in literature.
 
Finally, in \textbf{Conclusions}, the overall conclusions of the thesis, limitations and possible future directions are presented.


\newcommand{\keyword}[1]{\textbf{#1}}
\newcommand{\tabhead}[1]{\textbf{#1}}
\newcommand{\code}[1]{\texttt{#1}}
\newcommand{\file}[1]{\texttt{\bfseries#1}}
\newcommand{\option}[1]{\texttt{\itshape#1}}


%% file: Chapters/Chapter2.tex

\chapter{Theoretical Background of Related Concepts} 

\label{Chapter2} 


In this chapter, a theoretical background of the related to our study concepts is provided. First, we present a general description of the lower limb anatomy and muscle physiology. Additionally, we outline the risk factors of \gls{acl} injury in an effort to meticulously depict the involved mechanisms. Finally, we attempt a brief introduction into multi -  body dynamics and methods of trajectory optimization that are the main mathematical tools deployed throughout this work.

\section{Lower - limb  anatomy and biomechanics}

\begin{figure}[H]
    \centering
    \includegraphics[width=0.5\linewidth]{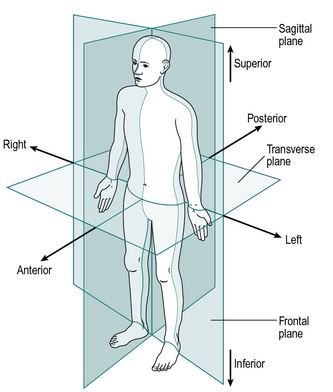}
    \caption{The anatomical position, with the three reference planes and six fundamental directions. Reproduced from \cite{anatomyplanes}.}
    \label{fig:reference_planes}
\end{figure}

In this section, we describe fundamental material regarding the anatomy, physiology and biomechanics of the lower human body. First, the anatomy of the relevant joints is presented, with specific focus on the knee joint and the involved ligaments. We also briefly describe the muscles that span the knee and provide an insight to muscle physiology and basic concepts of force generation. Moreover, to describe human body posture and understand the relative configuration between different bodies we use the notation of reference planes and directions presented in \autoref{fig:reference_planes}. Particularly, we denote the sagittal, transverse and  frontal planes. Also, six fundamental directions are described: the superior - inferior, anterior - posterior, and lateral - medial (or contra - lateral).

\begin{figure}[H]
    \begin{subfigure}[t] {0.5\textwidth}
    \centering
    \includegraphics[width=0.8\linewidth, height=7cm]{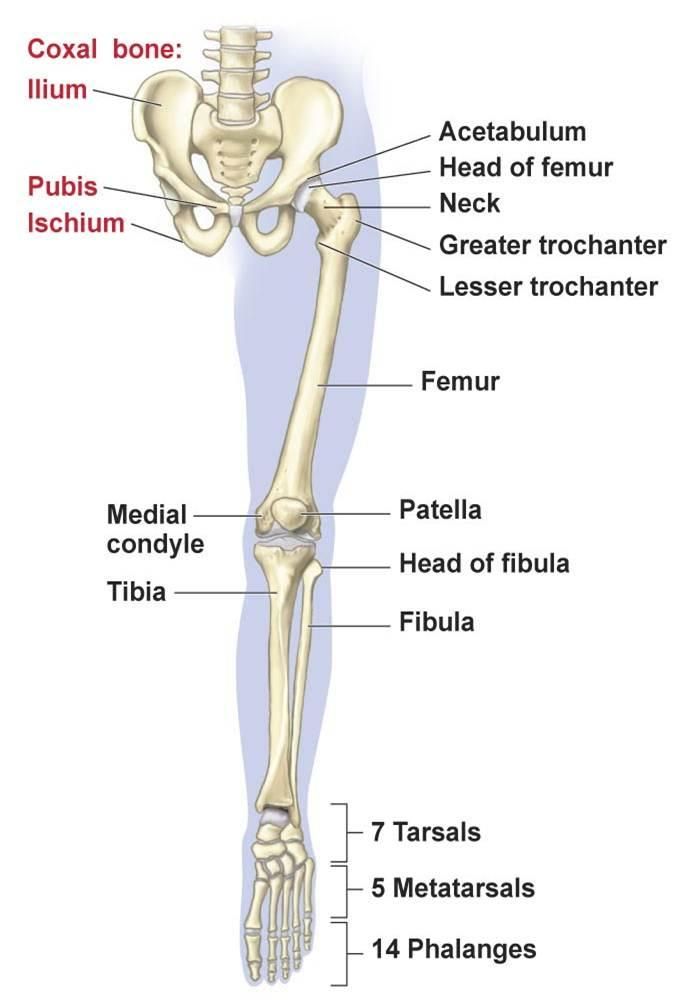}
    \caption{Anatomy of the lower human body}
    \end{subfigure}%
    \begin{subfigure}[t]{0.5\textwidth}
    \centering
    \includegraphics[width=0.8\linewidth, height=7cm]{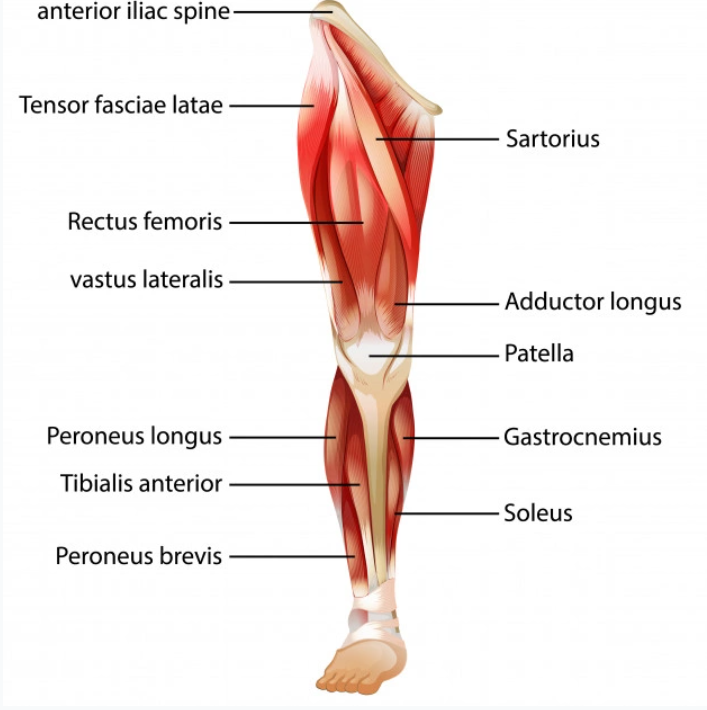}
    \caption{Muscles of the lower human body}
    \end{subfigure}
    \caption{Anatomy and physiology of the lower human body. Reproduced from \cite{howtorelief,KENHUB}.}
    \label{fig:bones_muscles}
\end{figure}

The lower extremity consists of the pelvis, femur, patella, tibia, fibula and the foot bones. These are connected through the hip, knee and ankle joints, each one featuring its own functionality and allowing for complex movement patterns, rotations and / or translations. These motions are coordinated by the central nervous system which orchestrates the activation of the muscles that span these joints. An overview of the key anatomical structures is presented in \autoref{fig:bones_muscles}.

\subsection{Hip joint}

\begin{figure}[H]
    \captionsetup{width=1\linewidth,justification=justified, singlelinecheck=false}
    \centering
    \includegraphics[width=0.3\linewidth, height=4cm]{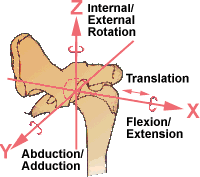}
    \caption{Presentation of hip joint \gls{dofs}. Reproduced from \cite{Smith1990}.}
    \label{fig:hip}
\end{figure}

The hip joint connects the pelvis and femur bodies and allows three rotational \gls{dofs}, namely flexion - extension, abduction - adduction, and internal - external rotation as shown in \autoref{fig:hip}.

\subsection{Knee joint}\label{subsec:knee_joint}

\begin{figure}[H]
    \captionsetup{width=1\linewidth,justification=justified, singlelinecheck=false}
    \centering
    \includegraphics[width=0.7\linewidth, height=8cm]{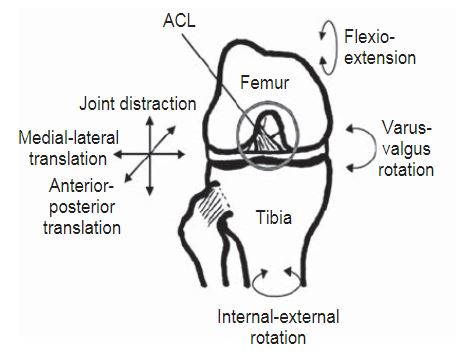}
    \caption{Presentation of knee joint \gls{dofs}. Reproduced from \cite{BioDigital}.}
    \label{fig:knee_joint}
\end{figure}

The knee joint comprises of two distinct joints, namely the tibio - femoral and patello - femoral joint, which work in conjunction to generate smooth and efficient knee movement. The joint names are derived by the bones that these connect. The tibio - femoral joint provide six \gls{dofs}, three relative translations (compression - distraction, anterior - posterior and medial - lateral translation) and three rotations (flexion - extension, internal - external rotation, and varus - valgus rotation) between the femur and tibia bones as depicted in \autoref{fig:knee_joint}. Physiological ranges for the rotational motions are \ang{3} - \ang{155} for flexion - extension, \ang{6} - \ang{8} for varus - valgus and \ang{25} - \ang{30} of internal - external rotation when the knee is flexed. Regarding translations, the anterior - posterior translation range is 5 - 10mm, medial - lateral is 1 - 2mm and compression-distraction is 2 - 5mm \cite{rom_knee}.    

\begin{figure}[H]
    \captionsetup{width=1\linewidth,justification=justified, singlelinecheck=false}
    \centering
    \includegraphics[width=0.8\linewidth, height=6cm]{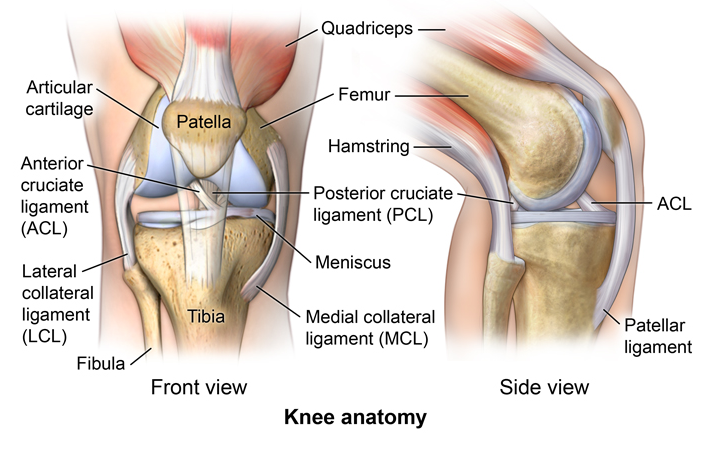}
    \caption{Overview of the knee anatomy. Reproduced from \cite{hopkinsmedicine}.}
    \label{fig:ligaments}
\end{figure}

Joint motion is dictated by muscles that span the knee articulation. The extensor muscles are the quadriceps and the flexors are the gastrocnemius, sartorius, popliteus, biceps femoris, semitendinosus and semimembranosus muscles. The role of the knee ligaments is fundamental in restraining the knee motion inside the above mentioned physiological limits by counterbalancing muscle - generated forces. An overview of the knee anatomical structures is presented in \autoref{fig:ligaments}.

Furthermore, the knee joint is surrounded by ligaments that are elastic bands of connecting tissue. Their main functionality is to act as stabilizers by restricting the anterior - posterior and varus - valgus knee forces and moments. The main ligaments are the \gls{acl}, \gls{pcl}, \gls{mcl} and the \gls{lcl}. Additional ligaments that play a supportive stabilizing role are the anterolateral ligament, the popliteal and meniscofemoral ligaments and the ligaments of the posterior capsule. 

\subsection{Ankle joint}

\begin{figure}[H]
    \captionsetup{width=1\linewidth,justification=justified, singlelinecheck=false}
    \centering
    \includegraphics[width=0.4\linewidth, height=6cm]{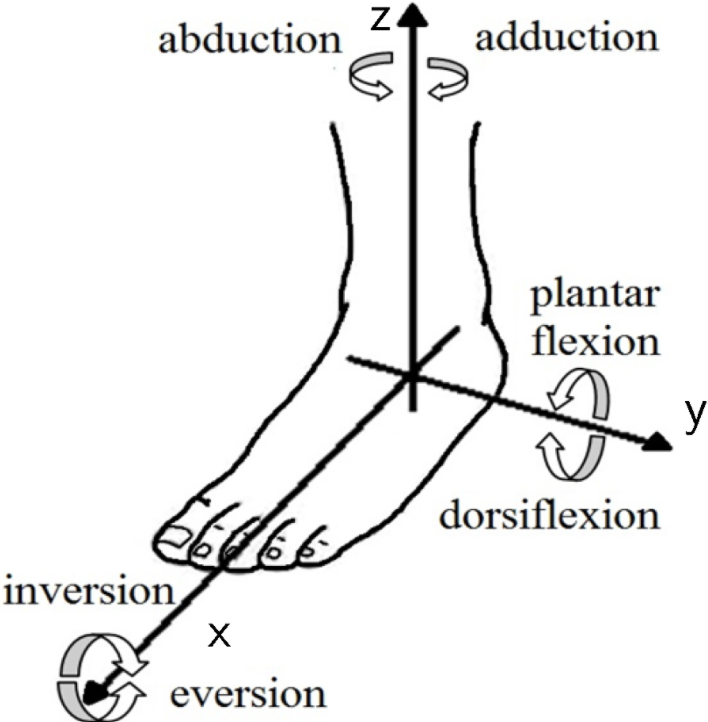}
    \caption{Ankle joint movements in three orthogonal planes.  Reproduced from \cite{Alcocer2012}.}
    \label{fig:ankle_joint}
\end{figure}

The ankle joint connects the shank and fibula with the talus bone and allows for plantar - flexion and dorsi - flexion, inversion - eversion, and internal - external rotation \autoref{fig:ankle_joint}.

\section{Muscle physiology, architecture and dynamics}

\begin{figure}[H]
    \captionsetup{width=1\linewidth,justification=justified, singlelinecheck=false}
    \centering
    \includegraphics[width=1\linewidth, height=10cm]{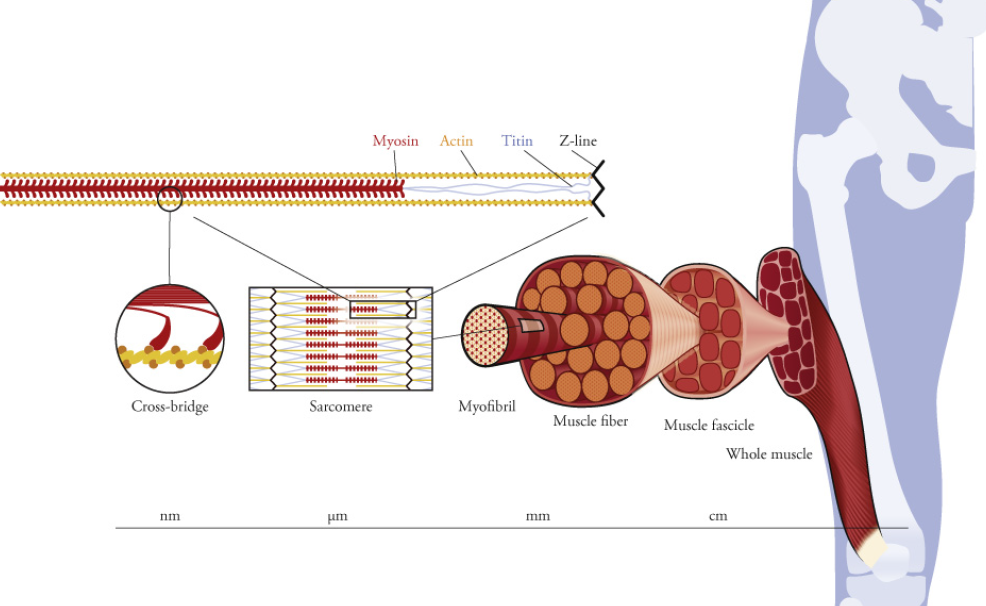}
    \caption{Hierarchical structure of skeletal muscle. Reproduced from \cite{Uchida2020}.}
    \label{fig:muscle_structure}
\end{figure}

In this section we present the fundamental theoretical concepts about muscle physiology. A brief description of the structure and function of skeletal muscles is necessary to clearly comprehend how motion is generated. An overview of the muscle structures and their hierarchical organization is presented in \autoref{fig:muscle_structure}.

By inspecting this figure from right to left we can see that the muscles are connected to bones via tendons on both ends. Tendons are responsible for storing and releasing energy and act as transmitters of muscle forces. As a whole, a muscle is formed by packed muscle fasicles that are shaped by arranged packs of muscle fibers (or myocytes). Subsequently, these comprise of multiple parallel myofibrils that are composed of the fundamental force generating units of a muscle, the sarcomeres. The sarcomeres are linked in parallel or are packed in series. Muscle forces are generated in this lowest level of muscle structure, mainly by two proteins. These are actin and myosin and they are arranged in parallel to form the cross - bridges, areas of a sacromere where these two proteins are linked. The length of these cross bridges affects the muscle force generation. A single muscle is capable of generating force of thousands of newtons during its contraction, but this force is generated by the summation of small peconewtons of force produced by its motors at the molecular level (myosin). It should be mentioned that in order for a muscle to generate force, it needs to be activated by the central nervous system that coordinates the action of trillions of actin and myosin proteins.

\begin{figure}[H]
    \captionsetup{width=1\linewidth,justification=justified, singlelinecheck=false}
    \centering
    \includegraphics[width=0.6\linewidth, height=6cm]{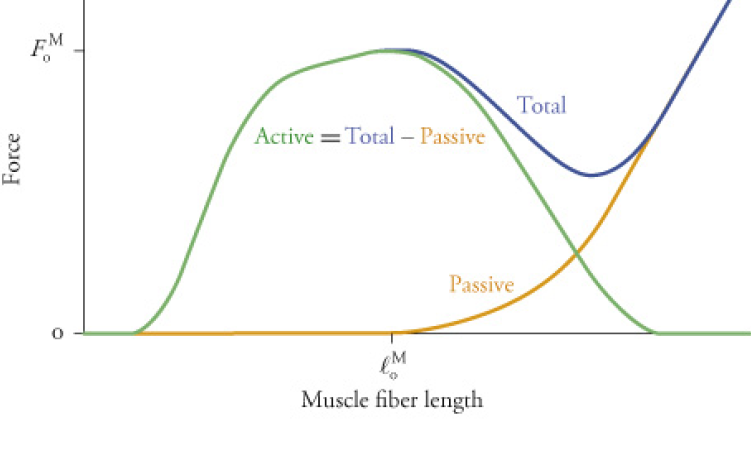}
    \caption{The active, passive and total force - length curves. The peak of active force $F_o^M$ happens at the optimal fiber length  $l_o^M$. Reproduced from \cite{Uchida2020}.}
    \label{fig:force_length_curve}
\end{figure}

The main parameters of a muscle are its optimal fiber length, the muscle fiber pennation angle at optimal fiber length, its maximum isometric force, its maximum contraction velocity and the tendon slack length. In \autoref{fig:force_length_curve} the force - length curve is presented describing the active, passive and total force. The optimal length $l_o^S$ is the length at which the sacromere develops its peak isometric force, $F_o^M$. This force is generated when a muscle is at its optimal fiber length and fully activated. In daily life activities, this parameter is tough to measure and our knowledge is based on cadaver and animal experiments. The fiber optimal length $l_o^M$ is accomplished when the sacromere reaches its optimal length and is equal to $l_o^M=nl_o^S$, if we assume equal length of all sacromeres, and \textit{n} is the number of sacomeres that form a muscle fiber. 

Moreover, the force - length curve is characterized by three different regions. The first is the ascending part where the force increases simultaneously with the length. Subsequently, the second region features an approximately constant force of maximum value (the plateau region) followed by a third region where the force decreases while the length of the sacromere is further increased (the descending region). The optimal range of sacromere length is detected in the plateau region. In this optimal range, the number of cross - bridges of actin and myosin are at a maximum. When muscles are stretched above the resting length, they still generate a passive force even if they are inactive (like a nonlinear spring). Overall, when a muscle is activated, the total force generated is the summation of active and passive forces.

\begin{figure}[H]
    \captionsetup{width=1\linewidth,justification=justified, singlelinecheck=false}
    \centering
    \includegraphics[width=0.6\linewidth, height=6cm]{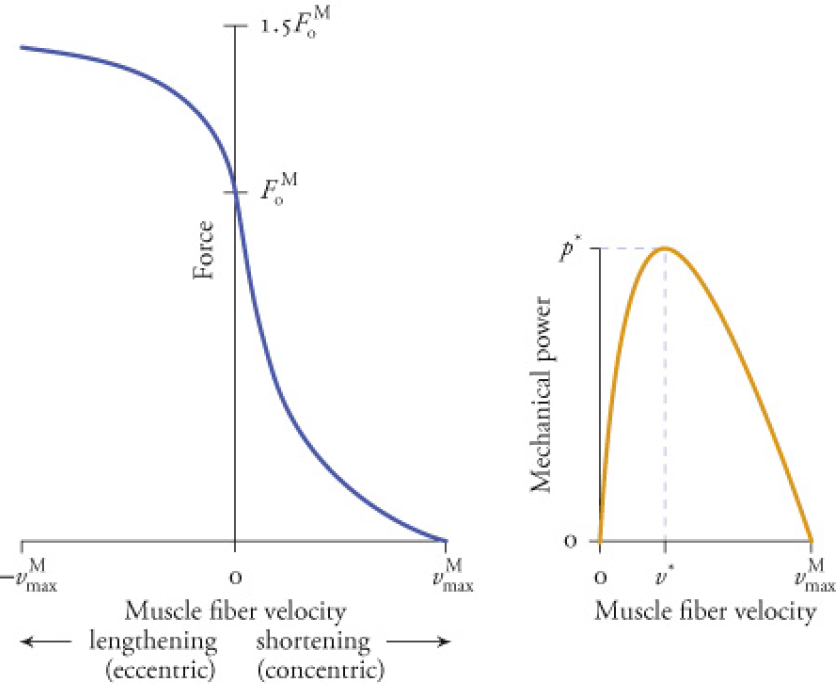}
    \caption{Muscle fiber force (left) and power (right) as functions of fiber velocity. Reproduced from \cite{Uchida2020}.}
    \label{fig:force_velocity_curve}
\end{figure}

The force that a muscle can generate also depends on the velocity that its length changes. This relationship is presented in \autoref{fig:force_velocity_curve}. Active force increases as fiber is lengthening and decreases as the fiber is shortening. For high shortening velocities no active force is developed while for high lengthening velocities the muscles can be damaged. In the right part of the figure we also present muscle mechanical power, which takes a maximum value at about 1/3 of maximum shortening velocity. 

\begin{figure}[H]
    \captionsetup{width=1\linewidth,justification=justified, singlelinecheck=false}
    \centering
    \includegraphics[width=1\linewidth, height=6cm]{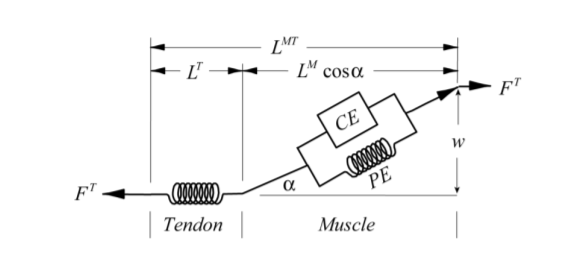}
    \caption{The Hill - type model of muscle - tendon complex. Reproduced from \cite{Anderson1999}.}
    \label{fig:hill_model}
\end{figure}

Study of muscle internal structure and functionality is hampered due to their complex nature. As we mentioned during the introduction, computational models can be developed to capture the features, properties and response of physiological systems. A widely adopted by research community muscle model, is the Hill - type muscle model. An overview of the model in the general case is presented in \autoref{fig:hill_model}. It is deployed in biomechanics studies to study muscle forces.

The main parameters of the Hill type muscle - tendon model are the following:

\begin{conditions*}
    \boldsymbol{CE} & Contractile Element\\
    \boldsymbol{PE} & Parallel Elastic Element\\
    \boldsymbol{L^{MT}} & Length\\
    \boldsymbol{L^M} & Muscle Fiber Length\\
    \boldsymbol{L^T} & Tendon Length\\
    \boldsymbol{α} & Pennation angle. Angle between the line of action of a muscle’s fibers and the line of action of its tendon
\end{conditions*}

The Hill - type model comprises of the following components: a tendon elastic element, an active contractile element and a passive elastic element. The last two components act in place of the active and passive force - generating properties of the muscle. The force that this muscle - tendon actuator can produce depends on the maximum isometric strength of the muscle and its corresponding pennation angle and fiber length, its maximum shortening velocity and tendons' resting length. 

\begin{figure}[H]
    \captionsetup{width=1\linewidth,justification=justified, singlelinecheck=false}
    \centering
    \includegraphics[width=0.6\linewidth, height=6cm]{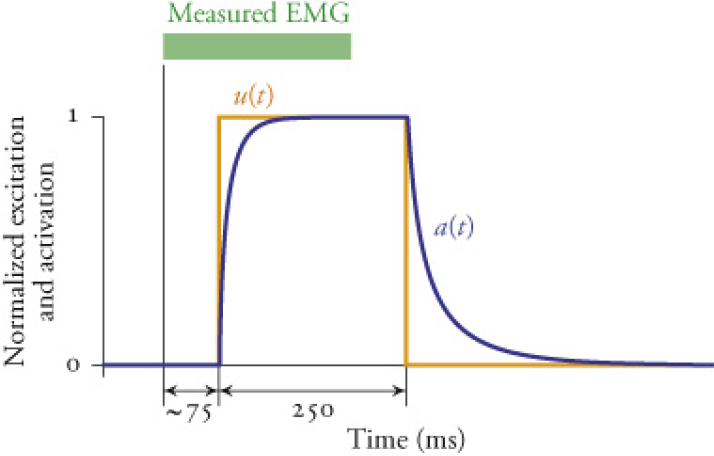}
    \caption{A computational model of activation dynamics that relates excitation (u(t)) to activation (a(t)).  Reproduced from \cite{Uchida2020}}
    \label{fig:activation_excitation}
\end{figure}

As mentioned previously, the muscles are excited by the central nervous system. In \autoref{fig:activation_excitation} a computational model of activation dynamics is presented where excitation \textit{u(t)} is related to activation \textit{a(t)}. This model relates the rate of change of muscle activation (i.e., the concentration of calcium ions within the muscle) to the muscle excitation (i.e., the firing of motor units). In the presented model, the activation can vary between 0 (no contraction) and 1 (full contraction). Furthermore, the excitation signal can vary continuously between 0 (no excitation) and 1 (full excitation).

Also, a muscle can neither relax nor generate force spontaneously. The production of force requires a complex chain of events. The delay between a motor unit action potential (the first step of these events) and the production of maximum force has been observed to vary from 5 to 50 milliseconds.

\section{ACL biomechanics, injury and treatment}

As already mentioned in \autoref{subsec:knee_joint}, \gls{acl} is one of the main ligaments that surround the knee joint. The \gls{acl} is placed at the midpoint of the knee. \gls{acl} functionality is fundamental as it prevents anterior sliding of the tibia relative to the femur, excessive tibial medial and lateral rotation and withstands varus and valgus stresses \cite{Petersen2007}. It is composed of two bundles, the antero - medial and postero - lateral bundle and extends from the lateral femoral intercondylar notch towards the tibial plateau and the intercondyloid eminence \cite{Duthon2006}. 

\begin{figure}[H]
    \captionsetup{width=1\linewidth,justification=justified, singlelinecheck=false}
    \centering
    \includegraphics[width=0.8\linewidth, height=6cm]{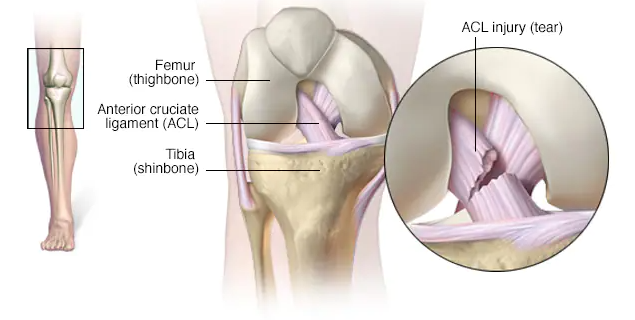}
    \caption{The anterior cruciate ligament (ACL) is one of the key ligaments that help stabilize the knee joint. The ACL connects the femur to the tibia. Reproduced from \cite{MayoClinic}}
    \label{fig:landing}
\end{figure}

\begin{figure}[H]
    \captionsetup{width=1\linewidth,justification=justified, singlelinecheck=false}
    \centering
    \includegraphics[width=1\linewidth, height=6cm]{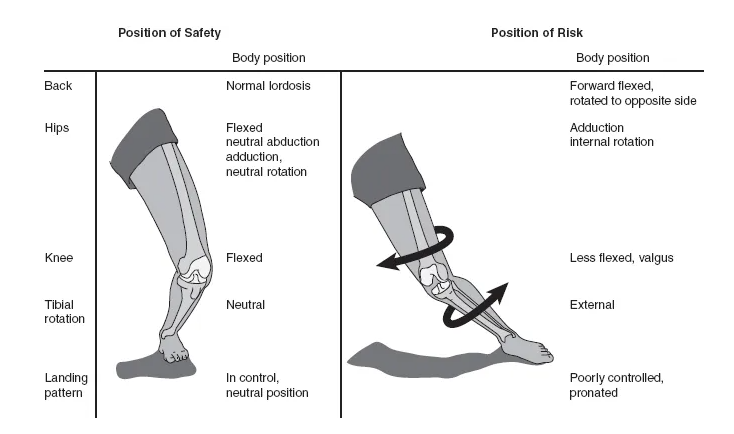}
    \caption{Mechanism of ACL rupture during change of direction. An increased injury is introduced with the foot planted, knee externally rotated or in valgus position and a varus or internal moment is applied. Reproduced from \cite{Brukner2016}.}
    \label{fig:acl_injury}
\end{figure}

\gls{acl} injuries are very common during dynamic movements when practising sport activities. These movements include sudden stops, abrupt changes of direction, landing after a jump and deceleration before changing direction \cite{Sugimoto2014,Boden2000}. Among the various types of injuries non - contact actions are the most significant cause of partially or completely \gls{acl} injury \cite{Marieswaran2018}. The risk factors for a non - contact ACL injury can be environmental, anatomical, hormonal and biomechanical – neuromuscular. From a biomechanical view, \gls{acl} can be injured from extreme anterior translation or by both valgus and internal rotation moments \autoref{fig:acl_injury} \cite{Georgoulis2010}. In the next section, significant factors that can contribute to an \gls{acl} injury when performing a landing after a jump motion are thoroughly presented.

Usually, rupture of the \gls{acl} require the total replacement of the injured tissue with grafts during a surgical procedure named as \gls{acl} reconstruction. Different techniques have been developed for the \gls{acl} reconstruction procedure, characterized by multiple parameters (like the graft material, the drilling location etc.). The regular workflow consists of  drilling sockets or tunnels into the tibia and femur,
collecting the graft from another anatomical structure (patellar tendon, hamstrings) and fixating it through the tunnels after replacing the damaged ligament. 



\subsection{Drop - Landing and potential ACL injury}

Landing is one of the most common movements associated with \gls{acl} injury. In \autoref{fig:landing} the motion of a double - leg landing is demonstrated. The landing phase is considered from foot strike to the time instant of maximum knee flexion angle. In \autoref{fig:landing} the landing phase begins from the drop and ends at the end of collision stage.

\begin{figure}[H]
    \captionsetup{width=1\linewidth,justification=justified, singlelinecheck=false}
    \centering
    \includegraphics[width=0.6\linewidth, height=5cm]{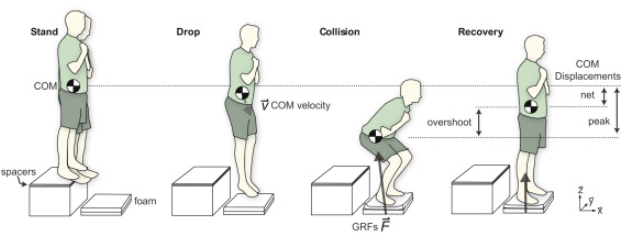}
    \caption{Representation of a drop - landing and displacement of the center of mass. The landing phase is considered from foot strike to the time instant of maximum knee flexion angle. The landing phase begins from the drop and ends at the end of collision stage. Reproduced from \cite{Skinner2015}}
    \label{fig:landing}
\end{figure}

 The correlation between \gls{acl} injury and the landing motion has been in the spotlight of multiple research studies. Initially, the height from which the landing is performed is a factor that affects potential \gls{acl} injury, as it is connected with the magnitude and direction of \gls{grf} \cite{Hewett2017,Cherry2018}. Furthermore, an association between \gls{acl} injury and landing is related to the body posture at time of ground contact. Most of the athletes, when they attempt to land in a natural way after a jump, they tend to land stiffly. This term refers to the limited flexion of hip, knee and ankle angles. The result of a stiff landing is that all impact is absorbed between ankle, knee, and hip joints. On the contrary, a soft landing lowers the impact forces, extends the life of joints and finally the injury risk is decreased \cite{Arnheim_book,Hewett2017,Cherry2018}. Additionally, flat - foot landings are considered of greater risk compared with natural and fore - foot landings \cite{Teng2020}. Furthermore, different energy dissipation strategies have been found for single - leg and double - leg landings. During single - leg landings the \gls{acl} injury risk is higher compared with double - leg landings \cite{Yeow2011}. 

Moreover, the alignment of the lower extremity has also been indicated by several researchers as a main risk factor. It has been reported by previous research studies that landings with lower knee flexion angle result on reduced absorption of the kinetic energy by the muscular system and on higher impact stress on other body tissues.  Nevertheless, Griffin et al. pointed out that in the kinematic chain, not only the knee is included, but also the hip, ankle and trunk should be taken into account \cite{Griffin2000}. The most common lower - limb posture associated with \gls{acl} injury are foot pronation, tibial internal rotation, and a valgus position of the knee joint \cite{Arnheim_book}. Knee valgus posture can be caused by rotations of the hip and knee joints \cite{Nguyen2015,Koga2017,Yasuda2016}. Also, combinations of knee joint internal rotation, knee joint abduction, knee joint adduction moment and hip joint internal rotation  can result in improper landings and \gls{acl} injury.  

Additionally, multiple research studies have examined the effect of different upper body postures on \gls{acl} injuries  \cite{Blackburn2008,Saito2020,Song2021,Jones2015,Donelon2020}. It has been observed that trunk flexion - extension, bending and rotation affect the kinematics of the joints of the lower body and the landing forces. 

Several studies have shown great correlations between tibio - femoral forces, \gls{grf} and muscle forces during landings \cite{Mokhtarzadeh2010, Mokhtarzadeh2012, Hernandez2013}. Along with the fact that the \gls{acl} restrain anterior tibial translation, the quadriceps and hamstrings muscles are also considered as a highly important pair of agonist - antagonist muscles related to \gls{acl} injuries because of their crucial role in anterior tibial translation. Additionally, at the peak of \gls{grf} during landing, absence or delay of muscle activation can cause \gls{acl} injury, while greater quadriceps/hamstrings ratio indicates greater anterior knee muscle force and increases the risk or ACL injury \cite{Mokhtarzadeh2012,Griffin2000}.

Other identified \gls{acl} injury risk factors are the axial tibial torque, the valgus moment and hyper    - extension moment. These factors can cause anterior tibial translation, valgus rotation or anterior tibial translation respectively and increase the risk of \gls{acl} excessive stretch \cite{Hernandez2013}. Also, the \gls{acl} strain can be increased by externally applied knee abduction  moments, anterior tibial shear force and knee internal rotation moments \cite{Shin2007,Shin2011,Bates2019,Bates2017}.
 

\section{Mathematics and physics concepts}

In this section we give a general description of multibody dynamics. Then, we introduce the trajectory optimization problem and outline the optimization methods adopted by the software tools deployed throughout the study workflow presented in this thesis.

\subsection{Multibody dynamics}\label{sec:multibody}

Mechanics studies the behavior of physical bodies subjected to displacements and forces \cite{Lindström2019}. It is distinguished in two branches, dynamics and statics. Statics study the forces and their effect on a system when it is in static equilibrium. On the contrary, dynamics examine bodies when in motion.

Dynamics are subdivided into kinematics and kinetics \cite{Mounts1970}. First, kinematics investigate the relative motion of the components of the system. They do not consider the forces that generate the motion, but rather provide knowledge about properties such as displacement, velocity, acceleration and their variation over time. On the contrary, kinetics examine the relation between the motion of the system and the cause of that motion (forces and torques).

In biomechanics, human musculoskeletal system is usually modeled as a multibody dynamics system, adopting many parts of the robotics theory \cite{Nordin2013,confluence}. These systems are represented as a series of rigid bodies called links, starting from a base link that is usually called ground and is considered fixed in space. Every link is added to the kinematic chain through connections that are called joints. These joints impose the constraints of the system, meaning that they permit only certain relative displacements (translation or rotation) between the rigid bodies they connect. Furthermore, force elements such as springs, actuators and dampers, are included along with any applied external force. This system model can also be mathematically described as the system of equations that model the physical response of a given system. In our case this response is the model's motion that is subject to the forces applied and the modeling assumptions we have adopted. A general description of the equations of motion for a multibody system is presented in \autoref{eq:eom}.

\begin{equation}\label{eq:eom}
    \tau = M (\theta) \ddot{\theta} + H(\theta, \dot{\theta})
\end{equation}

\begin{conditions*}
    \boldsymbol{\tau} & generalized forces and moments\\
    \boldsymbol{M} & symmetric positive definite mass matrix\\
    \boldsymbol{\theta} & generalized coordinates\\
    \boldsymbol{\dot{\theta}} & generalized velocities\\
    \boldsymbol{\ddot{\theta}} & generalized accelerations\\
    \boldsymbol{H} & centripedal, Coriolis, gravity, friction, external forces
\end{conditions*}


\subsection{Trajectory Optimization problems} 

Trajectory optimization problems refer to a set of mathematical methods that aim to produce an optimized trajectory for a dynamic system. This trajectory minimizes a given cost functional and has to satisfy the constraints of the system. A general scheme of this problem is displayed by the next formulas:\\
\newline
Optimal trajectory:
$$ {(x^*(t),u^*(t))} $$ 
System dynamics: 
$$ \dot{x} = f(t,x,u) $$ 
Constraints:  
$$ c_{min} < c(t,x,u) < c_{max} $$
Boundary Conditions:  
$$ b_{min} < b(t_0,x_0,t_f,x_f) < b_{max} $$ 
Cost Functional:  
$$ J=φ(t_0,x_0,t_f,x_f) + \int_{t_0}^{t_f} g(t,x,u) \,dt $$ 

where \textbf{x} refers to a state variable, a differentiated variable in the dynamics equation, and \textbf{u} refers to a control variable. 

The above description is a continuous problem that needs to be transformed into a non - linear one in order to be solved. This procedure is executed by transcription methods. Then, the problem can be solved by solvers like the IPOPT \cite{Wächter2006}. The new form of the problem is displayed by the next formula:

$$ \text{minimize}  \quad \Bar{J}(z), \quad z\in R^n $$ 
$$ \text{subject to}: \quad 1 \leq 
\begin{pmatrix}
    z\\
    \Bar{c}(z) \\
    Az
\end{pmatrix}  
\leq u $$

The transcription methods are divided in two main categories: the shooting methods and the simultaneous methods. Simultaneous methods enforce the dynamics at a series of points along the trajectory, while shooting use a simulation to explicitly enforce the system dynamics.

Also, the methods for trajectory optimization can be direct or indirect. Direct methods convert the trajectory optimization problem into a discrete nonlinear problem. Both shooting methods and collocation methods, that will be presented next, are direct methods.

\subsubsection{Shooting methods} 

Shooting methods are based on simulation and the discretization is executed using explicit integration schemes. Also these kind of methods performs better for problems with very simple controls and in absence of path constraints.

One of the simplest methods of transcribing an optimal control problem is the single - shooting method. If we examine the problem where a canon tries to hit a target, then we have two decision variables and one constraint. More specifically, the constraint is that the projectile has to hit the target and the decision variables are the mass of the powder and the firing angle. Also, the powder mass serves as the cost function. In general single - shooting methods make a guess for the decision variables and then assess the result. For our example, this method makes a guess about the mass of the powder and the angle. Based on these, the cannon fires the projectile and the method investigates whether the target was hit. If not, a new guess is created based on the previous result. This simulation is repeated until the target is finally hit.   

\begin{figure}[H]
    \centering
    \includegraphics[width=1.0\textwidth,height=8cm]{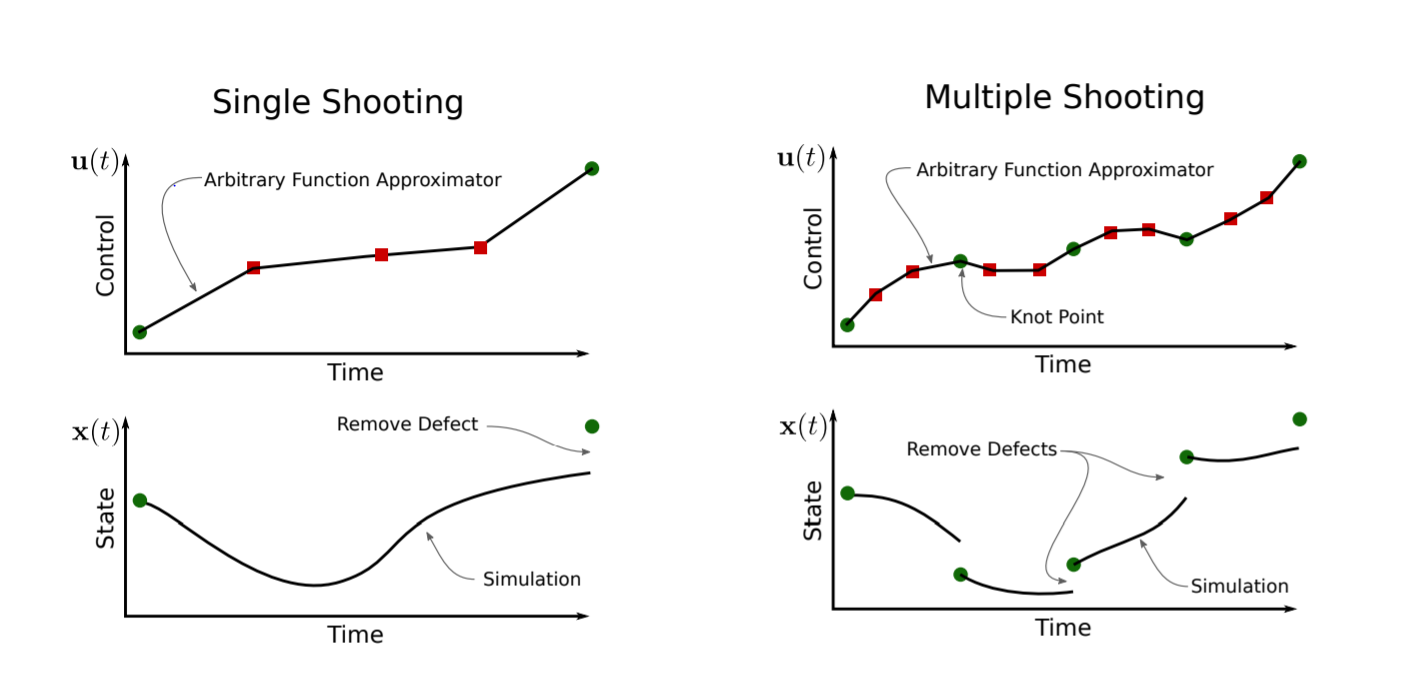}
    \caption{Single - Shooting and Multiple - Shooting methods. Reproduced from \cite{Kelly2017_1}.}
    \label{fig:shooting}
\end{figure}

On the other hand, in multiple - shooting the trajectory is split in segments and in each segment single - shooting is applied. These segments are not necessarily matched up. If we choose to solve for more segments, then the relation between the objective function and the decision variables becomes more linear. Next, in \autoref{fig:shooting} we display a comparison between single - shooting and multiple - shooting. We can notice that in multiple - shooting a constraint is added in order to create a continuous trajectory. 

In trajectory optimization problems, the dynamics constraints can be written in two different forms. The first form is the derivative method where the derivative of the state with respect to time should be equal to the dynamics function. The second method is the integral method where the state trajectory must be equal to the integral of the dynamics with respect to time. The shooting methods use the integral method. 

The formulas for the previously mentioned methods are presented next.

$$ \text{Derivative method: } \dot{x}= f(x,u) $$ 
$$ \text{Integral method: } x = \int f(x,u) \,dt $$ 

One of the main complications when using trajectory optimization methods is that they can become stuck in a local minimum in the space of trajectories. As a result, their response depends on an initial guess.It is extremely hard for these methods to detect a trajectory that is not homotopic to the initial trajectory and they cannot recover from an initial guess in a wrong homotopy class. For that reason, providing an appropriate initial guess to the solver can put it close to the "right" local minimum of the trajectory.

\subsubsection{Collocation methods}  

Collocation methods are based on function approximation. These methods use an implicit integrator to approximate system dynamics. Furthermore, they perform more efficiently in problems with complicated control and / or path constraints.

In direct collocation method the controls and states of the system are approximated as splines over the mesh of time points. The continuous trajectory optimization problem is discritized by approximating the continuous functions of the problem as polynomial splines. Subsequently, the optimizer solves for the knot points that lead the splines to comply with the dynamics of the system. The dynamics are enforced by demanding that at specified time points, the time derivative of the state splines match the derivative from the system differential equations. That way, a non linear program is created, where the states are the variables and the constraints that are enforced are the system dynamics. An important attribute of this method is that large problems can be tractable because the constraints at a given time point depend only on the variables near that specific time. 

An example of how direct collocation solves simple problems is presented next, adopted from \cite{Kelly2017}. Lets consider a model of a small block that consists of a point mass, that can translate only in one direction. If we try to simulate its movement between two points with initial and final states at rest in a defined amount of time, then the dynamics describe the way that the system moves. The input of the system (or the control) is the force applied on the point mass. \newline

\text{System dynamics}: \quad $\Dot{x}=v$, \quad $\Dot{v}=u$ \newline

where \textit{x} refers to position, \textit{v} to velocity and \textit{u} to control. \newline

\begin{figure}[H]
    \captionsetup{width=1\linewidth,justification=justified, singlelinecheck=false}
    \centering
    \includegraphics[width=0.8\linewidth, height=7cm]{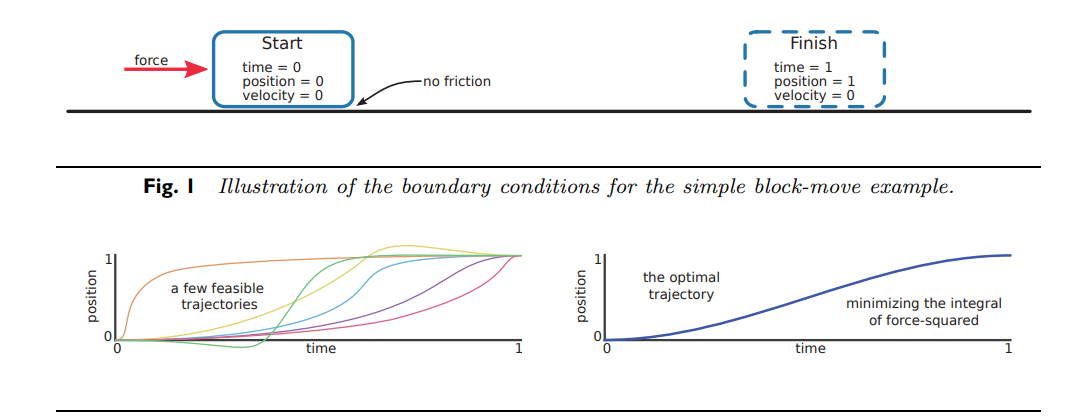}
    \caption{Feasible (left) and optimal (right) trajectories for a simple movement of a block using direct collocation method. Reproduced from \cite{Kelly2017}.}
    \label{fig:DirectCollocation_ex}
\end{figure}

Both initial and final position and velocity should be zero. The requirements concerning initial and final state along with the desire to move the point mass one unit of distance in one unit of time are presented next and schematically in \autoref{fig:DirectCollocation_ex}. \newline

\begin{equation}
    \text{Boundary conditions: } x(0)=0, \quad x(1)=0, \quad v(0)=0, \quad v(1)=0
\end{equation}

As seen in \autoref{fig:DirectCollocation_ex} (left), there are many feasible solutions to this problem. Trajectory optimization aims to find the best feasible trajectory or the optimal solution. The objective function describes the "best" solution based on an objective. Therefore, optimal trajectory is a feasible one that also minimizes the objective function. One example of an objective function can be the minimum of force squared which is provided by the next formula:

\begin{equation}
    \text{Minimum force squared: } \displaystyle \min_{u(t),x(t),v(t)} \int\limits_{0}^{1} u^2(τ)dτ
\end{equation}

The solution of this problem is given by:

\begin{equation}
    u^{\ast}(t)=6-12t , \quad x^{\ast}(t)=3t^2-2t^3  \newline
\end{equation}

Subsequently, we describe some basic concepts of ''trapezoidal'' and ''Hermite - Simpson'' collocation methods. The ''trapezoidal'' collocation method offers a finite set of decision variables. This is accomplished by expressing the position and velocity (which are continuous on time) by their values at exact time points. These time points are the collocation points. Also, the system dynamics needs to be converted into a set of constraints that can be applied for every collocation point at states and controls.

The ''trapezoidal'' rule is used in order to make the conversion of the system dynamics. The main idea is that the change in state between two points is equal to the integral of the system dynamics. Then, the integral is evaluated and the dynamics between all pairs of knot points is approximated. Also, the constraints and the boundary conditions are enforced on all segments and collocation points respectively. Finally, the objective function is estimated as a summation of each segment at the knot points using the control error. 

For the previous block example, we get a quadratic program (easier to solve compared to a nonlinear one), given that the constraints are linear and the objective quadratic. A  spline that interpolates the solution trajectory must be created given the set of positions, velocities and controls at the collocation points. In trapezoidal collocation, a linear spline is used for the control, while for the states a quadratic one is used.

The "Hermite - Simpson" collocation method is analogous to the trapezoidal but it offers a higher-order accurate solution. This is due to approximating the system dynamics and the objective function as quadratic functions and not linear. Also, the state trajectory is a cubic Hermite spline. To approximate the integrand, Simpson quadrature uses a quadratic segment. That segment fits three points uniformly aligned in space.


\section{Literature Review}

In this section, a review of research studies that follow an experimental and/or simulation approach regarding the identification of \gls{acl} injury risk factors during single - leg landings and subsequent cutting movements is conducted. Based on this review we will be able to present the methods of this study, the scientific improvements and limitations along with the contribution of this work. The order of presentation is based on each investigated parameter. Initially, a description of research studies that consider the correlation of initial landing height and \gls{acl} injury risk is conducted. Subsequently, we reference studies that explore lower limb static alignment and produced loads on the knee joint, following by studies that examine the relationship between trunk orientation and \gls{acl} injury. 

\subsection{Landing height}

Starting with the initial landing height, Mokhtarzadeh et al. conducted a study where single - leg and double - leg landings from different initial heights were performed by ten volunteers in a motion analysis lab. The researchers collected joint kinematics and \gls{grf} using six cameras and two force plates respectively. The acquired data were used in an inverse kinematics and dynamics analysis that utilized subject specific models through the OpenSim software. The authors reported the importance of the hamstring muscle due to its ability to posteriorly move the tibia bone and, therefore reducing the load on the \gls{acl}. Also, they stated that when the \gls{q/h} was greater than 1, the quadriceps caused the tibia to translate anteriorly which can increase the risk of \gls{acl} rupture. On the the other hand, when the ratio was less than 1, the increased hamstrings muscle group contribution caused a posterior tibia movement protecting the excessive and abrupt \gls{acl} extension. Furthermore, greater values for \gls{vgrf} were observed for landings initiated from greater heights. Regarding quadriceps and hamstrings muscle groups, greater forces were observed for landings from 60cm height at time of max \gls{vgrf} \cite{Mokhtarzadeh2010}. In a latter study, the contribution of ankle plantarflexors (gastrocnemius and soleus muscles) to the \gls{acl} load during single - leg landing was examined ~\cite{Mokhtarzadeh2013} . The parameter of interest was the set of knee joint reaction forces to predict the forces exerted on the \gls{acl}. The authors collected kinetic, kinematic, \gls{grf} and \gls{emg} data. Then, muscle forces were estimated using a processing procedure (\gls{ik}, \gls{rra}, \gls{so}) using again the OpenSim software platform. They observed that at peak \gls{grf}, quadriceps and soleus forces were greater as the initial height increased and that these two muscles acted as \gls{acl} agonist - antagonist. Finally, the forces of the muscles that act at the knee joint reached their maximum value after the \gls{grf} and \gls{acl} forces' peak.

In a similar study, Verniba et al. examined how drop height (22 and 44cm) and style of landing (natural, soft or stiff) affect knee compression force and external flexion moment. Towards this objective, they measured knee flexion angles and \gls{vgrf} during landings performed by 20 male participants. They observed that peak knee flexion angle was greater for soft landings and for higher heights. Also, they detected that \gls{vgrf} and knee joint compression were greater as height increased. The same was also true during stiff landings \cite{Verniba2017}.  Landing from different heights (32cm, 52cm and 72cm) was again the objective of an additional study \cite{Niu2018}. The authors observed that when the height was increased the vertical forces of all ankle, knee and hip joints were also incremented. Similarly, the forces of several muscles like rectus femoris, gluteus medius, vastii, biceps femoris and adductor magnus were also increased. The researchers also noticed that the iliopsoas, rectus femoris, gluteus minimus and soleus muscles were activated at an earlier time as the dropping height was increased, in contrast with the activation of tibialis anterior that exhibited a delayed activation. Finally, Heinrich et al. examined landings from different heights and compared the effect of landing height along with the effect of trunk leaning on \gls{acl} injuries. The simulation included a muscluloskeletal model of a skier and tracking experimental data of a jump landing. The results of the study showed that as the jump height and the backward lean were increased, the peak \gls{acl} force was also increased. The authors concluded that peak \gls{acl} force was sensitive to both variations of trunk leaning and landing height \cite{Heinrich2018}.

\subsection{Lower extremity alignment}

As it was previously mentioned, another risk factor of interest that has been associated with \gls{acl} injury is the static lower extremity alignment. This condition refers to the rotational alignment of the knee and hip joint during landings and that poses the knee in a valgus posture which is an \gls{acl} damage risk factor. Nguyen et al. evaluated combinations of hip and knee joint kinematics, kinetics and external moments during the landing phase of double - leg landings \cite{Nguyen2015}. The participants were separated in three groups: those with internally rotated hip and knee valgus posture, those with externally rotated knee and valgus knee posture and finally those with neutral posture. The authors observed that groups with internal  rotated hip experienced also greater knee valgus motion and smaller hip flexion moments compared to participants with neutral alignment. Also, participants with internal rotated hip had increased hip rotation moment that tend to place the knee joint in a valgus configuration.

Moreover, the association of ACL injury risk with stiff landings was the main interest of a study that included 171 female basketball and football players. The researchers observed that lower knee flexion and higher peak \gls{vgrf} during stiff landings are correlated with high risk of ACL injury \cite{Leppänen2016}. In a study conducted in 2018, Guy - Cherry et al. tested three different landing styles (stiff, self - selected and soft landings) in order to indicate which one leads to the most safe landing conditions. They used \gls{mocap} equipment with ten cameras to record the kinematics of the lower extremity and an \gls{emg} system to record muscle activity. It was observed that peak \gls{grf} were different for each style, with lower values for soft landings and higher for stiff landings \cite{Guy-Cherry2018}.

Koga et al. \cite{Koga2017}, attempted to describe the foot position along with hip and ankle kinematics based on ten videos of \gls{acl} injuries experienced by women athletes during basketball and baseball games. They observed that the hip was internally rotated during the interval between initial contact and the next forty milliseconds. The hip internal rotations during all injury cases ranged between \ang{18} and \ang{39}. The authors used a model - based image matching technique to reconstruct the \gls{3D} kinematics to estimate these ranges. It should be mentioned that in six out of ten cases a contact (no direct with the knee) with another player took place at the time of injury and three of the injuries occurred during single - leg landing, while the rest of the injuries took place during cutting movements.

In another study, Yasuda et al. investigated landings of handball players and observed that the \gls{acl} injured players demonstrated greater hip internal rotation than the uninjured players \cite{Yasuda2016}. Internal rotation dominance of the hip joint was observed in seven out of eight injured players. Their logistic regression analysis revealed that internal and external rotation and extension of the hip joint is correlated with injury of the \gls{acl}. These observations concerns the female participants, while for the male participants the results were not that obvious. Also, studies conducted about cutting or changing of direction motions identified hip internal rotation as a risk factor. Internal hip rotation translates the knee more medial to the \gls{grf} vector, thus increasing the moment arm of the force and intoducing a greater moment at the knee joint. This further results to an increase of the \gls{grf} and higher risk of injury \cite{Sigward2015}.

Peel et al. investigated lower extremity kinetics and kinematics when double - leg landing from 40cm with \ang{15} and \ang{30} of toe - in and toe - out conditions or self - selected \cite{Peel2020}. Their observations indicate that the toe - in landing position is connected with increased hip internal rotation, knee abduction, ankle inversion angles at \gls{igc} and peak knee \gls{irm}, knee \gls{abdm} and hip \gls{abdm}. This implies that toe - in landing position is of high risk concerning \gls{acl} tear. Landing with toe - out position was related to higher hip flexion and abduction, knee flexion and external rotation angles at \gls{igc} and peak hip \gls{em} and \gls{abdm}, knee \gls{addm}, ankle plantar - flexion and inversion moments. Also, the ankle inversion angle was reduced. These findings suggest that the toe - out landing position could reduce the risk of \gls{acl} injury. All these angles and moments were higher for the cases of \ang{30} compared with \ang{15}. Concerning the self - selected landings, it was observed that participants landed with a \ang{18} toe - out position. The authors concluded that toe - in position compared to self - selected was associated with increased risk.

\subsection{Trunk orientation}

In two studies conducted by Blackburn et al., the effect of trunk flexion - extension on the kinematics of the hip and knee joints and on the landing forces during drop landing tasks was examined \cite{Blackburn2008,Blackburn2009}. They used a \gls{mocap} system to record kinematic, kinetic and \gls{emg} data of a participant group. All participants were asked to perform two different landings. First, they attempted a natural landing (of their own preference) and subsequently to land with the upper body in a flexed position. These data were used in both of these studies. The authors observed that during the flexed landing the trunk was more flexed at both \gls{igc} and loading phase. Also, the knee and hip joints were more flexed. Moreover, they derived that the peak of \gls{pgrf}, \gls{vgrf} and the mean amplitude of the quadriceps \gls{emg} were all lower during the flexed landing. 

In a similar study conducted by Shimokochi et al., three different landing styles were evaluated \cite{Shimokochi2012}. These included a self - selected, one leaning forward and finally an upright position single - leg landing. The \gls{3D} biomechanics, muscle activities of the lower body and landing forces of these cases were recorded using motion analysis (3D electromagnetic tracking system), \gls{emg} and \gls{grf}. Regarding the upright landings, the authors observed greater peak \gls{vgrf} but lower values of medial and lateral gastrocnemius and lateral quadriceps activation. Also, lower values of plantar - flexion and hip extensor moments were observed, while the peak knee extensor moments were greater for the upright landings. Based on these observations they concluded that landing with an upright position can harm the \gls{acl}, while leaning forward can protect it.

In a review research conducted by Song et al., the trunk motion was associated with \gls{acl} loading variables by observing videos with \gls{acl} occurrences. First, the authors stated that limited trunk flexion was associated with increased \gls{acl} risk of injury. Then, increased lateral bending on both sides but mostly towards the injured leg and increased axial rotation towards the landing side were also associated with increased knee external abduction moment arm, which can potentially lead to a \gls{acl} injury. A similar study focused on trunk position during a single - leg landing \cite{Saito2020}. The following trunk orientations were assessed: neutral or preferred, in flexion, in extension and in right and left lateral flexion. Using a two - dimensional video analysis and \gls{emg} recording devices, the researchers recorded knee flexion, knee valgus angle values and activity of the rectus femoris and biceps femoris muscles. They stated that the knee valgus angle had greater value during the left lateral flexion trunk position (when leaning toward the opposite site of the landing side) compared to trunk right lateral flexion position. Both of these had greater values compared to the extension position. As conclusion, a trunk flexion can be less risky for \gls{acl} injury due to the higher knee flexion angle, lower knee valgus angle, lower muscle activity of the rectus femoris and higher muscle activity of the biceps femoris (hamstrings). Finally, regarding trunk bending, higher values of \gls{grf} were observed  when a contro - lateral lean of trunk in the opposite direction of travel was involved during \ang{90} cutting motions \cite{Jones2015}. 

\subsection{Knee joint surrounding muscles}

One additional factor of \gls{acl} injury during landing movement is the role of the knee surrounding  muscles. Morgan et al. conducted a dynamic simulation study with \gls{mocap} data in order to reveal the importance of the muscles spanning the knee joint in injuries during the weight acceptance phase of single - leg jump landings \cite{Morgan2014}. A model of 5 \gls{dofs} on the knee joint and an \gls{acl}, was used in OpenSim software. The authors observed greater maximum muscle forces for quadriceps, gastrocnemius and then hamstrings (in decreasing order). Two groups were created, high - risk  and low - risk based on the mean values of the shear \gls{acl} forces and joint compressive forces. It was observed that for the high - risk group (with greater mean \gls{acl} strain and force), the quadriceps and hamstrings yielded greater forces.

Mazumder et al. attempted to simulate the mechanism of \gls{acl} injuries using OpenSim software~\cite{Mazumder2019}. Their objective was to derive optimal solutions concerning muscles activation while landing in order to avoid injury. Data from 5 landing trials were collected. They used a musculoskeletal model that featured a knee joint with three \gls{dofs} and ligament modeling for their analysis. The authors pointed out that common causes of injury are increased abduction of the hip and knee joints and greater knee abduction loads, which are often caused by muscle fragility. They also evaluated and adjusted co - activation of quadriceps and gastrocnemius muscle groups through a controller. The optimal result (decreased \gls{acl} injury risk)  for  quadriceps / hamstrings activation ratio was equal to 0.3 whereas for gastrocnemius / tibialis anterior the activation ratio was equal to 0.2.

Heinrich eτ al. investigated landings of skiers and they observed that quadriceps force was also sensitive to trunk lean before \gls{igc}. Hamstrings force decreased with increased backward lean but the change was not significant \cite{Heinrich2018}.

Shin et al. \cite{Shin2007} developed a \gls{3D} subject - specific knee dynamic simulation model in order to test the impact of deceleration forces on the \gls{acl} strain during single - limb landing when running. The knee model was created based on MRI of a cadaver specimen. They demonstrated that the posterior deceleration force can protect the \gls{acl} by reducing the peak strain that the contraction of quadriceps muscles produce. Their findings indicated that a quadriceps force is a non likely mechanism for injury and other mechanisms like valgus rotation, or internal rotational can be more relevant. Internal knee extension moments, internal knee rotation moments and knee flexion moments can also be responsible for higher loads at the knee joint \cite{McBurnie2019}.

%% file: Chapters/Chapter3.tex

\chapter{Methods} 

\label{Chapter3} 

\begin{figure}[H]
    \centering
    \includegraphics[width=1\textwidth, height = 11cm]{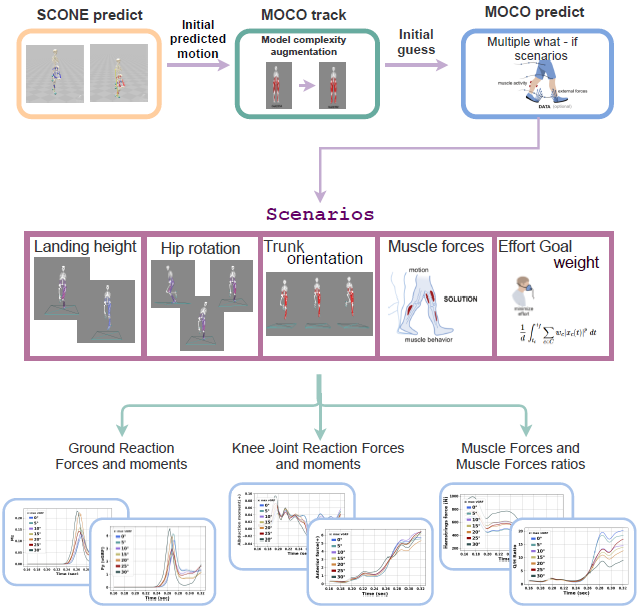}
    \caption{Overview of the proposed simulation pipeline of this thesis. In the first layer, the implemented methods and software tools used in this thesis are presented. Next, in the second layer all the examined scenarios are introduced and finally in the third layer, the exported results are presented.}
    \label{fig:pipeline}
\end{figure}

In this chapter we describe the methods used in this thesis. First, we present the software platforms and tools utilized to perform the simulations. These are the OpenSim software, the SCONE software and the MOCO tool. Also, a description of the musculoskeletal models is introduced along with the applied modifications. Then, we demonstrate the simulation setup for predicting a single - leg landing motion using the SCONE software. Finally, we describe the simulations conducted in MOCO in order to predict the motion of a single - leg landing using more complex models and the formulation of multiple what - if scenarios to evaluate \gls{acl} injury risk factors found in literature. All the steps of the simulation pipeline along with the examined scenarios and the exported results are presented in \autoref{fig:pipeline}.

\section{Dependencies}

Predictive simulation tools offer the opportunity to generate motion trajectories of a variety of tasks (gait, squat - to - stand, jump, landing) subject to high - level objectives without the need of experimental clinical data. Examples of objectives are solution stability, gait speed, energy efficiency, minimization of certain joint loads, pain avoidance or combination between them. That way, researchers can compose "what - if" scenarios and optimize them. In this section we present the software and tools used in this thesis and allow for simulations of biological motions. These are the OpenSim and SCONE software packages and the MOCO tool that is part of the OpenSim software.

\subsection{The OpenSim software}

OpenSim \cite{OpenSim_sortware} is a freely available software package that allows researchers to build and analyze musculoskeletal models and dynamic simulations of movements. Furthermore, it provides a user - friendly graphical user interface (GUI) allowing to use multiple tools of musculoskeletal analysis. OpenSim gives the opportunity to study a given motion in a high level, analyzing the joint angles, the joint torques and the external, muscle or joint reaction forces that are involved. In general, it requires data that are generated by human motion experimentally and are collected with a \gls{mocap} system. The availability of predictive simulation tools is an alternative source of motion data that can be used as an input into the standard OpenSim pipeline.

\begin{figure}[H]
    \centering
    \includegraphics[width=0.8\textwidth, height = 6cm]{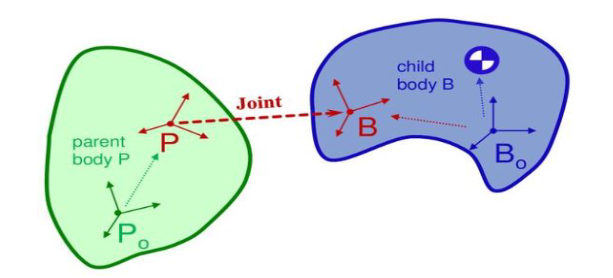}
    \caption{A joint connects two rigid body reference frames and is used to define the kinematic relationship between the rigid bodies. Reproduced from
    \cite{confluence}.}
    \label{fig:frames}
\end{figure}

The key block of the OpenSim workflow  is the OpenSim model. OpenSim models consist of multiple rigid bodies that are connected through joints. The joints define a parent - child kinematic relationship between the reference frames that are fixed to each body. A joint is described by its spatial transform. For a 6 \gls{dof} joint, three rotations and three translations comprise a joints' spatial transform. The transform describes the spatial position of body B in parent body P as a function of the joint coordinates as presented in \autoref{fig:frames}. The model’s posture at each time instant is described by the coordinates of all the model joints, also mentioned as the generalized coordinates. Additionally, point, weld or coordinate coupling constraints can be used to enforce constraints to the model. Point constraints restrain relative translations between two bodies. The weld constraint fixes the relative location and orientation of the bodies. Finally, coupling constraints links joint coordinates to any other model coordinate. Moreover, forces are used to actuate the motion of the model. These forces can be produced by either passive elements like springs, dampers and contacts or by active generators like torque actuators and muscles. As we see the OpenSim models encapsulate all key features of multi - body systems dynamics described in \autoref{sec:multibody}

Subsequently, the basic tools of OpenSim are concisely described, which are Scaling, \gls{ik}, \gls{so}, Analyze tool and \gls{jra}. In this study, the Analyze tool and \gls{jra} tool were mainly used to extract the results of the prediction simulations.

Firstly, the Scale tool allows to modify a generic musculoskeletal model, so it can accurately reflect a real subject based on its anthropometric properties. Usually, this is achieved by measuring the distance between experimental marker data placed on the subject during a static trial and virtual marker data placed on the model. Specific scaling weights are applied based on the confidence we have on our experimental data. At the end of the scaling process, each model segment is scaled accordingly. Moreover, the mass parameters (mass, inertia tensor) and the other components attached to the body segments (such as muscle actuators) are also scaled.

Next, the \gls{ik} tool attempts to find the posture of the model that closely resembles the acquired motion at each time frame. This is performed by identifying the pose for which the weighted squared errors between experimental data and the coordinates estimated by the \gls{ik} tool are minimized at each time frame. 

Furthermore, the \gls{so} tool, uses as input the movement extracted by the \gls{ik} tool and gives as output the unknown generalized forces based on the equations of motion. Therefore, it analyzes the joint moments into the forces that produce the motion at every time frame. 

Moreover, the Analyze tool enables to analyze a model or simulation based on multiple inputs. These inputs may be model states, controls, and external loads applied to the model. 

Finally, the \gls{jra} tool calculates the forces and moments at joints based on the loads acting on the model. These forces and moments account for the internal loads of the joint and are transferred between consecutive bodies. They should not be interpreted as the generalized forces estimated by the \gls{so} tool that act along or about the joint axes. The \gls{jra} loads act on the center of the joint and can be expressed on parent, child or ground frame. These loads represent the summation of all contributions for each joint structure (including the non - modeled structures like ligaments and cartilage contact) that would produce the desired joint kinematics.

\subsubsection{OpenSim models}\label{susec:opensim_models}

In this section, we present the OpenSim models that were deployed in our simulations. We utilized three different models. The first one, was a simple model in terms of total muscles and \gls{dofs} and it was used to predict the motion in SCONE. Then, for the simulations in MOCO we used two musculoskeletal models of augmented complexity. 

\begin{figure}[H]
    \centering
    \includegraphics[width=\textwidth, height = 6cm]{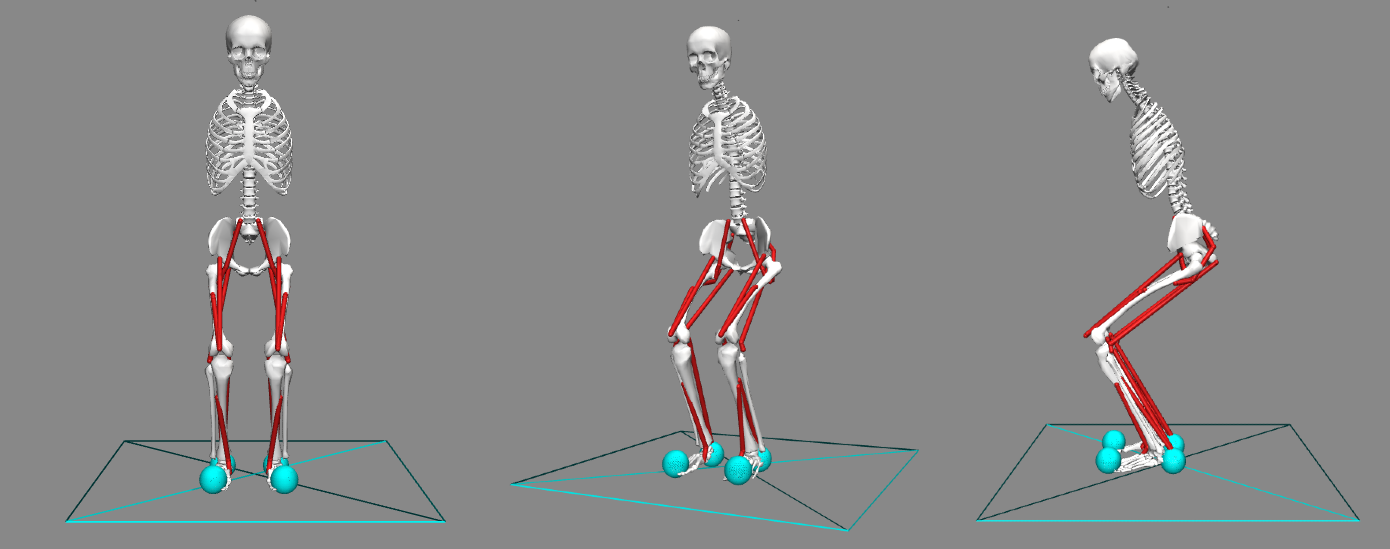}
    \caption{The musculoskeletal model developed by Ajay Seth used for the predictive simulation of single - leg landing in SCONE. It has 9 \gls{dofs} and contact spheres to estimate \gls{grf}.}
    \label{fig:human0914}
\end{figure}

The musculoskeletal model adopted for the simulations conducted in SCO\-NE was "Human0916". It is a 2D planar reduced gait model with 9 \gls{dofs} and 16 muscles created by Ajay Seth, closely based on an older model by Delp \cite{Delp1990}. The model resembles a subject of 1.8m height and 75.16kg weight. The movement of the model is constrained in the sagittal plane. Also, it consists of a planar joint with 3 \gls{dofs} between the pelvis and the ground and 6 more \gls{dofs}, 3 at each lower limb, one at the ankle, one at the knee and one at the hip. Furthermore, the model contains two spheres at each foot to establish a contact model in order to estimate the \gls{grf} when the feet touch the ground. These contact spheres have a radius of 4cm and are attached at the anatomical reference of the right and left calcaneus and toes. Moreover, the model contains eight muscle - tendon units of Hill - type at each leg: biarticular hamstrings (HAMS), gluteus maximus (GMAX), iliopsoas (ILPSO), rectus femoris (RF), vasti (VAS), gastrocnemius (GAS), soleus (SOL), and tibialis anterior (TA). The musculoskeletal model along with the contact spheres is presented in \autoref{fig:human0914}. The simplicity of this model offered us the ability to perform quick simulations to obtain initial guesses for the subsequent MOCO predictions.

The musculoskeletal models used for the simulations conducted in Moco were "Gait2354" and "Gait2392". These are \gls{3D} models of the human body with 23 \gls{dof} created by Darryl Thelen (University of Wisconsin - Madison), Ajay Seth, Frank C. Anderson, and Scott L. Delp (Stanford University).

\begin{figure}[H]
    \centering
    \includegraphics[width=0.8\textwidth, height = 10cm]{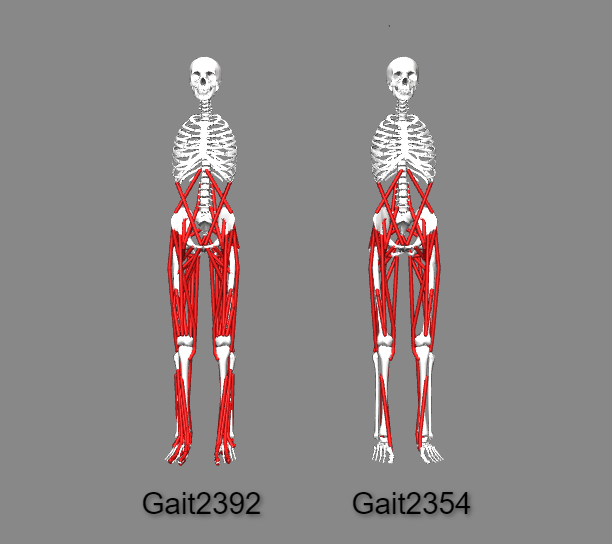}
    \caption{"Gait2392" and "Gait2354" OpenSim musculoskeletal models used for the simulations conducted in Moco. Both models were developed by Darryl Thelen, Ajay Seth, Frank C. Anderson, and Scott L. Delp, used for the predictive simulation of single-leg landing in Moco.}
    \label{fig:Gait2354-2392}
\end{figure}

\begin{figure}[H]
    \centering
    \includegraphics[width=\textwidth, height = 10cm]{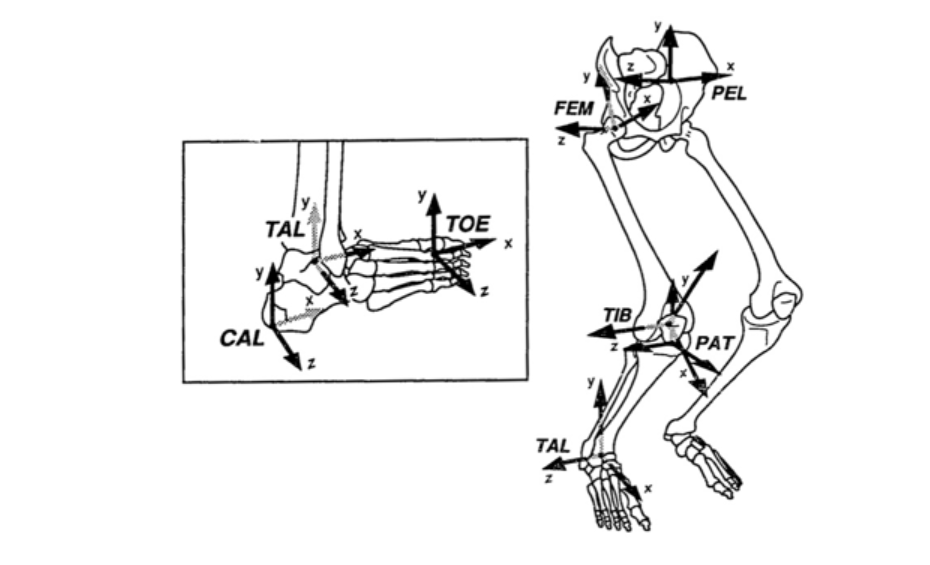}
    \caption{Location of the body - segmental reference frames. Reproduced from \cite{Delp1990}}
    \label{fig:reference_frames}
\end{figure}

The "Gait2392" musculoskeletal model consists of 92 musculotendon actuators in place of 76 muscles of the two lower limbs and torso. In "Gait2354" model the number of muscles is reduced in order to increase speed of simulations and simplify it for educational purposes. Also, Anderson took out the patella from "Gait2354" to avoid kinematic constraints. Both models represent a subject with 75.16kg mass and 1.8m of height. The models described previously are presented in \autoref{fig:Gait2354-2392}. Both models feature lower limb joint definitions from \cite{Delp1990}, low back joint and anthropometry adopted from Anderson and Pandy (1999), and a planar knee model adopted from Yamaguchi and Zajac (1989). There are 7 body segments for each leg: toes, foot, talus, tibia, patella, femur and pelvis. Each segment has each own reference frame as shown in \autoref{fig:reference_frames}.

The reference frame of the pelvis is located at the middle of the line connecting the two anterior superior iliac spines. In the neutral position, the reference frame of the pelvis and the ground frame are aligned, meaning that model's pelvic tilt is zero, with respect to the ground. Next, the femoral frame is located at the center of the femoral head, while the reference frame of the tibia is fixed at the center of the line between the medial and lateral femoral epicondyles. Also, the reference frame of the toe is fixed at the base of the second metatarsal. Moreover, the calcaneal frame is fixed at the most interior lateral point on the posterior surface of the calcanus. Finally the patellar frame is fixed at the most distal point of the patella, while the reference frame of talus is located at the center of the line between the apices of the medial and lateral malleoli. 

\begin{figure}[H]
    \centering
    \includegraphics[width=0.8\textwidth, height = 8cm]{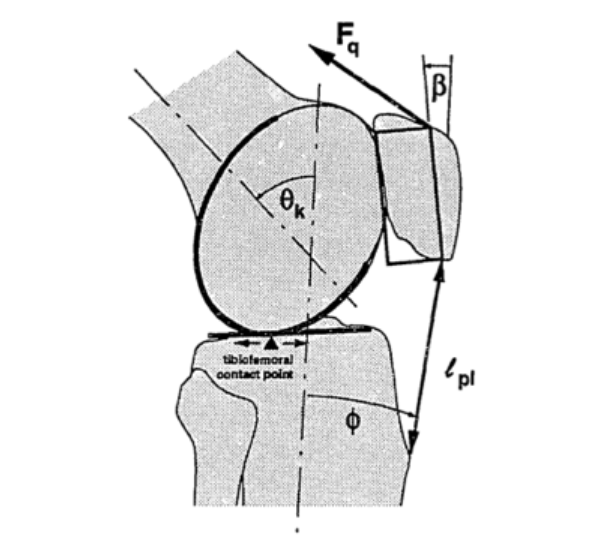}
    \caption{Geometry of the knee. Reproduced from \cite{Delp1990}.}
    \label{fig:knee}
\end{figure}

 The hip joint is modeled as a ball - and - socket joint, where transformations between the femoral reference frame and the pelvic reference frame allow rotations of the femoral frame about the 3 axes located in the femoral head. Next, the ankle, subtalar, and metatarsophalangeal joints are modeled as frictionless revolute joints. Regarding the knee joint as already mentioned it consists of 3 bones and it features a very complicated motion. For that reason Yamaguchi et al. provided a 1\gls{dof} model of the knee to represent the kinematics of not only the tibiofemoral joint, but also the patellofemoral joint. The tibiofemoral contact point depends on the knee angle and is specified according to data reported by Nisell et al. (1986). The geometry of the knee as described previously is presented in \autoref{fig:knee}.

Concerning the muscle - tendon units, each muscle - tendon path is described by line segments. Muscles can wrap over bones or can be constrained by retinacula. In OpenSim, intermediate via points can be  used to describe the muscle path more accurately and model their real - life behavior \cite{confluence}. 
 
\begin{figure}[H]
    \centering
    \includegraphics[width=0.6\textwidth, height = 10cm]{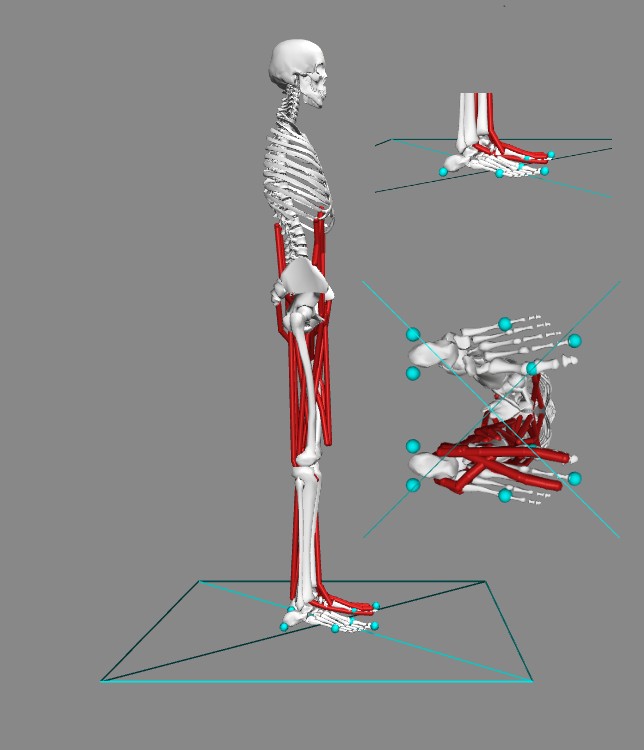}
    \caption{Schematic diagram of the musculoskeletal model with only the left side muscles. Foot - ground contact was modeled with five contact spheres per foot.}
    \label{fig:spheres}
\end{figure}

Foot - ground interaction for the "Gait2392" was modeled using compliant contact force model similar to "HuntCrossleyForce" by Hunt – Crossley \cite{Serrancolí2019}, except that this model applies force even when not in contact. Five spheres were placed under the foot of the model to stimulate the interaction of the foot and the ground adopted by \cite{confluence_skyhigh}. One sphere was placed under the toes and four under the hind - foot (\autoref{fig:spheres}). For static and dynamic friction coefficients a value of 0.9 was selected. Also, in order to speed up and simplify the simulations, all muscles of the right lower limb were deleted from the model, since we are interested only in studying left - leg landing.

\subsection{The SCONE software} \label{sec:scone_info}

SCONE \cite{scone_software} is an optimization and control software framework for predictive simulations that builds on OpenSim. It is an open source software that allows performing predictive simulations of biological motion through a user - friendly environment. Also, it allows users with a Lua scripting interface. In SCONE, simulations are performed through scenarios. 

Each scenario is composed of several building blocks. First, it contains a model of a human, an animal or a robot. Also, it includes the objective that defines the goals that the user wants to optimize (this can be a weighted combination of different goals). Moreover, the controller that generates the input for the actuators or muscles of the model is included in the scenario. Last but not least, it consists of the optimizer that takes the free parameters of the problem and optimize them for the objective chosen by the user. 

The models used in SCONE can be \gls{2D} or \gls{3D} musculoskeletal or other type OpenSim models. Nevertheless, there is a requirement that the model contains torque - driven actuators or muscle - tendon units. Also, the model needs collision shapes and contact models to simulate contact reaction forces and limit forces for simulating forces generated by bones and tendons when joints reach their limits. 

Concerning the controller, there are two already implemented controllers. First, the feed - forward controller that creates fixed patterns based on a function. Also, the feedback controller that is based on sensor information and generates inputs for actuators. In this thesis, we adopted the reflex controller proposed by Ong et al. \cite{Geijtenbeek2019}. This controller was purposed for the gait motion. The function of the controller was based on the positive feedback from the Golgi tendon organs, and on the length and velocity feedbacks. Moreover a Proportional Derivative (PD) helped to maintain body balance by adjusting hip muscles to limit trunk's lean \cite{Russo2020}.
Next, we provide the equations for the simulation of muscles:

Feed-forward simulation:

\begin{equation}
    \label{u_c}
    u_c = k_c
\end{equation}

Length feedback:

\begin{equation}
    \label{u_l}
    u_L =k_L\cdot\max(0,(\tilde{l}(t-t_d)-l_0)
\end{equation}

Velocity feedback:

\begin{equation}
    \label{u_v}
    u_V = k_V\cdot\max(0,(\tilde{v}(t-t_d))
\end{equation}

Force feedback:

\begin{equation}
    \label{u_f}
    u_f = k_f\cdot\tilde{f}(t-t_d)
\end{equation}

where $k_c$ is a constant, $k_L$,$k_V$ and $k_F$ are the gains of the reflex controller and $l_0$ is the threshold value for the muscle length and is defined as the length offset of the stretch response. \\

PD balance controller:

\begin{equation}
    \label{u_pd}
    u_{PD} = k_p(θ(t-t_d)-θ_0)+k_v\cdot{θ}(t-t_{d})
\end{equation}

where $k_p$, $k_V$ are the proportional and derivative gains of the controller, and $θ_0$ the desired forward lean angle. \\

In all equations, $k_p$ is the time delay depending on the muscle proximity to the vertebral column. Also, the muscle length $l$, contraction velocity $v$ and force $f$ are all normalized to optimal length ($l_opt$), and maximum isometric force ($f_{max}$) as described by:

\begin{equation}
    \label{norm}
    \tilde{l} = \frac{l}{l_{opt}}, \quad \tilde{v} = \frac{v}{l_{opt}}, \quad \tilde{f} = \frac{f}{f_{max}},
\end{equation}

The simulation that the controller provides is associated to muscle activation through first-order dynamics:

\begin{equation}
    \label{norm_2}
    \frac{\mathrm d α}{\mathrm d t} = \frac{u-α}{τ}
\end{equation}

where $α$, $u$, $τ$ are the muscle activation, muscle stimulation and dynamic time constant (with  value of 0.01s) accordingly. \\

Finally, he optimizer identifies the parameters in order to minimize or maximize the objective function by executing the simulation multiple times and each time adjusting the parameters based on the result. The optimizer repeats until the number of maximum iterations is met or there is no more improvement of the objective function evaluation. The Covariance Matrix Adaptation Evolutionary Strategy (CMA-ES) method is used to optimize the objective of each scenario. The parameters of the optimizer are the samples per iteration and the maximum number of iterations.

\subsubsection{Objective functions}

The objective function in a predictive simulation can be the \gls{com} position, the energy consumption, the gait speed or weighted combination of them. The goal is to create a motion that optimally performs that task. The parameters that are being optimized are the reflex gains and offsets.
The basic components of a cost function used in SCONE are the following:

\begin{itemize}
    \item \textbf{A parameter that penalizes the model when falling based on the termination height:} This parameter is defined as the ratio between the \gls{com} height to the initial state: 
            \begin{equation}
            \label{stability_penalty}
            \frac{COM_{height}}{{initial COM}_{height}} < th
        \end{equation}
    The penalty value is give by the next formula: 
        \begin{equation}
            \label{stability_penalty_2}
            p_{stability} = w_{stability}\cdot(\frac{time_{max}-time_{sim}}{time_{max}})
        \end{equation}
        where $th$ is the threshold for height, $time_{max}$ is the maximum duration of the simulation, $time_{min}$ is the time after which the simulation is terminated and $w_{stability}$ the weight of that penalty.
    
    \item \textbf{A penalty that minimizes the effort:}The metabolic model of Umberger \cite{Umberger2003,Umberger2010}, with updates from Uchida is used in order to calculate the effort \cite{Uchida2016} as described by the next formulas:
    
        \begin{equation}
            \label{effort_penalty}
            p_{effort} = w_{effort}\cdot effort
        \end{equation}
        
        \begin{equation}
            \label{effort}
            effort = \frac{basalEnergy + \sum_{n=1}^{N_{muscles}} (totHeatRate_{i} + mechWorkRate_{i})\cdot mass_i }{distance\cdot mass_{model}}
        \end{equation}
        where $totHeatRate$ is the total heat rate and $mechWorkRate$ the mechanical work rate. 
    
    \item \textbf{Penalties for exceeding desired joint ranges}. This penalty evaluates the differences between the desired angles and the actual ones.
    
        \begin{equation}
            \label{p_joints}
            p_{joints} = w_{joints} \cdot |{jointsRanges_{model}-jointsRanges_{}target}|
       \end{equation}
        
        where "model" refers to values recorded in simulation and "target" refers to desired ranges.

\end{itemize}

The total cost function is given by the summation of all components of the objective function:
\begin{equation}
    \label{p_joints_2}
    costFunction = p_{stability} + p_{effort} + p_{joints}
\end{equation}

\subsection{The MOCO tool}

\begin{figure}[H]
    \captionsetup{width=1\linewidth,justification=justified, singlelinecheck=false}
    \centering
    \includegraphics[width=0.9\linewidth, height=10cm]{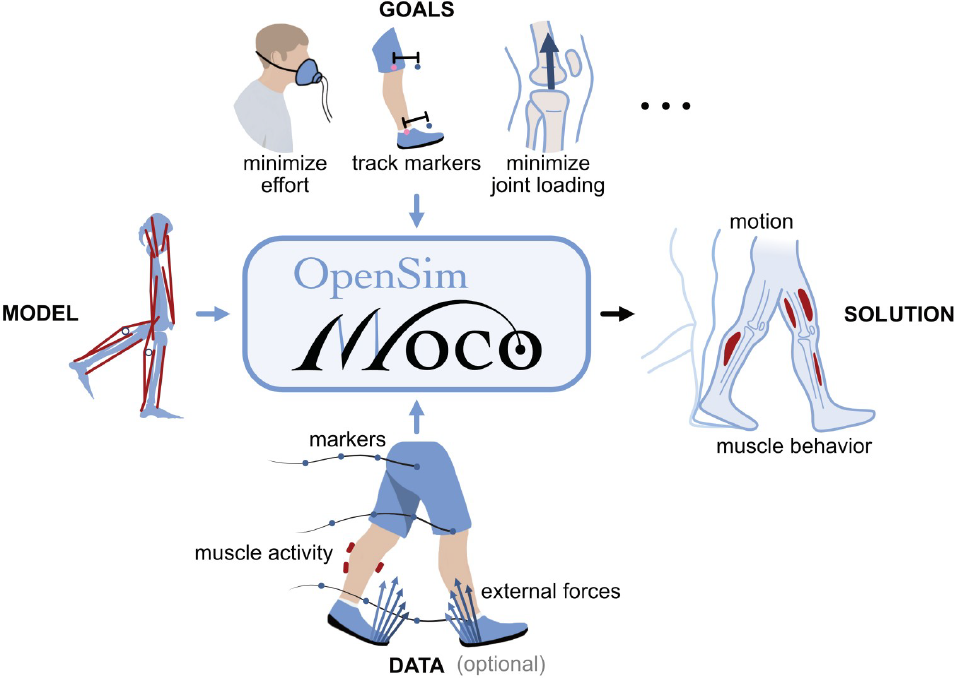}
    \caption{Overview of OpenSim Moco. It produces the optimal motion and muscle behavior for an OpenSim musculoskeletal model, given goals to achieve during the motion and/or reference data. Reproduced from \cite{Dembia2020}.}
    \label{fig:moco_1}
\end{figure}

OpenSim Moco is a software toolkit for optimizing the motion and control of musculoskeletal systems \cite{MOCO}. It uses the direct collocation method in order to solve quickly problems of trajectory optimization. Users can use Moco to solve not only motion tracking problems, but also parameter optimization and motion prediction problems, model fitting, electromyography-driven simulations, and simulations for device design using any OpenSim model and already implemented cost functions. The core library of Moco is writen in C++, but users can also work with Python, Matlab and XML interfaces \cite{Dembia2020}. An overview of the tool is presented in \autoref{fig:moco_1}.

Moco is built as a compilation of classes. The user can define the problem within the "MocoProblem" class and use the "MocoSolver" class to solve it. These two classes are nested into the "MocoStudy" class. Also, there is the "MocoSolution" class, which allows the user to plot and visualize the solution of the study created. The organization of the cornerstone Moco classes is presented in \autoref{fig:moco_2}.

\begin{figure}[H]
    \captionsetup{width=1\linewidth,justification=justified, singlelinecheck=false}
    \centering
    \includegraphics[width=0.9\linewidth, height=10cm]{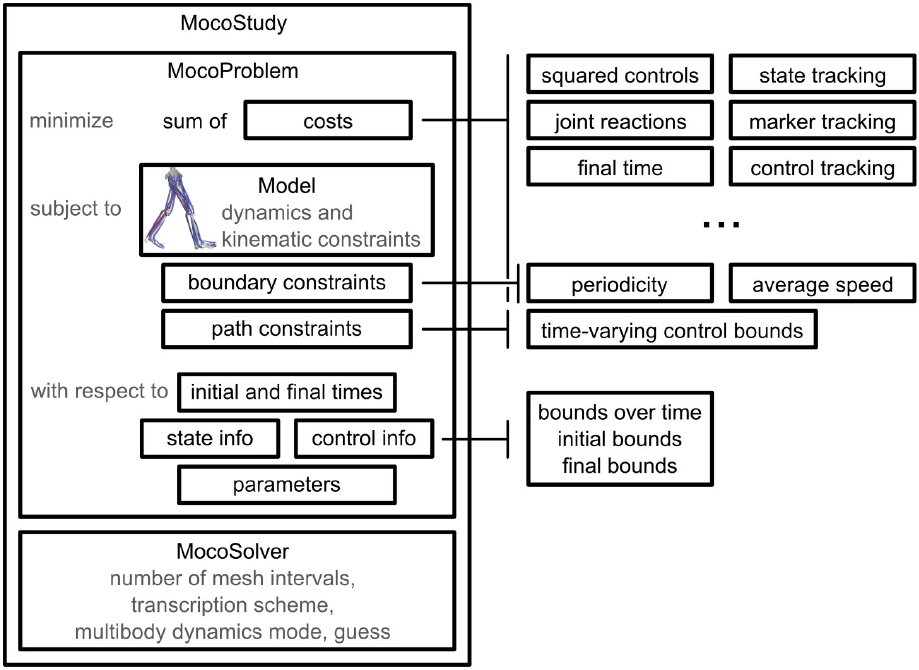}
    \caption{Organization of basic Moco classes. Overview of MocoStudy. Reproduced from \cite{Dembia2020}.}
    \label{fig:moco_2}
\end{figure}

A "MocoProblem" is a list of "MocoPhases". Nevertheless, Moco supports only single - phase problems. The "MocoProblem" class contains the following elements (as shown on \autoref{fig:moco_2}):
\begin{itemize}
    \item \textbf{cost terms}: Users can choose from multiple costs (duration of motion, deviation from an observed motion, effort, joint reaction loads, etc.) and the weighted sum of the squared controls is being minimized. All goals can be found in the "MocoGoal" class.
    \item \textbf{multibody dynamics, muscle dynamics, kinematic constraints}: Moco uses the musculoskeletal models from OpenSim to obtain multibody dynamics, muscle dynamics and kinematic constraints. The equations are provided by  Simbody multibody dynamics library \cite{SimBody}.
    \item \textbf{boundary constraints}: Users can enforce constraints concerning the initial and final state in order to impose preriodicity, average speed or symmetry.
    \item \textbf{path constraints}: A function of time (for example for  muscle excitation) can be constrained in specified range using the "Moco Control Bound Constraint" class. 
    \item \textbf{parameter optimization}: Any property of the model can be optimized.
    \item \textbf{bounds on variables}: Values of initial and final time, states and controls can be bounded. 
\end{itemize}

Moco includes a continuous and differentiable muscle model named "DeGrooteFregly2016Muscle" \cite{DeGroote2016} sutiable for the optimization method. Moreover "SmoothSphereHalfSpaceForce" contact model is used to model compliant contact forces \cite{Serrancolí2019}.

\subsubsection{Objective functions}

Moco provides a library of implemented goals like minimizing effort, joint loading, or deviation from marker data. The "Moco Goals" used in this thesis were the "MocoStateTrackingGoal" and the "MocoControlGoal".

"MocoStateTrackingGoal" is used for tracking joint angles, activations etc. This goal is translated as the squared difference between a reference state variable value and a state variable value, summed over the state variables for which a reference is provided, and integrated over the phase. 

"MocoControlGoal" is used for minimizing the sum of the absolute value of the controls raised to an exponent (greater or equal to 2, default value is 2) over the phase. In our case squared muscle activation is minimized. The goal is computed by the following formula:

\begin{equation}
    \label{effort_goal}
    \frac{1}{d} \int\limits_{t_i}^{t_f} \sum_{c \in C}^{} w_c |x_c(t)|^p dt
\end{equation}

where: \newline
$d$: the displacement of the system, \newline
$C$: the set of the control signals,  \newline
$w_c$: the weight for the control C, \newline
$x_c$: the control signal c, and  \newline
$p$: the exponent.

\section{Predictive simulation of landing with SCONE}

In this section we demonstrate the simulation setup in SCONE. This simulation was conducted in order to acquire a motion of a single - leg landing and use it as initial guess for the upcoming Moco simulations.

\subsection{Simulation setup}\label{sec:Scone_pipeline}

First the model was imported to OpenSim to apply an initial configuration and define the desired initial and final states of the motion. The initial state of the model for landing motion was defined by assigning specific values to the available \gls{dofs}. The initial height was set to 30cm height. To achieve this, the \gls{dof} corresponding to the pelvis vertical translation was set to 1.25.

Pelvis is connected with the ground with a floating joint of 6 \gls{dofs} with an initial translation value of 0.95 m along the vertical axis. The difference between this value and the one we imposed is the desired 30 cm of initial height. Subsequently, we modified the \gls{dofs} of both lower - limbs to achieve a single - leg landing with the left foot. The left hip flexion angle was set to \ang{5}, the left knee flexion angle was set to \ang{12} and the left ankle angle was set to \ang{34}. Furthermore, the \gls{dofs} of the right - limb were locked to specific values to prevent it from touching the ground at contact phase. The right hip flexion was locked at \ang{25}, the right knee angle was locked at \ang{120} and the right ankle angle at \ang{120}. Moreover, all \gls{dofs} of the upper body were set at their default values with only the  exception of pelvis vertical translation as previously mentioned. The initial state of the model is displayed in \autoref{fig:scone_initial_state}.

\begin{figure}[H]\ContinuedFloat
    \begin{subfigure}[t]{0.49\textwidth}
        \centering
        \includegraphics[width=0.8\linewidth,height=7cm]{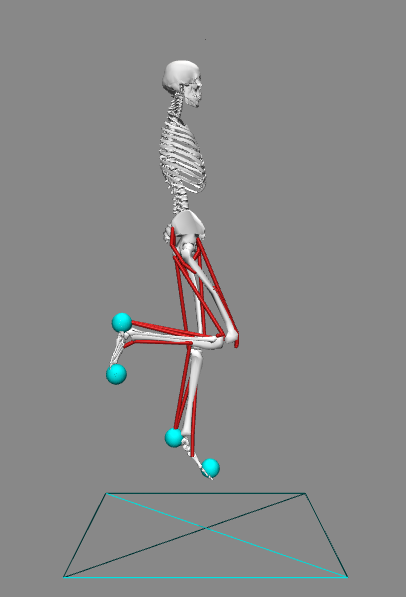}
    \end{subfigure}
    \hspace*{\fill}
    \begin{subfigure}[t]{0.49\textwidth}
        \centering
        \includegraphics[width=0.8\linewidth,height=7cm]{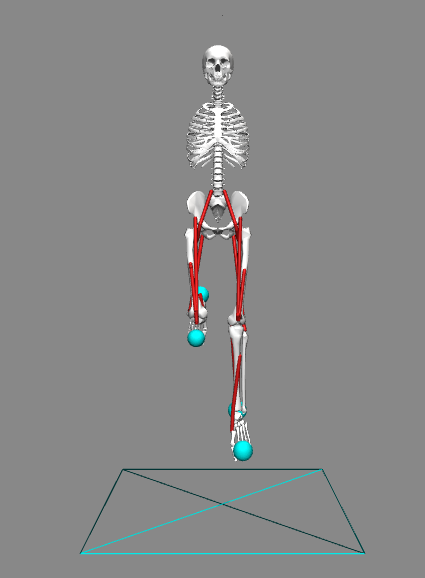}
    \end{subfigure}
    \caption{Initial state of the musculoskeletal model "Human0916" used for the predictive simulations in SCONE. Sagittal (right) and Frontal (left) planes.}
    \label{fig:scone_initial_state}
\end{figure}

Next, we will take a look into the Optimizer and its components. The Optimizer contained the previously described model, a reflex controller and a composite measure which were presented in \autoref{sec:scone_info}. The reflex controller included "MuscleReflex" entries that simulated proprioceptic reflexes and a "BodyPointReflex" that simulated the vestibular reflexes. Moreover, each reflex defined a reflex arc with specific gain, offset and delay. Also, the composite measure contained several measures. The first measure was the "ReactionForceMeasure" that was used to minimize the \gls{grf} contact with the ground. Furthermore, a range of accepted values for the \gls{grf} was defined and if their values were out of that range a penalty was applied. Then, a "BalanceMeasure" was added  which was used for keeping the model from falling. Finally several "DofMeasures" for ankle and knee angles were defined to achieve the desired landing (a single - leg soft landing). More specifically, we defined ranges of accepted values for the angles. Penalties were applied when those ranges were violated.

The components of the cost function that were included in our study are described next:

\begin{itemize}
    \item \textbf{A parameter that penalized the model when falling based on the termination height} 
    \item \textbf{A penalty that minimized the effort}
    \item \textbf{Penalties for exceeding desired joint ranges}. The desired joint range for the ankle was: [ \ang{25}, \ang{35}] (positive dorsiflexion). For the knee  was: [\ang{-60}, \ang{-30}] (positive flexion) and for the pelvis tilt was: [ \ang{-50}, \ang{0}].
\end{itemize}

Next, we present a general overview of the script used for this simulation setup and its basic components.
   
\begin{lstlisting}[language=Python]
Optimizer {
	# definition of model used in simulation
	OpenSimModel { }
	Controller {
		type = ReflexController
		# Muscle length reflexes
		MuscleReflex {}
		# Vestibular reflexes
		BodyPointReflex {}
	CompositeMeasure {		          
			# measure to minimise GRF
			ReactionForceMeasure {}	
			# measure for not falling
			BalanceMeasure {}
			# measute for desired joint ranges
			DofMeasure {}
   }
}
\end{lstlisting}

\section{Predictive simulation of landing with MOCO}

In this section we present the simulation performed in MOCO in order to create case studies of single - leg landings using as an initial guess the predicted motion from SCONE. First, we describe the basic simulation setup in MOCO. Then, we introduce the pipeline for augmenting model complexity. Finally we examine one by one the examined case studies that answer to multiple what - if scenarios. 

\subsection{Simulation setup}\label{sec:Moco_pipeline}

Here we present the main outline of the Moco study that is deployed to assess any of the subsequent what - if - scenarios. This is accomplished with Python scripting taking advantage of the Python bindings available for Moco. The order of the commands follows the same hierarchy as presented in \autoref{fig:moco_2}. Therefore, we created a "MocoProblem" object and a "MocoSolver" object as parts of the high - level "MocoStudy" object. The following subsections are dedicated to describe each distinct block of the overall workflow.

\subsubsection{Model operations}
At first, certain modifications where applied to the musculoskeletal model. The muscle model was selected as the "DeGrooteFregly" muscle type, since this model is compliant with Moco. The passive fiber forces were ignored and the active fiber force width for all muscles was scaled with a factor of 1.5. Ideal torque actuators were appended to the \gls{dofs} that were not actuated by muscles. Also, reserve actuators were added to all \gls{dofs} to act supplementary to the already present muscles. The maximum torque was set to 100 Nm.

\begin{lstlisting}[language=Python]
import opensim as osim

model = osim.ModelProcessor("Gait2392.osim")
# turn off tendon compliance
model.append(osim.ModOpIgnoreTendonCompliance())
# replace with "DeGrooteFregly" muscle type
model.append(osim.ModOpReplaceMusclesWithDeGrooteFregly2016())
# turn off passive fiber forces 
model.append(osim.ModOpIgnorePassiveFiberForcesDGF())
# Scale the active fiber force curve width 
model.append(osim.ModOpScaleActiveFiberForceCurveWidthDGF(1.5))
# add reserve actuators to the model
model.append(osim.ModOpAddReserves(100))

\end{lstlisting}

As seen in \autoref{fig:pipeline} our workflow is divided in two parts. First, we performed tracking studies using the MocoTrack interface and the previously adjusted model. Then the MocoStudy class was used to predict the what-if scenarios.

\subsubsection{Augmenting model complexity - MocoTrack}

The aim of the tracking study was twofold. First, we used it to create initial guesses for our optimization studies. Secondly, this helped us to chain three different simulations where each new one features a higher complexity musculoskeletal model and takes as input the optimal solution of the previous one. Model complexity refers to its total number of \gls{dofs} and total number of available muscle - tendon units. As we already described in \autoref{sec:Scone_pipeline} the objective of the first simulation using SCONE was to obtain an initial guess for the upcoming Moco studies. Therefore, for the first MocoTrack we used as reference trajectory the SCONE simulation output. The selected model was the Gait2354 model. The output of this tracking study served as input for an additional tracking study, this time with the Gait2392 model.

The simulation initial and final time were identical to these of the SCONE trajectory. The track tool instance was connected to the problem. Instantly the "MocoStateTrackingGoal" was added to the tracking study. Some states of the model were edited to further assist the trajectory solution. The bounds for these \gls{dofs} were set based on the initial and final states of the predicted motion from SCONE simulation. A detailed overview of these bounds is presented in \autoref{tab:dofs1}. For the \gls{dofs} that are not included in this  table no bounds were set.
Moreover, we applied bounds for the initial state and the entire motion for the activation of all muscles. At the first time instant the activation of all muscles was set to zero. Furthermore, both models consists of identical \gls{dofs} and the following description concerns both of them. 
An overview of the commands that were used to setup the analysis is presented next:

\begin{lstlisting}[language=Python]

# set MocoTrack tool
track = osim.MocoTrack()

# define model
track.setModel()
# set table containing values of model state variables
track.setStatesReference()
# set initial time
track.set_initial_time()
# set final time
track.set_final_time()
    
# initialize MocoTrack in the study
study = track.initialize()
# access  MocoProblem within the study.
problem = study.updProblem()

# set bounds for all state variables for phase 0
problem.setStateInfoPattern('/forceset/.*/activation', osim.MocoBounds(0,1), osim.MocoInitialBounds(0)))
problem.setStateInfo("/jointset/../../value",
                         osim.MocoBounds(), osim.MocoInitialBounds(), osim.MocoFinalBounds())

\end{lstlisting}

\begin{table}[H]
    \begin{center}
    \normalsize
    \begin{threeparttable}[t]
      \begin{tabular}{c c c c} 
        \toprule
        \textbf{\gls{dofs}} & \textbf{Bounds} & \textbf{Initial Bounds} & \textbf{Final Bounds}\\
        \midrule
        pelvis\_ty & (0.7, 1.25) & 1.25 & (0.75, 0.85)\\
        pelvis\_\tnote{\textdagger} & (-0.01, 0.01) & 0 & -  \\
        lumbar\_\tnote{\textdagger} & (-0.01, 0.01) & 0 & -  \\
        hip\_flexion\_l & (0.08, 0.5) & 0.087 & -  \\
        hip\_\tnote{\textdagger} & (-0.01, 0.01) & 0 & -  \\
        subtalar\_angle\tnote{\ddag} & (-0.01, 0.01) & 0 & -  \\
        mtp\_angle\tnote{\ddag} & (-0.01, 0.01) & 0 & -  \\
      \bottomrule
      \end{tabular}
      \begin{tablenotes}
      \item[\textdagger] This stands for all \gls{dofs} of the joint if they are not explicitly defined.
      \item[\ddag] This stands for both right and left joints. 
      \end{tablenotes}
    \end{threeparttable}
      \caption{Bounds for the \gls{dofs} of Gait2354 and Gait2392 OpenSim models for MOCO Track tool.}
      \label{tab:dofs1}
    \end{center}
\end{table}

\subsubsection{Predictive simulation settings - MocoStudy}

Regarding prediction in Moco, we created a new study with the same commands as previously. Since this class was used to execute all the what - if scenarios, the settings for each case study will be presented in the following sections. The common setting for all cases was the initial guess wich was set as the MocoTrack output.

\begin{lstlisting}[language=Python]
# set MocoStudy
study = osim.MocoStudy()

# access  MocoProblem within the study.
problem = study.updProblem()
#Set the model
problem.setModel()
#Set time bounds 
problem.setTimeBounds()

# set bounds for all state variables for phase 0
problem.setStateInfoPattern('/forceset/.*/activation', osim.MocoBounds(0,1), osim.MocoInitialBounds(0)))
problem.setStateInfo("/jointset/../../value", osim.MocoBounds(),  osim.MocoInitialBounds(), osim.MocoFinalBounds())

\end{lstlisting}

\subsubsection{Post - processing operations}

\begin{figure}[H]
    \centering
    \includegraphics[width=0.8\textwidth, height = 8cm]{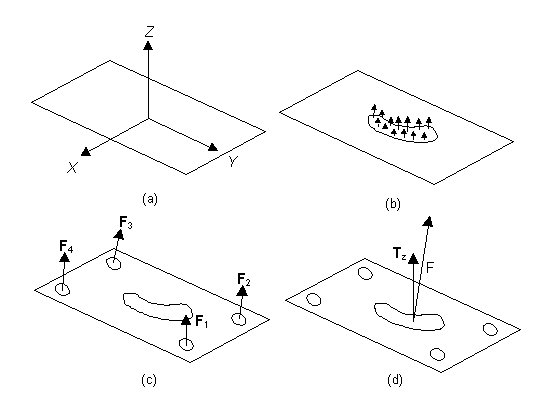}
    \caption{Computation of the Center of Pressure. Reproduced by \cite{KwonYH}.}
    \label{fig:cop1}
\end{figure}

Moreover, we developed scripts to perform post - processing operations on each estimated by Moco predicted trajectory. The parameters of interest were normalized tendon forces, \gls{grf}, \gls{kjrf} and \gls{kjrm}. Initially, we used the OpenSim Analyze Tool to obtain the muscle tendon forces. Also, \gls{jra} was implemented for the left knee joint. The \gls{jra} pipeline requires muscle forces and all external forces applied to the model. The external forces were acquired from the estimated solution and the contact forces estimated from the sphere contact model expressed in the ground frame. Since force is a bound vector the point of application is also required to appropriately estimate joint reaction loads. Towards this direction, we deployed a method for estimating the \gls{cop} \cite{KwonYH}, as it serves the actual \gls{grf} application point. As depicted in \autoref{fig:cop1}, the contact between the foot and the ground leads to the development of contact forces between these two parts. These forces can be summed into a ground force vector \textbf{F} and a free torque vector \textbf{$T_z$}, which is applied at the \gls{cop}.  

\begin{figure}[H]
    \centering
    \includegraphics[width=0.6\textwidth, height = 6cm]{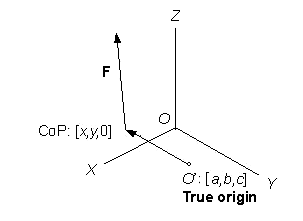}
    \caption{True origin of the \gls{cop}.Reproduced by \cite{KwonYH}.}
    \label{fig:cop2}
\end{figure}

Assuming that the true origin is the \textit{O'} at \textit{(a,b,c)} in \autoref{fig:cop2}, then the \textit{Z} component is 0. Thus, the moment caused by \textbf{F} about the true origin plus \textbf{$T_z$} is equal to the moment measured from the plates. These are described by the next formulas: \\

\begin{equation}
    \label{M_1}
    M = \begin{bmatrix} x-a, & y-b, & -c \end{bmatrix} x \begin{bmatrix} F_x, & F_y, & F_z \end{bmatrix} + \begin{bmatrix} 0, & 0, & T_z \end{bmatrix}
\end{equation}

\begin{equation}
    \label{M_2}
     \begin{bmatrix} M_x \\ M_y \\ M_z \end{bmatrix} = \begin{bmatrix} 0 & c & y-b \\-c & 0 & -(x-a) \\-(y-b) & (x-a) & 0 \end{bmatrix}  \begin{bmatrix} F_x\\ F_y\\F_z \end{bmatrix} + \begin{bmatrix} 0\\0\\T_z \end{bmatrix} 
\end{equation}

or \\

\begin{equation}
    \label{M_3}
    \begin{bmatrix} M_x \\ M_y \\ M_z \end{bmatrix} = \begin{bmatrix} (y-b)F_z+cF_y \\-cF_x-(x-a)F_z \\(x-a)F_y-(y-b)F_x+T_z \end{bmatrix}
\end{equation}

Finally:\\

\begin{equation}
    \label{xy}
    x = - \frac{M_y+cF_x}{F_z} + a ,\quad  y = \frac{M_x-cF_y}{F_z} + b
\end{equation}

\begin{equation}
    \label{T_z}
     T_z = M_z-(x-a)F_z+(y-b)F_x
\end{equation}

Based on the previous analysis, the \gls{cop} can be calculated from the values of the \gls{grf}: \textbf{F\_x}, \textbf{F\_y} and \textbf{F\_z}, the moment caused by the \gls{grf}: \textbf{M\_x}, \textbf{M\_y} and \textbf{M\_z}, and finally the location of the true origin \textbf{a}, \textbf{b} and \textbf{c}.

A Python code snippet that executes all these post - processing operation is presented in the following block:

\begin{lstlisting}[language=Python]

#export normalized tendon forces from the Moco solution using Analyze tool
outputPaths.append('.*tendon_force')
outputTable = study.analyze(solution, outputPaths)

# export GRF expressed in ground frame
externalForcesTableFlat = osim.createExternalLoadsTableForGait()
# exprees GRF in CoP
express_GRF_inCop('GRF.sto')                      

# JRA for the knee joint
JR_paths.append('.*reaction_on_child')
states = solution.exportToStatesTable()
controls = solution.exportToControlsTable()
JR_outputs_table = osim.analyzeSpatialVec(model, states, controls, JR_paths).flatten()
                                          
# express JRA in Child frame
express_jra_inChild()

\end{lstlisting}

\subsection{Initial height case study}\label{sec:height_info}

In this section, we describe the simulation setup of drop - landing from different initial heights. The model used was \textit{Gait2392.osim}. The \textit{pelvis\_ty} value was modified as presented in \autoref{tab:dofs2} in order to achieve landings from  30, 35, 40, 45, 50 and 55 cm of height. For every height value a new study was created with the parameters described previously. The solution acquired with the track tool was used as an initial guess for the solver. Also, the  "MocoControlGoal" goal was added to the problem with a weight of 0.001.

\begin{table}[H]
    \begin{center}
    \normalsize
      \begin{tabular}{c c} 
        \toprule
        \textbf{Height} & \textbf{Pelvis\_ty} \\
        \midrule
        30 & 1.25 \\
        35 & 1.30 \\
        40 & 1.35 \\
        45 & 1.40 \\
        50 & 1.45 \\
        55 & 1.50 \\
        \bottomrule
      \end{tabular}
      \caption{Moco values for \textit{pelvis\_ty} based on landing height as displayed in OpenSim.}
      \label{tab:dofs2}
    \end{center}
\end{table}
 
\begin{figure}[H]
    \centering
    \includegraphics[width=1\textwidth, height = 6cm]{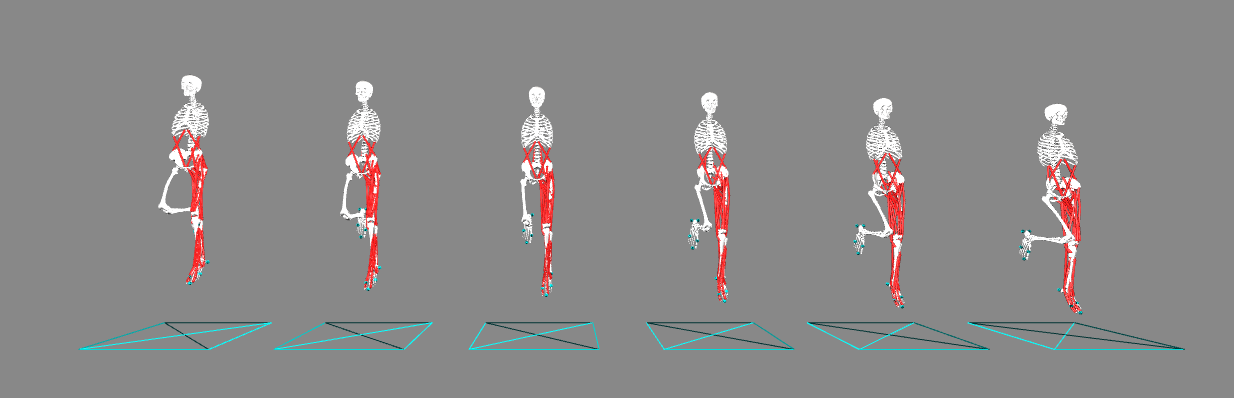}
    \caption{Initial state of the model for different drop - landing heights.}
    \label{fig:heights}
\end{figure}

\autoref{fig:heights}, we demonstrate the initial pose of the model from these different drop - landing heights. As it can be observed, apart from the \textit{pelvis\_ty} value which was adjusted in order to achieve multiple initial landing heights, all the other \gls{dofs} remained identical for the initial state in all scenarios. It should be mentioned that a deviation of 0.01cm of the selected landing height was allowed in all cases, as with all \gls{dofs} of the model.
 
\subsection{Hip rotation case study} \label{sec:hip_info}

\begin{figure}[H]
    \centering
    \includegraphics[width=1\textwidth, height = 6cm]{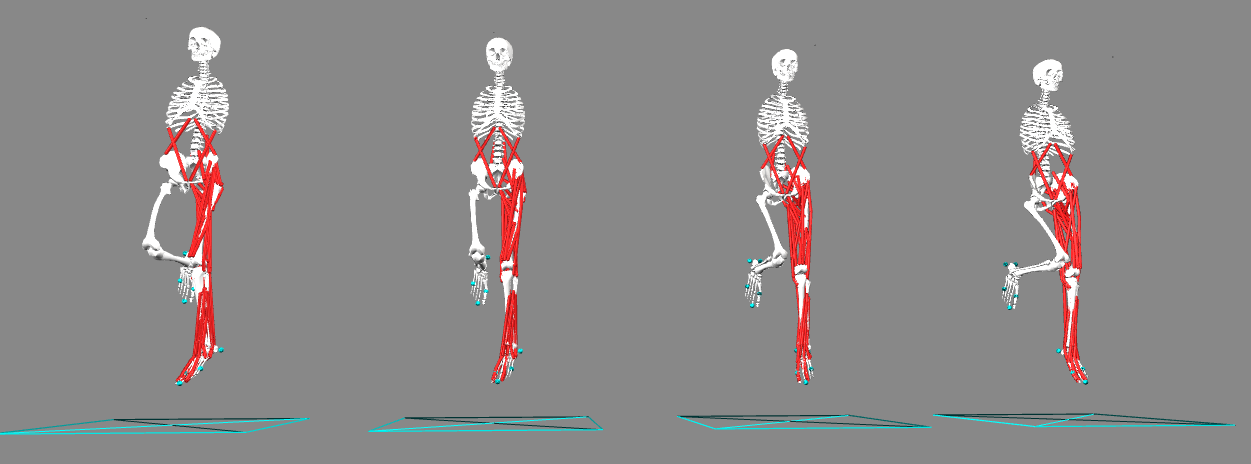}
    \caption{Initial state of the model for \ang{0}, \ang{10}, \ang{20} and \ang{30} of hip internal rotation.}
    \label{fig:hip_int_rots}
\end{figure}

\begin{figure}[H]
    \centering
    \includegraphics[width=1\textwidth, height = 6cm]{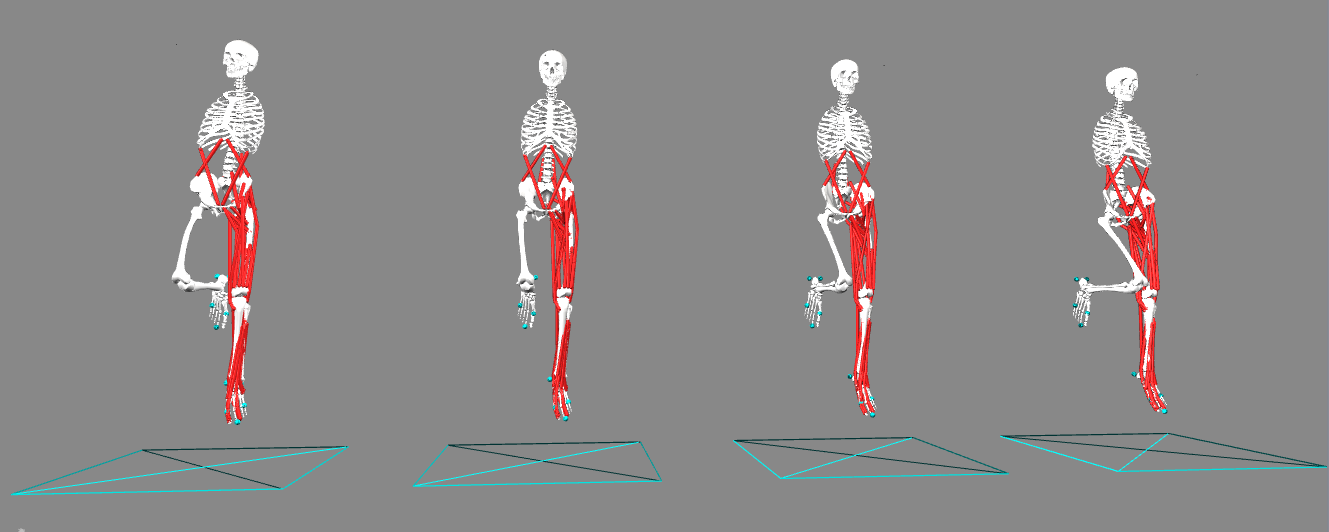}
    \caption{Initial state of the model for \ang{0}, \ang{10}, \ang{20} and \ang{30} of hip external rotation.}
    \label{fig:hip_ext_rots}
\end{figure}

In this section we describe the analysis parameters used to conduct several studies for drop - landings with different values of hip rotation for the landing leg. We examined cases where the hip was internally or externally rotated for the entire trajectory. In total we created 13 studies, 6 for hip internal rotation, 6 for hip external rotation and 1 for no hip rotation. The angles of external and internal rotation  examined were \ang{5}, \ang{10}, \ang{15}, \ang{10}, \ang{25} and \ang{30}. The model in its initial configuration with internally and externally rotated hip (\ang{0}, \ang{10}, \ang{20}, \ang{30}) is presented in \autoref{fig:hip_int_rots} and \autoref{fig:hip_ext_rots} respectively. 
 
\begin{table}[H]
    \begin{center}
    \normalsize
    \begin{threeparttable}[t]
      \begin{tabular}{c c c c c c c c c c c c c c c c c} 
        \toprule
        \textbf{\gls{dofs}} & \multicolumn{7}{c}{Angle (degrees)} & Bounds \\
        \midrule
        hip\_internal\_rotation &  0 & 5 & 10 & 15 & 20 & 25 & 30  \\
        hip\_external\_rotation & 0 & 5 & 10 & 15 & 20 & 25  & 30  \\
        \bottomrule
      \end{tabular}
     \end{threeparttable}
      \caption{Overview of investigated values for hip joint rotation.}
      \label{tab:trunk_dofs_23}
    \end{center}
\end{table}

 Again, we used the \textit{Gait2392} OpenSim model, and the previously tracked solution was used as an initial guess for the solver. Also, the "MocoControlGoal" was added to the problem with a weight of 0.001. In \autoref{tab:hip_dofs} we display the bounds assigned to all \gls{dofs} in the Moco studies.

\begin{table}[H]
    \begin{center}
    \normalsize
      \begin{tabular}{c c c c} 
        \toprule
        \textbf{DoFs} & \textbf{Initial bounds} & \textbf{Final bounds} & \textbf{Bounds}\\
        \midrule
        pelvis\_ty & 1.25 & (0.75, 0.85) & (0.7, 1.25)  \\
        pelvis\_*\tnote{\textdagger} & 0 & - & (-0.57, 0.57)  \\
        hip\_flexion\_r & \ang{30} & - & (\ang{29},\ang{30}) \\
        hip\_rotation\_r & \ang{0} & - & (\ang{-0.01},\ang{0.01}) \\
        hip\_adduction\_r & \ang{0} & - & (\ang{-0.01},\ang{0.01}) \\
        knee\_angle\_r & \ang{-120} & - & (\ang{-121},\ang{-119}) \\
        ankle\_angle\_r & \ang{0} & - & (\ang{-0.57}, \ang{0.57}) \\
        subtalar\_angle\_r & \ang{0} & - & (\ang{-0.57}, \ang{0.57}) \\
        mtp\_angle\_r & \ang{0} & - & (\ang{-0.57},\ang{0.57}) \\
        hip\_flexion\_l & \ang{5} & - & (\ang{4.5},\ang{28.5}) \\
        hip\_rotation\_l & - & - & (v-\ang{0.57},v+\ang{0.57}) \\
        hip\_adduction\_l & - & - & - \\
        knee\_angle\_l & \ang{-11.5} & - & (\ang{-57},\ang{0}) \\
        ankle\_angle\_l & \ang{-34} & - & (\ang{-34},\ang{45}) \\
        subtalar\_angle\_l & - & - & (\ang{-2.85},\ang{2.85}) \\
        mtp\_angle\_l & \ang{0} & - & (\ang{-5.73}, \ang{5.73}) \\
        \bottomrule
      \end{tabular}
      \begin{tablenotes}
      \item[\textdagger] This stands for all \gls{dofs} of the joint if they are not explicitly defined. 
      \end{tablenotes}
      \caption{Values for \gls{dofs} of the model for different values of left hip rotation scenarios }
      \label{tab:hip_dofs}
    \end{center}
\end{table}

For \textit{hip\_adduction} of the left lower limb we did not assign bounds because it is highly related to \textit{hip\_rotation} and we wanted to examine how it will respond to different conditions of hip rotation.

\subsection{Trunk orientation case study}\label{sec:trunk_info}

In this section we outline the settings for the scenarios that evaluate deviations of the trunk \gls{dofs}. This is achieved through modifying the lumbar joint. It consists of three \gls{dofs}, namely flexion - extension, right - left bending and external - internal rotation. The assessed flexion angles were: \ang{5}, \ang{10}, \ang{15}, \ang{20}, \ang{25} and \ang{30} degrees, while for extension were: \ang{5}, \ang{10}, \ang{15} and \ang{20}. Moreover, the angles studied for right - left bending and external - internal rotation were: \ang{0}, \ang{5}, \ang{10}, \ang{15}, \ang{20}, \ang{25} and \ang{30}. Also, an upright position with no flexion, bending or rotation was considered. These scenarios are presented in \autoref{tab:trunk_dofs_23}. We also include the bounds for the \gls{dofs} of interest for the entire prediction pipeline. It should be stated that these bounds are active only for the lumbar joint generalized coordinate under consideration, while all other lumbar \gls{dofs} are equal to zero. 

\begin{table}[H]
    \begin{center}
    \normalsize
    \begin{threeparttable}[t]
      \begin{tabular}{c c c c c c c c c c c c c c c c c} 
        \toprule
        \textbf{\gls{dofs}} & \multicolumn{7}{c}{Angle (degrees)} & Bounds \\
        \midrule
        lumbar\_flexion & 0 & 5 & 10 & 15 & 20 & 25 & 30 & (v-0.57,v+0.57)\tnote{\textdagger}\\
        lumbar\_extension &  0 & 5 & 10 & 15 & 20 & - & -  & " \\
        lumbar\_rotation & 0 & 5 & 10 & 15 & 20 & 25  & 30 & " \\
        lumbar\_bending & 0 & 5 & 10 & 15 & 20 & 25  & 30 & " \\
        \bottomrule
      \end{tabular}
      \begin{tablenotes}
      \item[\textdagger] v is the value of the generalized coordinate under consideration for each scenario.
      \end{tablenotes}
     \end{threeparttable}
      \caption{Overview of investigated values for lumbar joint \gls{dofs}. We also describe the bounds for each value.}
      \label{tab:trunk_dofs_23}
    \end{center}
\end{table}

\begin{figure}[H]
    \centering
    \includegraphics[width=0.6\textwidth, height = 8cm]{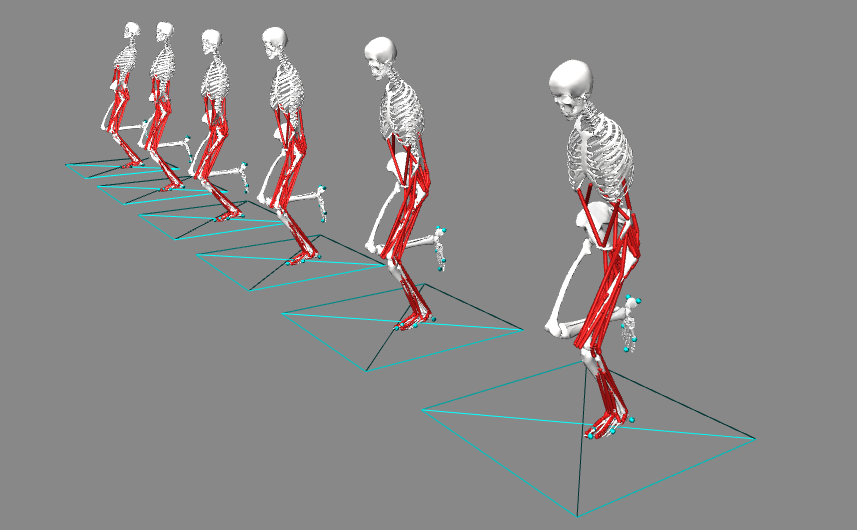}
    \caption{Initial state of the OpenSim model for different values of lumbar flexion - extension.}
    \label{fig:trunk_flex}
\end{figure}

\begin{figure}[H]
    \centering
    \includegraphics[width=1\textwidth, height = 6cm]{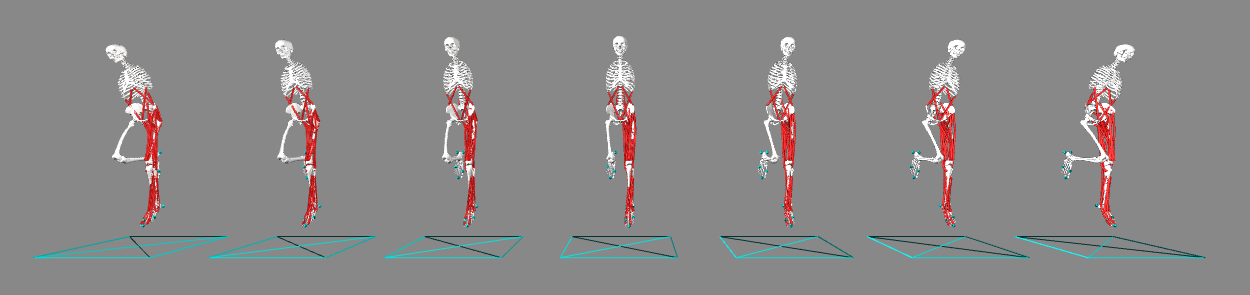}
    \caption{Initial state of the OpenSim model for different values of lumbar right - left bending.}
    \label{fig:trunk_ben}
\end{figure}

\begin{figure}[H]
    \centering
    \includegraphics[width=1\textwidth, height = 6cm]{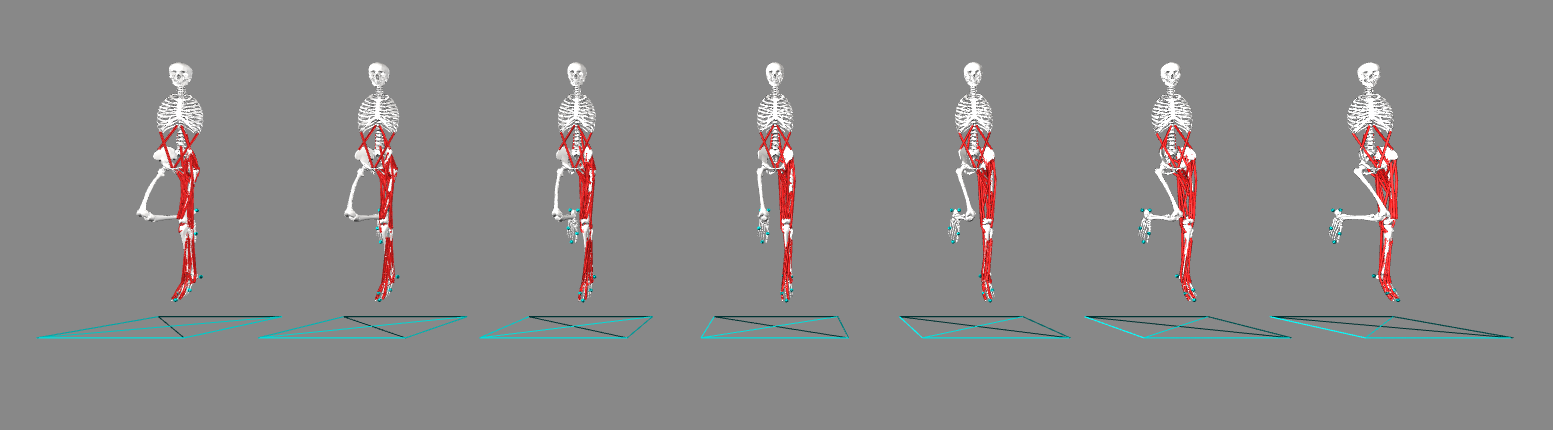}
    \caption{Initial state of the OpenSim model for different values of lumbar internal - external rotation.}
    \label{fig:trunk_rot}
\end{figure}

In \autoref{fig:trunk_flex}, \autoref{fig:trunk_ben} and \autoref{fig:trunk_rot} we display the initial state of the model for selected angles of trunk flexion - extension, right - left bending and internal - external rotation respectively.

For all of the above cases, different studies were produced using the initial guess described previously and certain bounds for the initial and final states that are presented in \autoref{tab:trunk_dofs}.

\begin{table}[H]
    \begin{center}
    \normalsize
    \begin{threeparttable}[t]
      \begin{tabular}{c c c c} 
        \toprule
        \textbf{DoFs} & \textbf{Initial bounds} & \textbf{Final bounds} & \textbf{Bounds}\\
        \midrule
        pelvis\_ty & 1.25 & (0.75, 0.85) & (0.7, 1.25)  \\
        pelvis\_*\tnote{\textdagger} & \ang{0} & - & (\ang{-0.57}, \ang{0.57})  \\
        hip\_flexion\_r & \ang{30} & - & (\ang{29},\ang{30}) \\
        hip\_rotation\_r & \ang{0} & - & (\ang{-0.01},\ang{0.01}) \\
        hip\_adduction\_r & \ang{0} & - & (\ang{0.01},\ang{0.01}) \\
        knee\_angle\_r & \ang{-120} & - & (\ang{-121},\ang{-119}) \\
        ankle\_angle\_r & \ang{0} & - & (\ang{-0.57}, \ang{0.57}) \\
        subtalar\_angle\_r & \ang{0} & - & (\ang{-0.57}, \ang{0.57}) \\
        mtp\_angle\_r & \ang{0} & - & (\ang{-0.57}, \ang{0.57}) \\
        hip\_flexion\_l & \ang{5} & - & (\ang{4.5},\ang{28.5}) \\
        hip\_rotation\_l & \ang{0} & - & (\ang{-0.01}, \ang{0.01}) \\
        hip\_adduction\_l & \ang{0} & - & (\ang{-0.01}, \ang{0.01}) \\
        knee\_angle\_l & \ang{-11.5} & - & (\ang{-57},\ang{0}) \\
        ankle\_angle\_l & \ang{-34} & - & (\ang{-34},\ang{45}) \\
        subtalar\_angle\_l & - & - & (\ang{-2.86},\ang{2.86}) \\
        mtp\_angle\_l & \ang{0} & - & (\ang{-0.57}, \ang{0.57}) \\
        \bottomrule
      \end{tabular}
      \begin{tablenotes}
      \item[\textdagger] This stands for all \gls{dofs} of the joint if they are not explicitly defined. 
      \end{tablenotes}
     \end{threeparttable}
      \caption{Bounds for the \gls{dofs} of the model for the trunk orientation case study.}
      \label{tab:trunk_dofs}
    \end{center}
\end{table}

For all the trunk cases, both the track tool and the prediction with initial guess were examined in order to compare them. Moreover different weights for the \textit{effort goal} were examined and compared for several cases in order to observe deviations of the \gls{grf} and muscle activation values.

\subsection{Muscle force of knee joint agonists and antagonists case study}\label{sec:muslces_info}

In this section we present the case studies that were performed to investigate how permutations regarding the max isometric force of the knee joint agonists and antagonists muscle groups affect the \gls{acl} injury indicators. Muscle fatigue is defined as the decrease in maximum force \cite{Bigland1995,confluence}. Towards this objective, we edited the values of max isometric forces to simulate weakness and strength of specific muscle groups, namely the quadriceps and the hamstrings muscles. To simulate strong muscles we enlarged max isometric force by 35 percent and to simulate weak muscles we reduced their max isometric force by 35 percent in accordance with \cite{Afschrift2014,Jing2017,Daniel2017}.

The model used for these scenarios was \textit{Gait2392}. The quadriceps muscles include rectus femoris, vastus medialis, vastus lateralis and vastus intermedialis. Hamstrings muscle group combines the semimembranosus, semitendinosous, biceps femoris long head and biceps femoris short head.

In \autoref{tab:muscles} we display the value of the standard max isometric force fore each muscle, along with the value when it is weakened or strengthened.

\begin{table}[H]
    \begin{center}
    \normalsize
      \begin{tabular}{c c c c} 
        \toprule
        \textbf{Muscle} & \textbf{max Isometric force} & \textbf{Weak} & \textbf{Strong}\\
        \midrule
        Rectus femoris & 1169 & 760 & 1578  \\
        Vastus intermedialis & 1365 & 887 & 1842  \\
        Vastus lateralis & 1871 & 1216 & 2526 \\
        Vastus medialis & 1294 & 841 & 1746 \\
        Semimembranosus & 1288 & 837 & 1739 \\
        Semitendinosus & 410 & 267 & 554 \\
        Bicep femoris long head & 896 & 582 & 1210 \\
        Bicep femoris short head & 804 & 523 & 1085 \\
        \bottomrule
      \end{tabular}
      \caption{Values for max isometric force for all muscles in hamstrings ans gastrocnemius muscle groups. Also, the modified max isometric forces for theses muscles for their strengthening or weakening are included.}
      \label{tab:muscles}
    \end{center}
\end{table}

The bounds used for the initial and final states were identical to those used in previous cases for the right lower limb. The trunk and the left lower limb \gls{dofs} were restricted based on the initial guess and small deviations were allowed. A "MocoControlGoal" was used with a weight of 0.002.

Based on \autoref{tab:muscles}, eight cases were simulated with different combinations of normal, weak and strong muscles. These cases are demonstrated in \autoref{tab:muscles_cases}. 

\begin{table}[H]
    \begin{center}
    \normalsize
      \begin{tabular}{c c c} 
        \toprule
        \textbf{case} & \textbf{Quadriceps} & \textbf{Hamstrings} \\
        \midrule
        1 & normal & normal   \\
        2 & normal & strong \\
        3 & normal & weak \\
        4 & strong & normal \\
        5 & weak & normal \\
        6 & strong & weak \\
        7 & weak & strong  \\
        8 & weak & weak \\
        \bottomrule
      \end{tabular}
      \caption{Demonstration of the eight cases  with different combinations of normal, weak and strong muscles.}
      \label{tab:muscles_cases}
    \end{center}
\end{table}

\subsection{Moco Control goal weight case study}\label{sec:effort_info}

In this section, we examined how the "Moco Control goal" affects the outcome of the study. For this, we tested different weight values: 0, 0.1, 0.2, 0.5, 1, 2, 5 and 10. Again, the selected model was the \textit{Gait2392} model, and the tracked solution was the initial guess. Also, the \gls{dofs} were bounded as described in previous sections. All parameters were identical for all cases except for the goal weight.

%% file: Chapters/Chapter4.tex
\chapter{Results} 

\label{Chapter4} 


In this chapter, we present the results of our case studies that reflect parameters associated with \gls{acl} injuries. These parameters concern kinematics, \gls{grf}, \gls{grm}, muscle forces and muscle force ratios of the landing lower limb. Also, we include \gls{kjrf} and \gls{kjrm} plots. Finally, tables with values for the previously mentioned parameters at the peak \gls{vgrf} time instant are introduced. This instant was selected since it is considered the moment of \gls{acl} injury occurrence. The kinematics are presented in degrees, the \gls{grf}, \gls{grm} and \gls{jrm} are presented with normalized values to \gls{bw} of the musculoskeletal model. The muscle forces are presented in Newtons. Moreover, one may notice deviations between the plots and the resulted values on the tables. This is due to an applied filter on the plots. This is only for visualization purposes and does not modify the comparison between the multiple scenarios.

The experiments were conducted using a computer machine with an  i7-9700 Intel(R) Core(TM) processor @3.00 GHz, a memory of 16GB, and a Windows 10 64bit operating system. Every simulation scenario required about 45 to 60 minutes to complete regarding the above specifications.
\section{Initial height case}\label{sec:height_results}

In this section, we present results concerning single - leg landings from 30, 35, 40, 45, 50 and 55 cm of height as described in \autoref{sec:height_info}. 

\begin{figure}[H]
    \centering
    \includegraphics[width=1\textwidth, height = 4cm]{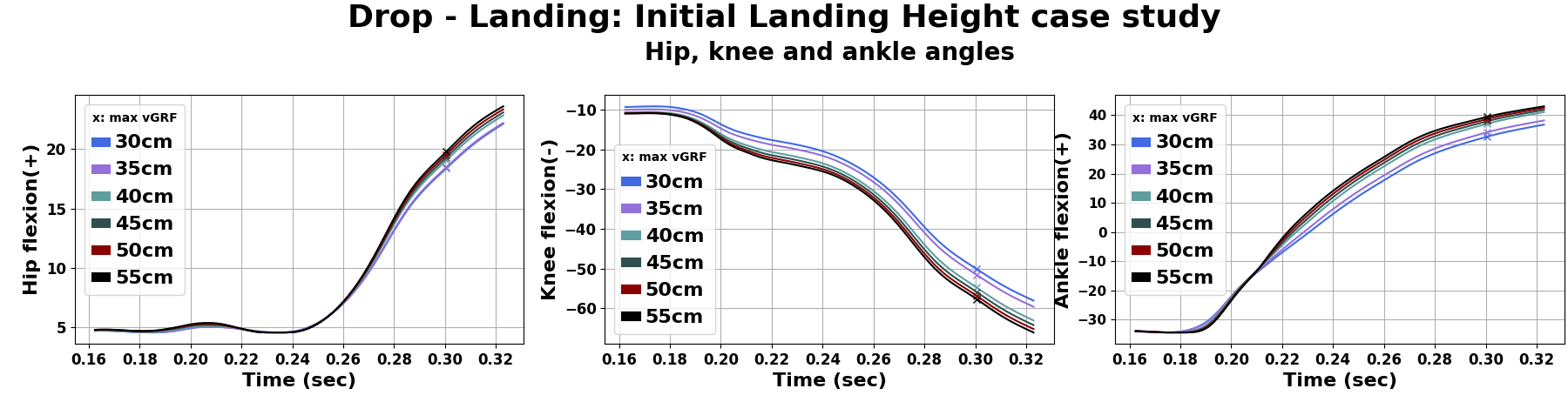}
    \caption{Hip, knee and ankle kinematics for the initial height case.}
    \label{fig:predict_height_effort0.001_sol}
\end{figure}

Initially, we present the predicted kinematics for the hip, knee and ankle joint angles for all examined scenarios (\autoref{fig:predict_height_effort0.001_sol}). As mentioned in \autoref{sec:height_info}, these joints' \gls{dofs} were restricted to their respective bounds, although we can observe that there are small deviations among the different cases. First, we can detect that the hip and ankle flexion angles at peak \gls{vgrf} are increased as the drop - landing height is raised. On the contrary, the knee flexion angle is decreased as the landing height is raised. It should be mentioned that the deviations for these angles between the examined scenarios are less than \ang{10}. 

\begin{figure}[H]
    \centering
    \includegraphics[width=1\textwidth, height = 8cm]{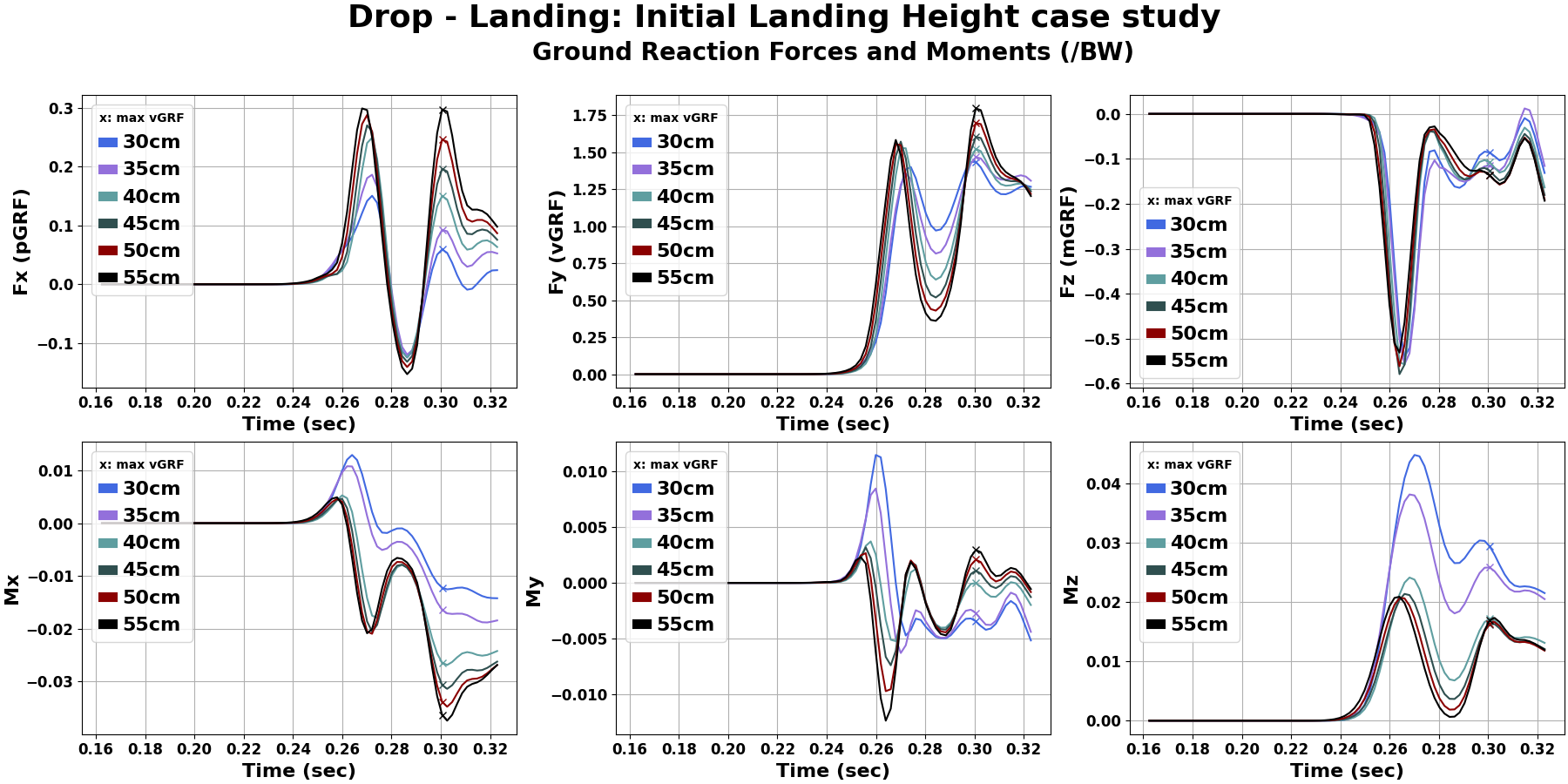}
    \caption{\gls{grf} and \gls{grm}  for the initial height case study..}
    \label{fig:predict_height_effort0.001_GRF}
\end{figure}

Next, we present \gls{grf} and \gls{grm} of single - leg landings from the same landing heights (\autoref{fig:predict_height_effort0.001_GRF}). We can observe that for greater landing heights the \gls{pgrf} and \gls{vgrf} values at max \gls{vgrf} time also increase. 

\begin{figure}[H]
    \centering
    \includegraphics[width=\textwidth, height = 8cm]{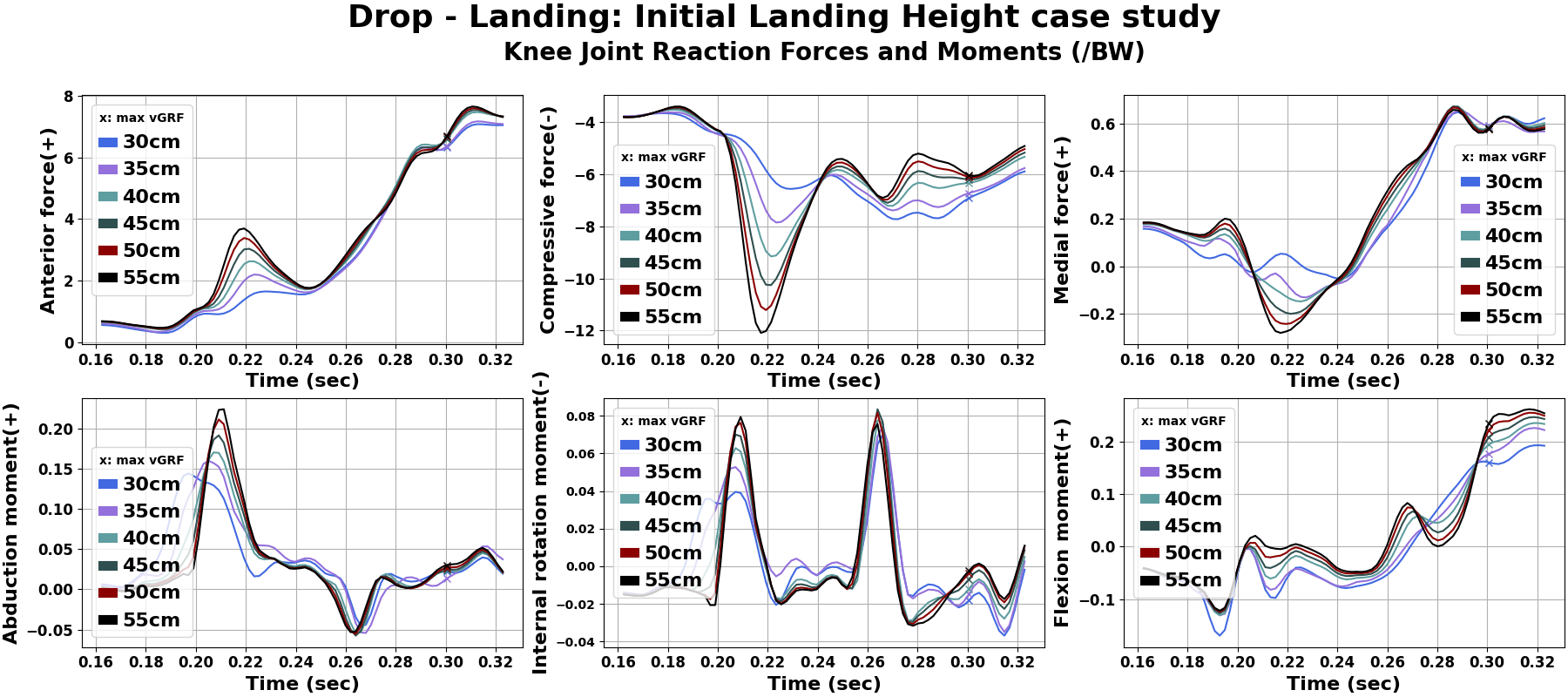}
    \caption{Knee joint \gls{jrf} and \gls{jrm} for multipe drop - landing heights.}
    \label{fig:predict_height_effort0.001_JRA}
\end{figure}

Subsequently, in \autoref{fig:predict_height_effort0.001_JRA} we demonstrate \gls{jrf} and \gls{jrm} for the knee joint for this case study. As we can observe, the \gls{af} reaches a first peak after \gls{igc} and then a greater one after the  max \gls{vgrf} time instant for all cases. Both peaks are greater as the landing height is increased. Nevertheless, at max \gls{vgrf} no great differences can be detected for the \gls{af} among the scenarios. Regarding \gls{cf}, we can see that the first peak occurs right after \gls{igc} with greater values for increased initial heights. Although, at max \gls{vgrf} time \gls{cf} is decreased as the landing height is increased. As for the \gls{mf}, a peak value is observed at \gls{igc} with greater values for greater landing heights.

Additionally, \gls{abdm} obtains a peak value for all examined scenarios around \gls{igc}. This peak value is greater for the landings from higher landing heights. Moreover, two \gls{erm} peaks are spotted for all cases, one after \gls{igc} and a second one between \gls{igc} and max \gls{vgrf}. Furthermore, lower \gls{irm} values are observed for higher initial heights with few exceptions.Also, \gls{fm} at max \gls{vgrf} time is greater for greater initial height scenarios. 

\begin{figure}[H]
    \centering
    \includegraphics[width=\textwidth, height = 8cm]{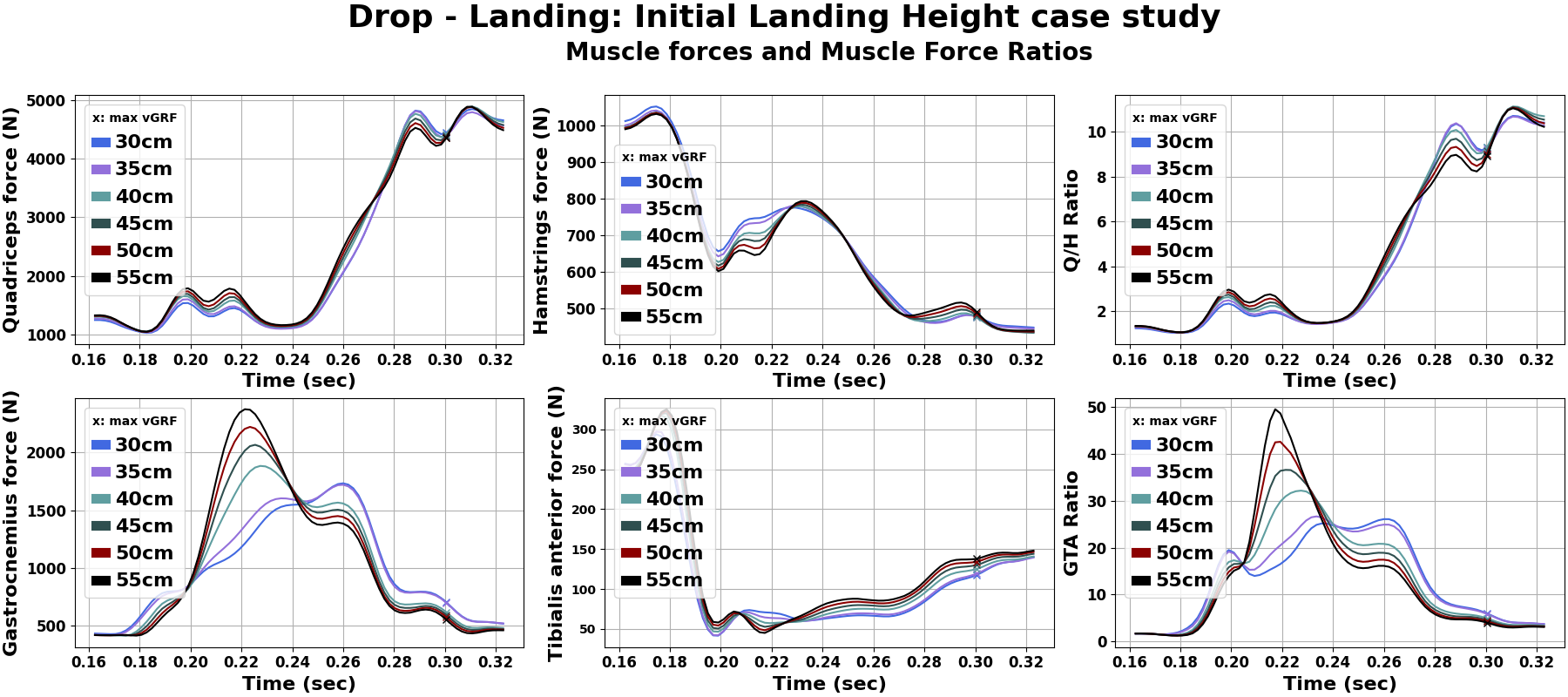}
    \caption{Muscle forces and muscle force ratios for multiple drop - landing heights.}
    \label{fig:predict_height_effort0.001_muscle_forces}
\end{figure}

Next, in \autoref{fig:predict_height_effort0.001_muscle_forces} we demonstrate  muscle forces and muscle force ratios for the studied scenarios. First, we observe that the quadriceps, hamstrings and tibialis anterior forces presents almost identical patterns for all examined scenarios. As a result the \gls{q/h} is almost similar for all cases. At \gls{igc}, \gls{q/h} is approximately 3 with lower values for lower initial heights. Then, it obtains greater values (around 10). Moreover, before max \gls{vgrf} time we observe that for lower heights  the ratio obtains greater values, while after that time it receives greater values for increased heights. Also, \gls{gta} appears a peak value after \gls{igc}, with increased values for greater heights. At max \gls{vgrf} no great deviations among different landing heights are observed.

In  \autoref{tab:height_1} and \autoref{tab:height_2}, values of the previously plotted parameters at the maximum \gls{vgrf} time instant for all cases are presented. In these tables and in order to compare more easily the different scenarios and validate our observations, the values correspond at time of maximum \gls{vgrf}.

We can observe that \gls{pgrf} and \gls{vgrf} are increased as the landing height is increased. Moreover, \gls{gta} is decreased as the landing height is increased. Also, \gls{q/h} obtains lower values for the landings from 55 and 50 cm.

\begin{table}[H]
    \begin{center}
        \begin{tabular}{c c c c c c} 
             \toprule
              Case & pGRF & vGRF & mGRF & Q/H\_ratio & GTA\_ratio \\  
             \midrule
             30 & 0.066 & 1.436 & -0.092 & 9.243 & 6.036  \\ 
             35 & 0.099 & 1.467 & -0.118 & 9.167 & 5.914 \\ 
             40 & 0.160 & 1.523 & -0.109 & 9.250 & 4.969  \\ 
             45 & 0.210 & 1.604 & -0.125 & 9.083 & 4.591  \\ 
             50 &  0.263 & 1.698 & -0.131 & 8.951 & 4.289  \\
             55 &  0.319 & 1.797 & -0.130 & 8.850 & 4.004  \\
             \bottomrule
        \end{tabular}
        \caption{\gls{grf}, \gls{q/h} and \gls{gta} at the time instant of peak \gls{vgrf} for multiple drop - landing heights..}
        \label{tab:height_1}
    \end{center}
\end{table}

In \autoref{tab:height_2} we can clearly see that as the landing height is increased, the \gls{af} is also increased at peak \gls{vgrf} time instant. On the other hand, \gls{cf} decreases as the initial landing height increases. Also, \gls{abdm} obtains the greatest value for the landing from 55cm and the lowest one for the landing from 35cm. Moreover \gls{irm} is decreased as the landing heights is greater. Finally, \gls{fm} is also greater for higher initial heights at max \gls{vgrf} time.

\begin{table}[H]
    \begin{center}
        \begin{tabular}{c c c c c c c} 
             \toprule
             Case & AF(+) & CF(-) & MF(+) & AbdM(+) & IRM(-) & FM(+) \\
             \midrule
             30 & 6.300 & -6.907 & 0.577 & 0.021 & -0.017 & 0.160 \\ 
             35 & 6.334 & -6.721 & 0.587 & 0.014 & -0.014 & 0.178  \\ 
             40 & 6.577 & -6.339 & 0.582 & 0.023 & -0.011 & 0.197  \\ 
             45 & 6.612 & -6.225 & 0.580 & 0.025 & -0.007 & 0.213 \\
             50 & 6.652 & -6.149 & 0.577 & 0.029 & -0.004 & 0.229  \\ 
             55 & 6.699 & -6.093 & 0.575 & 0.033 & -0.003 & 0.245  \\ 
             \bottomrule
        \end{tabular}
        \caption{Knee joint \gls{jrf} and \gls{jrm} at the time instant of peak \gls{vgrf} for multiple drop - landing heights.}
        \label{tab:height_2}
    \end{center}
\end{table}

It should be mentioned that in all figures presented in the current section the results for all scenarios were shifted in time. This was performed in order to synchronize the \gls{igc} of all scenarios. 

\section{Hip rotation case} \label{sec:hip_results}

In this section, we present the results concerning single - leg landings with different values of internal and external hip rotation of the landing leg. These angles were set at \ang{0}, \ang{5}, \ang{15}, \ang{20}, \ang{25} and \ang{30} for both cases as described in \autoref{sec:hip_info}.

\begin{figure}[H]
    \centering
    \includegraphics[width=1\textwidth, height = 8cm]{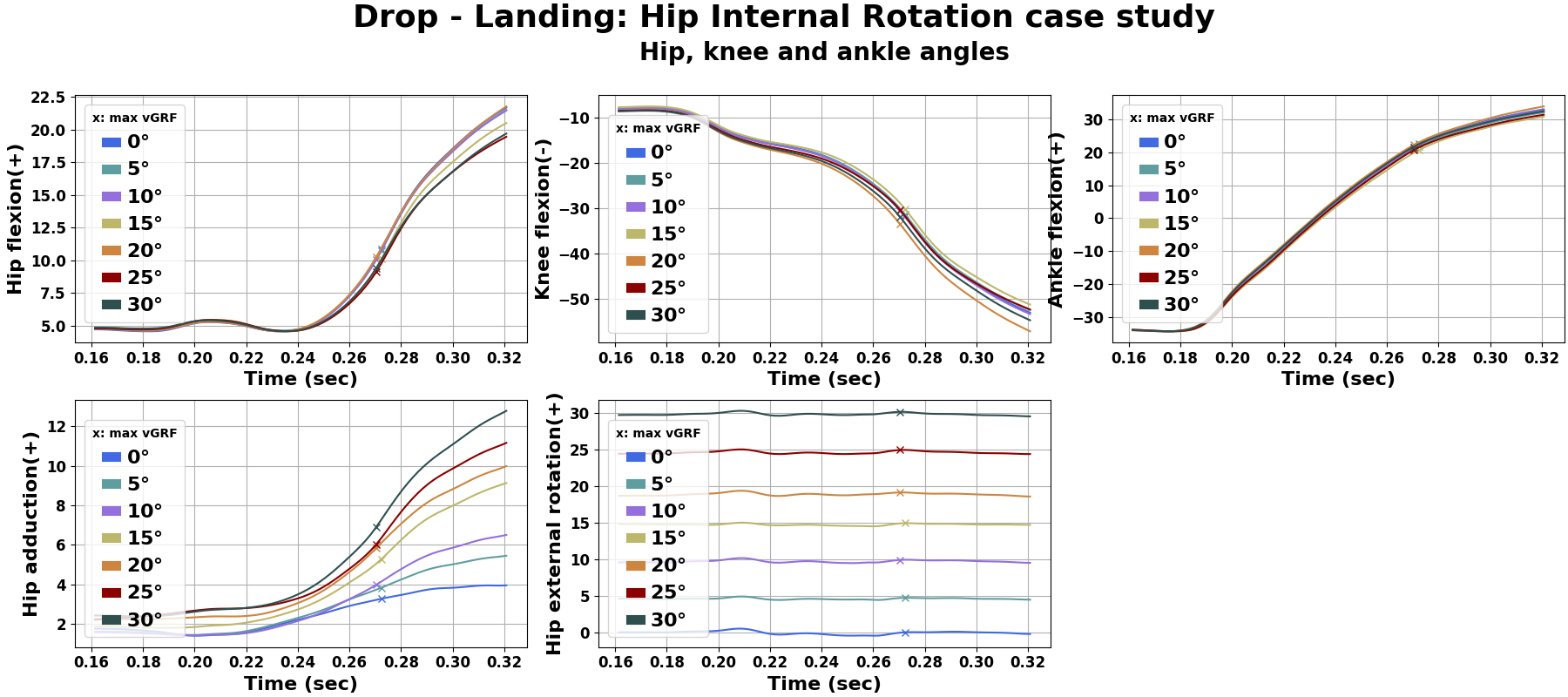}
    \caption{Hip, knee and ankle kinematics for various hip internal rotation angles.}
    \label{fig:predict_hip_effort0.001_internal_sol}
\end{figure}

Initially, in \autoref{fig:predict_hip_effort0.001_internal_sol} we provide the hip, knee and ankle flexion angles along with hip adduction and hip rotation angles for the hip internal rotation scenarios. As displayed in \autoref{tab:hip_dofs}, these \gls{dofs} were restricted to their respective bounds. For hip, knee and ankle flexion we can detect small deviations between the cases. Nevertheless, hip adduction's bounds were more relaxed and we can identify deviations between the cases. It is noticeable that as we impose greater hip internal rotation angles, the adduction of the hip is also increased.  

\begin{figure}[H]
    \centering
    \includegraphics[width=1\textwidth, height = 8cm]{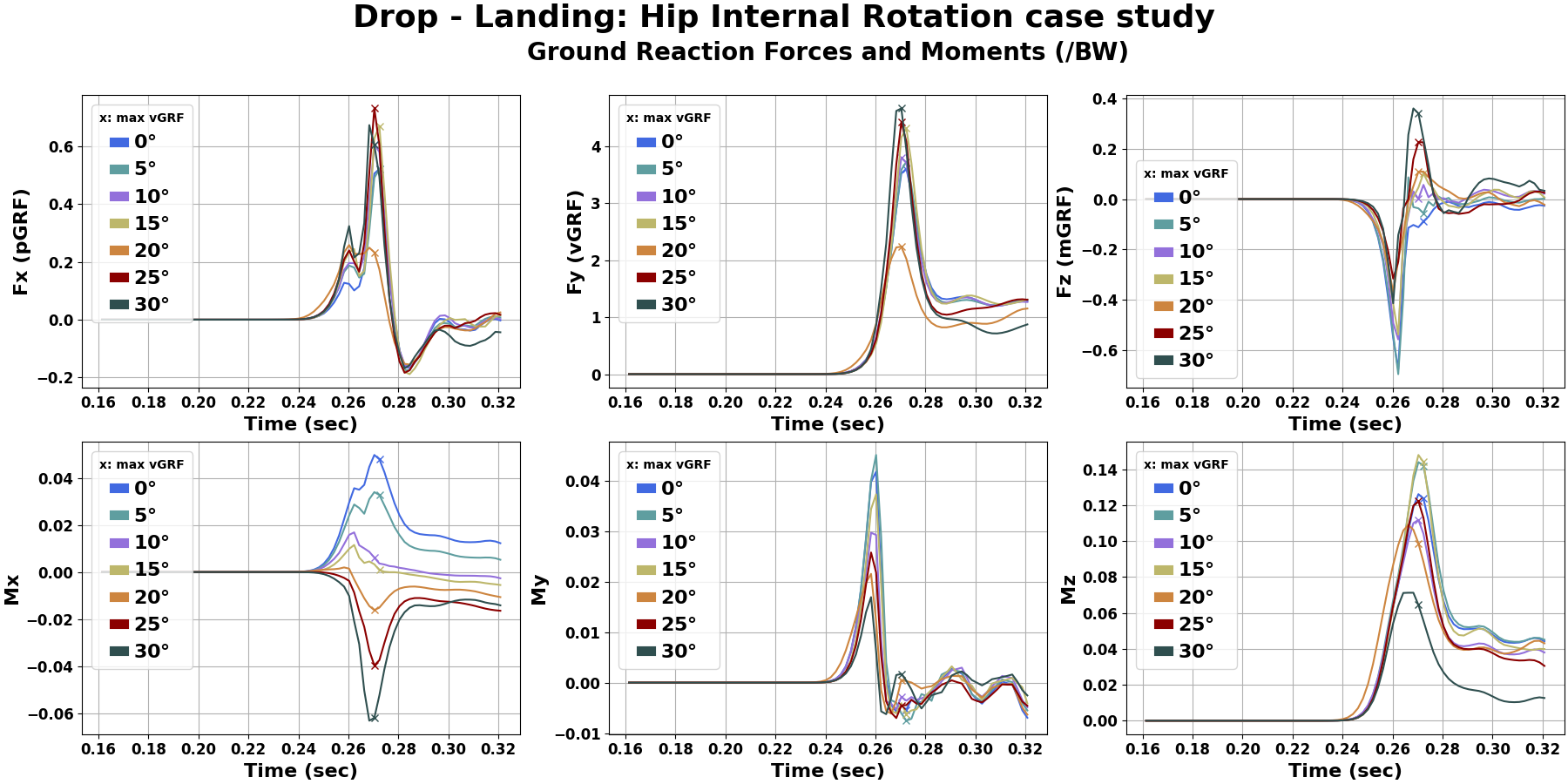}
    \caption{\gls{grf} and \gls{grm} for various hip internal rotation angles.}
    \label{fig:predict_hip_effort0.001_internal_GRF}
\end{figure}

Subsequently, in \autoref{fig:predict_hip_effort0.001_internal_GRF} we demonstrate \gls{grf} and \gls{grm} for scenarios with different hip internal rotation angles. We can observe that as the hip internal rotation is increased, the \gls{vgrf} amplitude is higher and for the \ang{30} case it obtains a value greater than 4 times \gls{bw}. The only exception is for \ang{20} where \gls{vgrf} derive a significantly smaller value. The same pattern is detected for \gls{pgrf} at max \gls{vgrf}.

\begin{figure}[H]
    \centering
    \includegraphics[width=1\textwidth, height = 8cm]{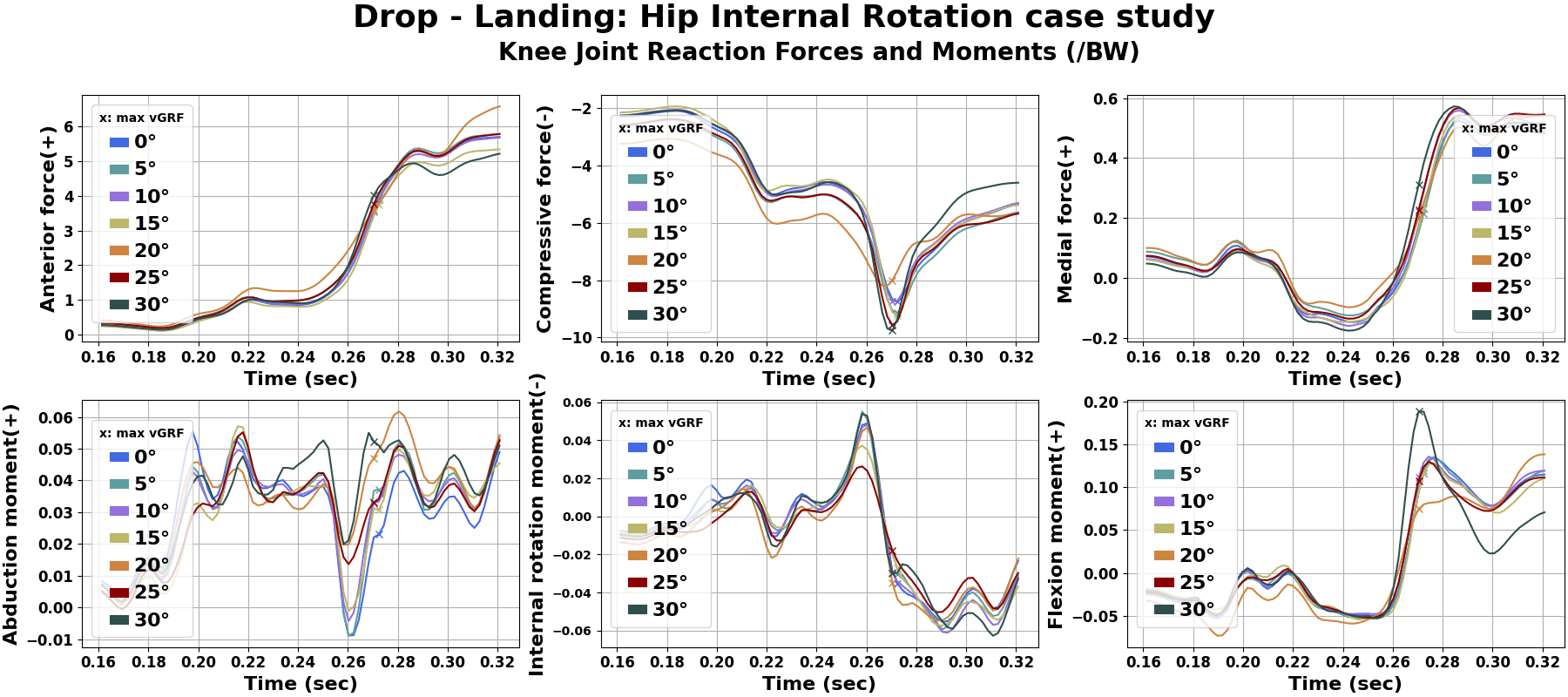}
    \caption{Knee joint \gls{jrf} and \gls{jrm} for various hip internal rotation angles.}
    \label{fig:predict_hip_effort0.001_internal_JRA}
\end{figure}

Moreover, in \autoref{fig:predict_hip_effort0.001_internal_JRA} we demonstrate  \gls{kjrf} and \gls{kjrm} for all cases. No great differences can be observed for the \gls{af} among the different cases. Also, regarding \gls{cf} the lower value is detected for the \ang{20} hip internal rotation case around its max \gls{vgrf}. However, we can notice that the general pattern of this graphs depicts that as the hip internal rotation increases \gls{cf} also increases. For \gls{abdm} and \gls{erm} no comparative observation can be easily made. Nevertheless, \gls{fm} obtains a peak value around peak \gls{vgrf}. This peak obtains the greatest value for the \ang{30} hip internal rotation and the lowest for the \ang{20} hip internal rotation scenario.

\begin{figure}[H]
    \centering
    \includegraphics[width=\textwidth, height = 8cm]{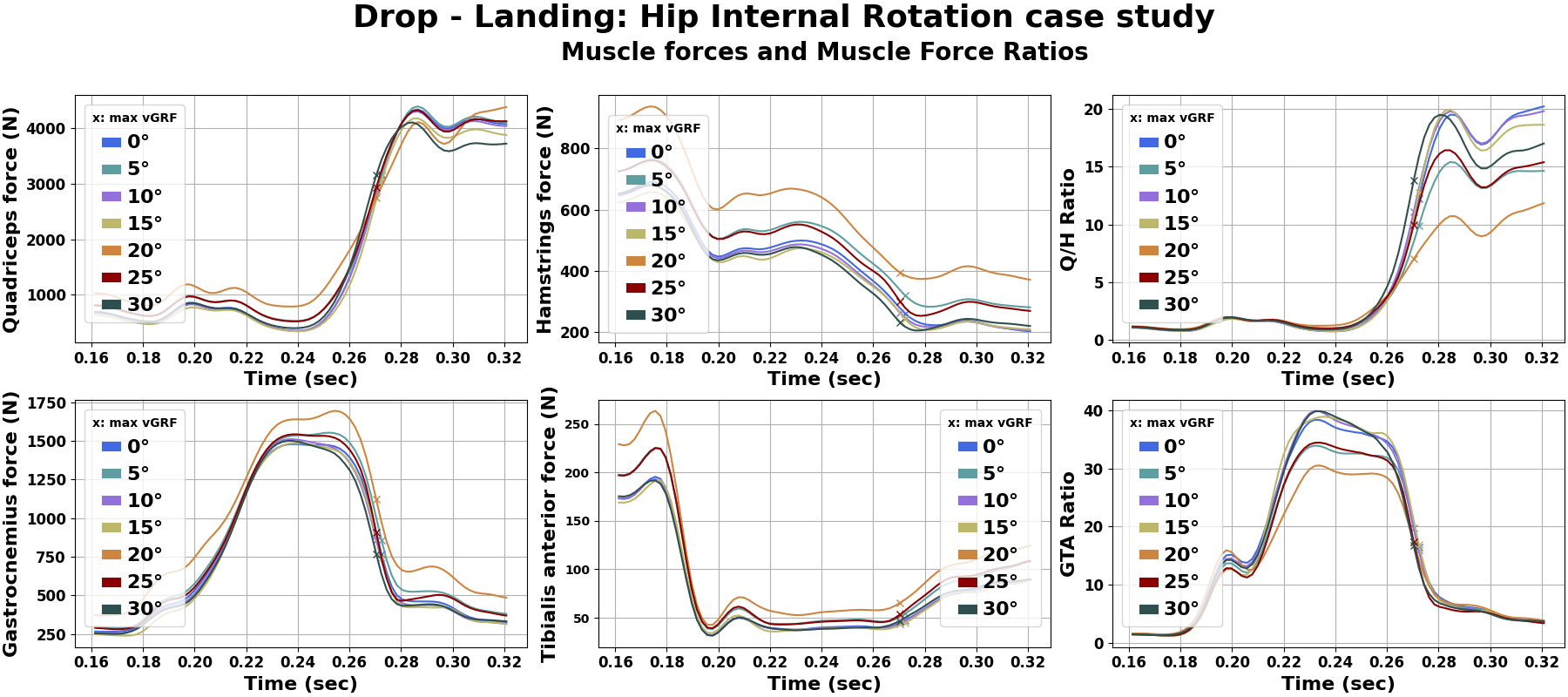}
    \caption{Muscle forces and muscle force ratios for various hip internal rotation angles.}
    \label{fig:predict_hip_effort0.001_internal_muscle_forces}
\end{figure}

Next, in \autoref{fig:predict_hip_effort0.001_internal_muscle_forces} we demonstrate  muscle forces and muscle force ratios for the various hip internal rotation angles. We can detect greater hamstrings force for the \ang{20} hip internal rotation. Correspondingly, \gls{q/h} obtain lower value for that scenario at max \gls{vgrf}. However, for the other scenarios we can not derive a clear description pattern.

Finally, in \autoref{tab:hip_internal_1} and \autoref{tab:hip_internal_2}, values for all examined parameters at maximum \gls{vgrf} for all cases are presented. We can state that the greatest \gls{vgrf} value is detected for the \ang{30} hip internal rotation scenario, while the lowest one is observed for the \ang{20} scenario. If we exclude the \ang{20} scenario we can claim that as the hip rotation is increased, the peak \gls{vgrf} is also increased.

\begin{table}[H]
    \begin{center}
        \begin{tabular}{c c c c c c}
             \toprule
             Case & pGRF & vGRF & mGRF & Q/H\_ratio & GTA\_ratio \\   
             \midrule
             0 & 0.522 & 3.602 & -0.088 & 12.264 & 15.933 \\ 
             5 & 0.521 & 3.722 & -0.055 & 9.911 & 15.583 \\ 
             10 & 0.607 & 3.795 & 0.000 & 11.007 & 20.074  \\ 
             15 & 0.671 & 4.310 & 0.102 & 12.775 & 16.568 \\ 
             20 & 0.233 & 2.234 & 0.108 & 7.016 & 17.291 \\ 
             25 & 0.732 & 4.417 & 0.226 & 9.860 & 17.736 \\ 
             30 & 0.605 & 4.658 & 0.340 & 13.880 & 16.145\\ 
              \bottomrule
        \end{tabular}
        \caption{\gls{grf}, \gls{q/h} and \gls{gta} at the time instant of peak \gls{vgrf} for various hip internal rotation angles.}
        \label{tab:hip_internal_1}
    \end{center}
\end{table}

Also, in \autoref{tab:hip_internal_2} we can observe that all \gls{kjrf} obtains the greatest value for the \ang{30} case at maximum \gls{vgrf}. Also, \gls{abdm} for no internal rotation is lower compared to the other cases, while \gls{irm} is lower for \ang{25} of hip internal rotation at the same time instant. Regarding the \ang{20} hip internal rotation with the lowest peak \gls{vgrf}, we can detect the greatest \gls{abdm} and \gls{irm} values at peak \gls{vgrf} compared with the other scenarios.

\begin{table}[H]
    \begin{center}
        \begin{tabular}{c c c c c c c} 
             \toprule
             Case & AF(+) & CF(-) & MF(+) & AbdM(+) & IRM(-) & FM(+) \\   
             \midrule
             0 & 3.927 & -8.921 & 0.231 & 0.020 & -0.034 & 0.120 \\ 
             5 &3.982 & -9.333 & 0.207 & 0.031 & -0.022 & 0.125 \\ 
             10 & 3.578 & -9.037 & 0.208 & 0.021 & -0.030 & 0.120 \\  
             15 & 3.787 & -9.508 & 0.231 & 0.028 & -0.027 & 0.120 \\  
             20 & 3.605 & -8.089 & 0.201 & 0.044 & -0.043 & 0.080 \\  
             25 & 3.809 & -10.050 & 0.213 & 0.027 & -0.019 & 0.116 \\  
             30 & 4.068 & -10.133 & 0.310 & 0.052 & -0.028 & 0.217 \\ 
            \bottomrule
        \end{tabular}
        \caption{Knee joint \gls{jrf} and \gls{jrm} at the time instant of peak \gls{vgrf} for various hip internal rotation angles.}
        \label{tab:hip_internal_2}
    \end{center}
\end{table}

Next, we exhibit results regarding cases with different hip external rotation angles. As mentioned before, the angles examined were \ang{0}, \ang{5}, \ang{15}, \ang{20}, \ang{25} and \ang{30} hip external rotation. 

\begin{figure}[H]
    \centering
    \includegraphics[width=\textwidth, height = 8cm]{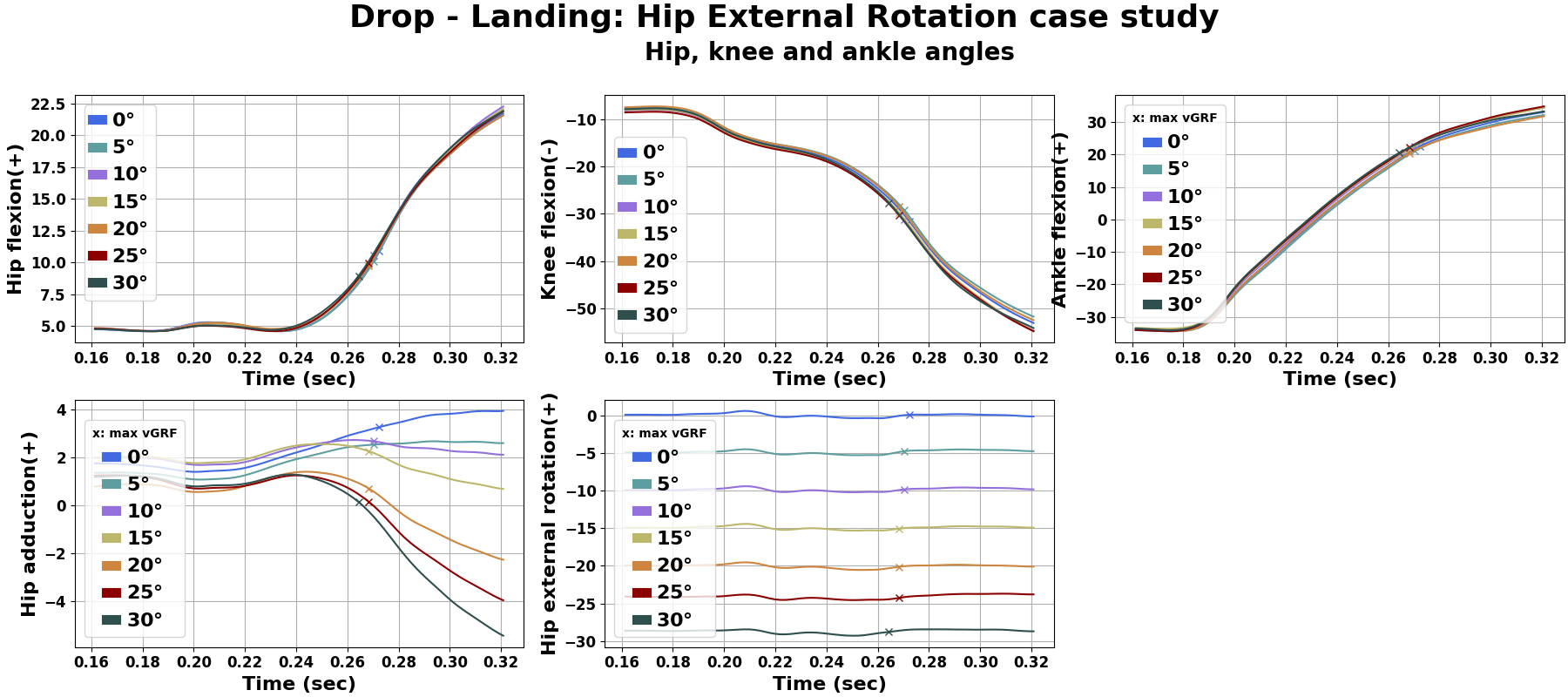}
    \caption{Hip, knee and ankle kinematics for various hip external rotation angles.}
    \label{fig:predict_hip_effort0.001_external_sol}
\end{figure}

First, in \autoref{fig:predict_hip_effort0.001_external_sol} we provide the hip, knee and ankle flexion angles for the examined scenarios. Also angles for hip adduction and external rotation are plotted. Again, hip adduction bounds are more relaxed and we can see that at max \gls{vgrf} for lower hip external rotation values (\ang{0}, \ang{5}, \ang{10} and \ang{15}), the hip is more adducted, while for the \ang{20}, \ang{25} and \ang{30} cases hip is less adducted.  

\begin{figure}[H]
    \centering
    \includegraphics[width=\textwidth, height = 8cm]{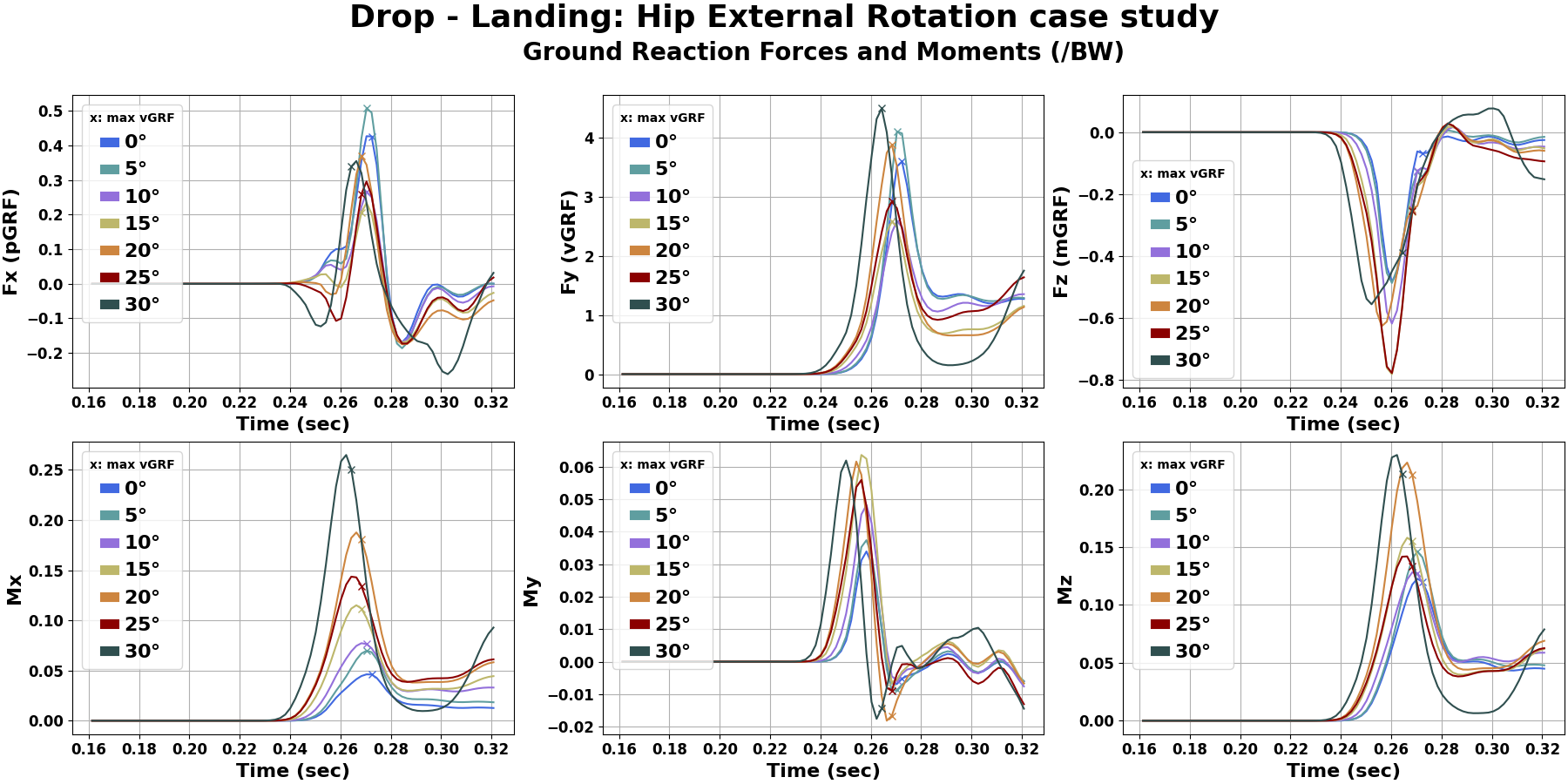}
    \caption{\gls{grf} and \gls{grm} for various hip external rotation angles.}
    \label{fig:predict_hip_effort0.001_external_GRF}
\end{figure}

Next, in \autoref{fig:predict_hip_effort0.001_external_GRF} we demonstrate \gls{grf} and \gls{grm} for scenarios with different hip external rotation angles. In ascending order of  \gls{vgrf} peak value the examined cases are: \ang{15}, \ang{10}, \ang{25}, \ang{0}, \ang{20}, \ang{5} and \ang{30} respectively. We notice that the minimum peak \gls{vgrf} values are for \ang{15} and \ang{10} hip external rotation. 

\begin{figure}[H]
    \centering
    \includegraphics[width=\textwidth, height = 8cm]{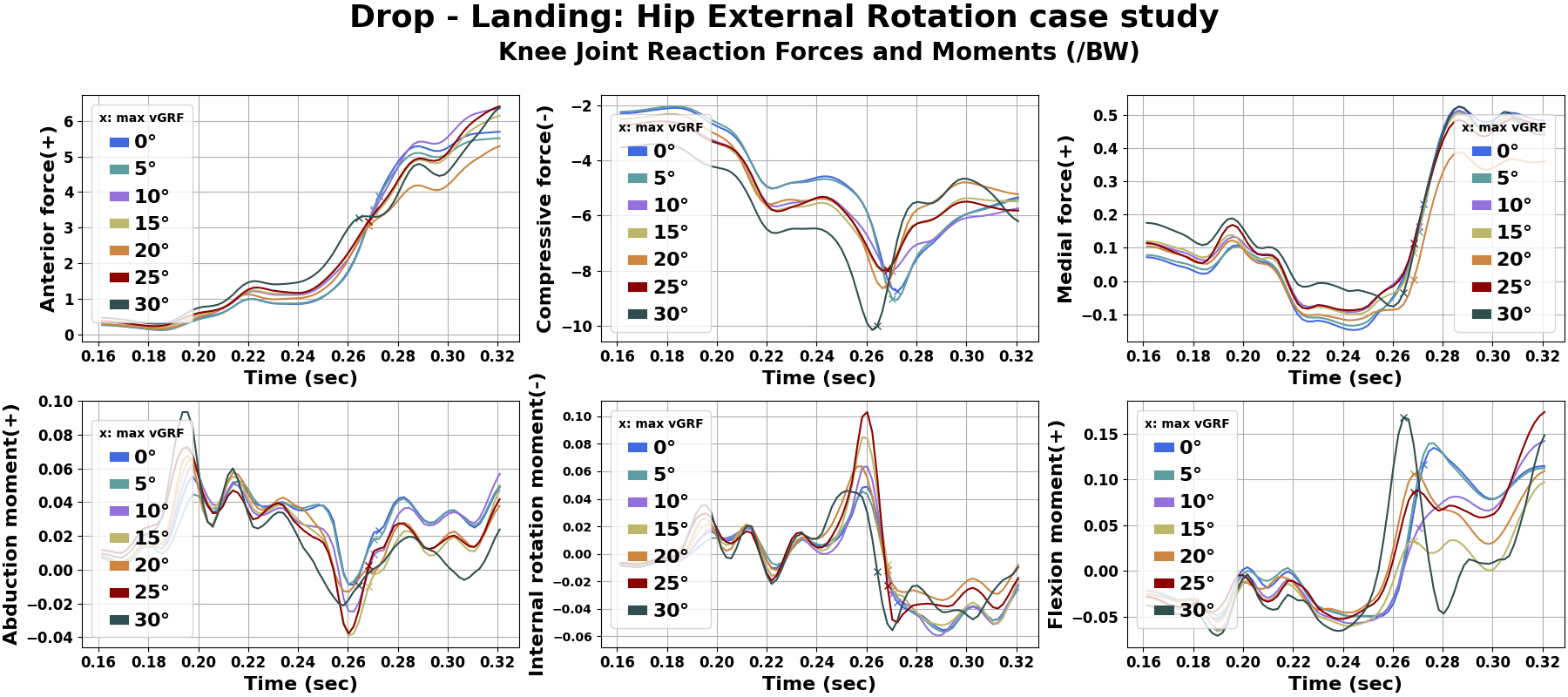}
    \caption{\gls{kjrf} and \gls{kjrm} for various hip external rotation angles.}
    \label{fig:predict_hip_effort0.001_external_JRA}
\end{figure}

Moreover, in \autoref{fig:predict_hip_effort0.001_external_JRA} we demonstrate \gls{kjrf} and \gls{kjrm} for different cases with externally rotated hip. No great differences are observed in moments and forces at the knee joint among the studied scenarios with a few exceptions. For example, \gls{cf} is higher for the \ang{30} cases close to maximum \gls{vgrf} time. Moreover, in the majority of the landing phase an \gls{erm} is introduced.

\begin{figure}[H]
    \centering
    \includegraphics[width=\textwidth, height = 8cm]{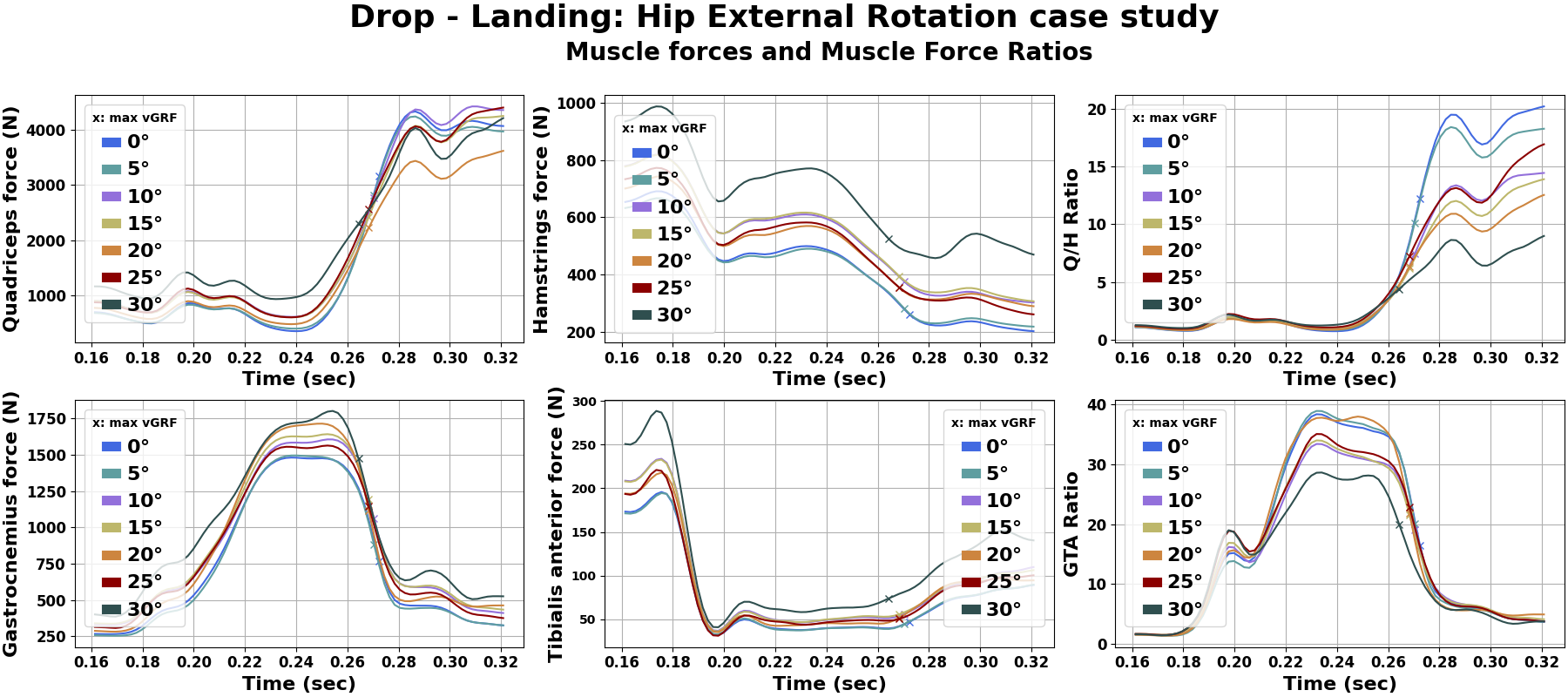}
    \caption{Muscle forces and muscle force ratios for various hip external rotation angles.}
    \label{fig:predict_hip_effort0.001_external_muscle_forces}
\end{figure}

Following, we demonstrate muscle forces and muscle force ratios for the examined cases in \autoref{fig:predict_hip_effort0.001_external_muscle_forces}. It is obvious that for \ang{30} hip external rotation the hamstrings, gastrocnemius and tibialis anterior forces are greater compared with the remaining scenarios at max \gls{vgrf}. That results in lower \gls{q/h} and \gls{gta} values at the same time instant.

Also, in \autoref{tab:hip_external_1} and \autoref{tab:hip_external_2}, values of parameters at maximum \gls{vgrf} for all cases are presented. We can notice that as the hip external rotation angle  is increased, the \gls{q/h} is decreased. Also, the lowest peaks of \gls{vgrf} are detected for \ang{10} and \ang{15} hip external rotation following by the case with no hip rotation. 

\begin{table}[H]
    \begin{center}
        \begin{tabular}{c c c c c c}
             \toprule
            Case & pGRF & vGRF & mGRF & Q/H\_ratio & GTA\_ratio \\   
             \midrule
             0 & 0.522 & 3.602 & -0.088 & 12.264 & 15.933 \\ 
             5 & 0.618 & 4.108 & -0.221 & 10.065 & 20.442 \\ 
             10 & 0.309 & 2.593 & -0.180 & 7.399 & 19.044  \\ 
             15 & 0.241 & 2.584 & -0.212 & 6.141 & 21.490 \\ 
             20 & 0.434 & 3.888 & -0.299 & 6.307 & 21.426 \\ 
             25 & 0.320 & 2.924 & -0.209 & 7.250 & 22.630  \\ 
             30 & 0.386 & 4.501 & -0.418 & 4.382 & 19.546 \\ 
              \bottomrule
        \end{tabular}
        \caption{\gls{grf}, \gls{q/h} and \gls{gta} at the time instant of peak \gls{vgrf} for various hip external rotation angles.}
        \label{tab:hip_external_1}
    \end{center}
\end{table}

Regarding \gls{af}, the lowest value is spotted for the scenario with \ang{20} hip external rotation, following by \ang{15} while the highest value is detected for the case with no hip rotation. The highest value for \gls{addm} is detected for the \ang{30} case at max \gls{vgrf} while the highest value for the \gls{irm} is spotted for the scenario with no rotation. 

\begin{table}[H]
    \begin{center}
        \begin{tabular}{c c c c c c c} 
             \toprule
             Case & AF(+) & CF(-) & MF(+) & AbdM(+) & IRM(-) & FM(+) \\
             \midrule
             0 & 3.927 & -8.921 & 0.231 & 0.020 & -0.034 & 0.120 \\ 
             5 & 3.554 & -9.356 & 0.129 & 0.005 & -0.024 & 0.114 \\ 
             10 & 3.486 & -8.123 & 0.158 & 0.002 & -0.027 & 0.050 \\
             15 & 3.167 & -8.051 & 0.088 & -0.003 & -0.012 & 0.035 \\ 
             20 & 3.069 & -8.694 & -0.005 & -0.002 & -0.010 & 0.108 \\ 
             25 & 3.203 & -8.071 & 0.119 & 0.010 & -0.033 & 0.088 \\  
             30 & 3.298 & -10.178 & -0.046 & -0.008 & -0.009 & 0.171 \\
            \bottomrule
        \end{tabular}
        \caption{Knee joint \gls{jrf} and \gls{jrm} at the time instant of peak \gls{vgrf} for various hip external rotation angles.}
        \label{tab:hip_external_2}
    \end{center}
\end{table}

\section{Trunk orientation case}\label{sec:trunk_results}

In this section we demonstrate results regarding single - leg landings corresponding to different trunk orientations. First, we will present simulation results regarding trunk leaning forward and backward. This corresponds to lumbar flexion and extension respectively. Then, we will continue with results concerning trunk right and left bending (lumbar bending). Finally, we will demonstrate simulation results for landings with the trunk internally and externally rotated (lumbar internal and external rotation respectively).  

\subsection{Trunk forward - backward leaning}

Starting from different lumbar flexion - extension angles, we examine the upright position (without any flexion, bending or rotation angle), landings with lumbar flexion (\ang{0}, \ang{5}, \ang{10}, \ang{15}, \ang{20}, \ang{25} and \ang{30}) and landings with lumbar extension (\ang{0}, \ang{5}, \ang{10}, \ang{15} and \ang{20}).

\begin{figure}[H]
    \centering
    \includegraphics[width=1\textwidth, height = 4cm]{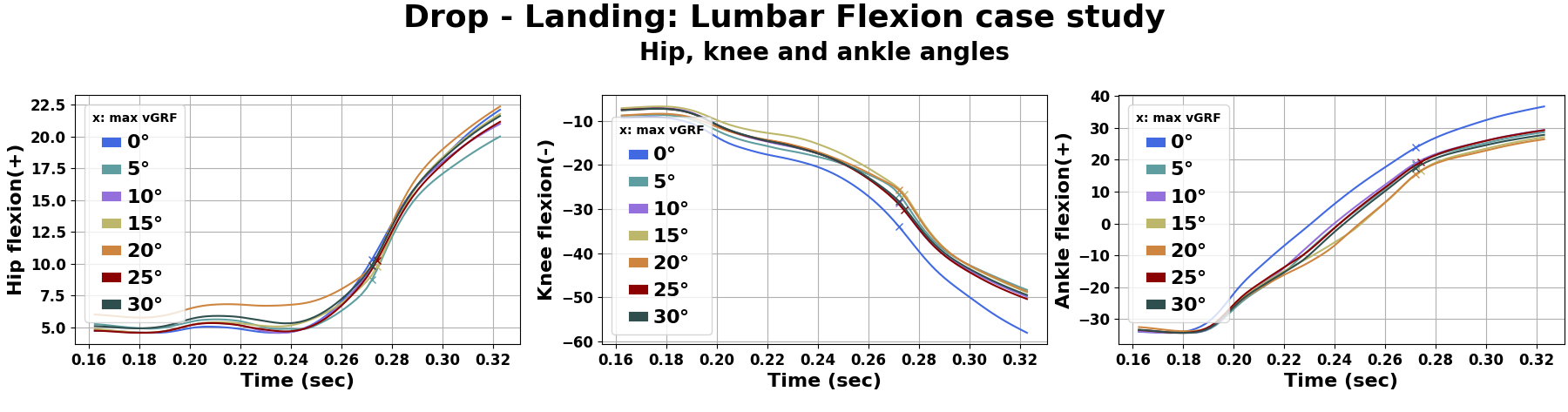}
    \caption{Hip, knee and ankle kinematics for multiple lumbar flexion angles.}
    \label{fig:predict_trunk_flexion_sol}
\end{figure}

In \autoref{fig:predict_trunk_flexion_sol} we display hip, knee and ankle joint angles for the scenarios with lumbar flexion. We can notice that for the upright case knee flexion is greater and ankle flexion is lower compared with the remaining cases. Also, the cases with \ang{15} and \ang{20} trunk flexion presents the lowest knee flexion for the entire motion.

\begin{figure}[H]
    \centering
    \includegraphics[width=1\textwidth, height = 8cm]{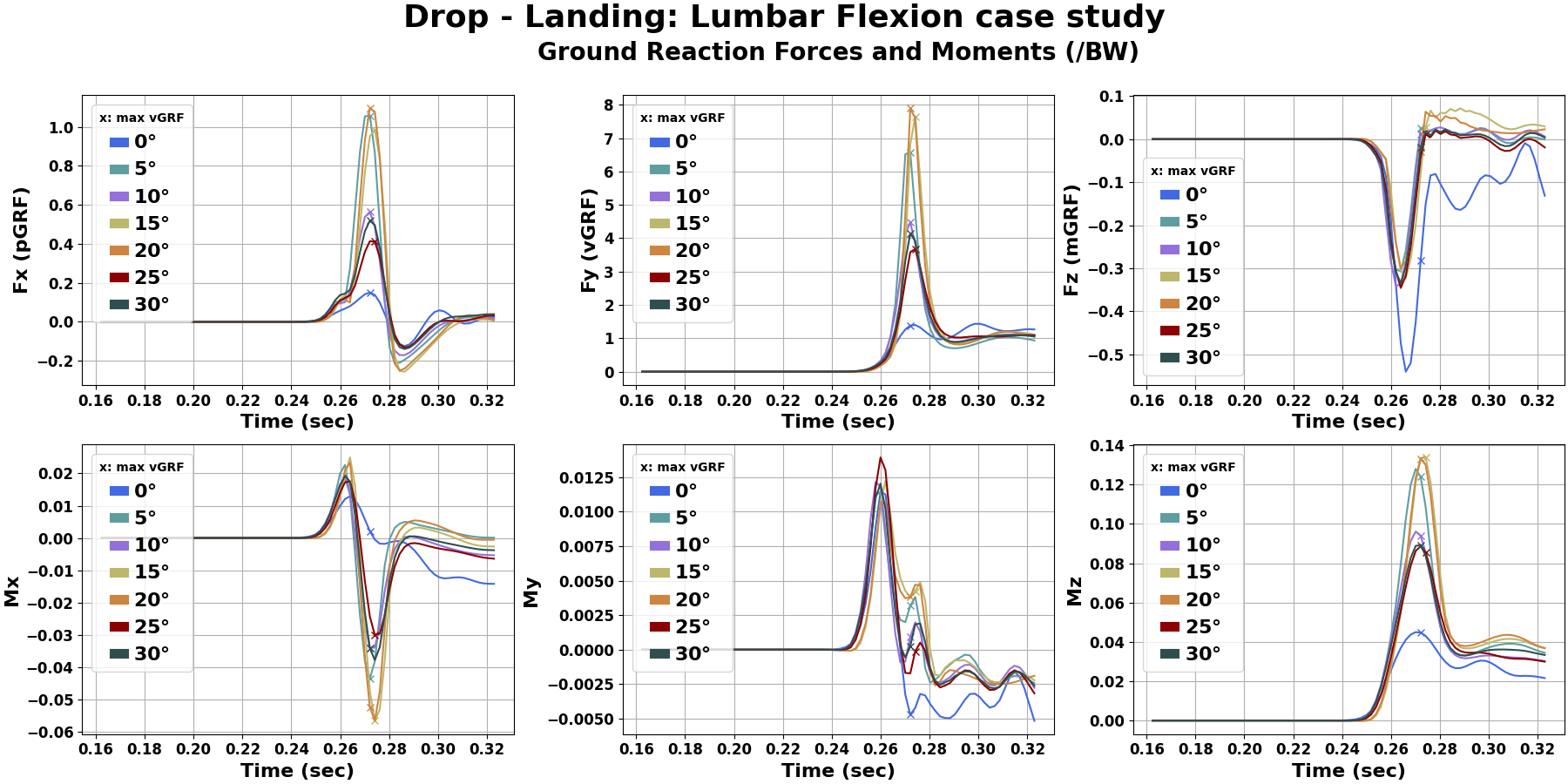}
    \caption{\gls{grf} and \gls{grm} for multiple lumbar flexion angles.}
    \label{fig:predict_trunk_flexion_GRF}
\end{figure}

Next, in \autoref{fig:predict_trunk_flexion_GRF} we demonstrate \gls{grf} and \gls{grm} for cases with multiple trunk flexion angles. We can mark that the lower \gls{vgrf} peak is spotted for the upright position of the upper body following by the case with \ang{25} trunk flexion and then the cases with \ang{30}, \ang{10}, \ang{5}, \ang{15} and \ang{20} in ascending order. Almost the same pattern is detected for the \gls{pgrf} at maximum \gls{vgrf}.

\begin{figure}[H]
    \centering
    \includegraphics[width=\textwidth, height = 8cm]{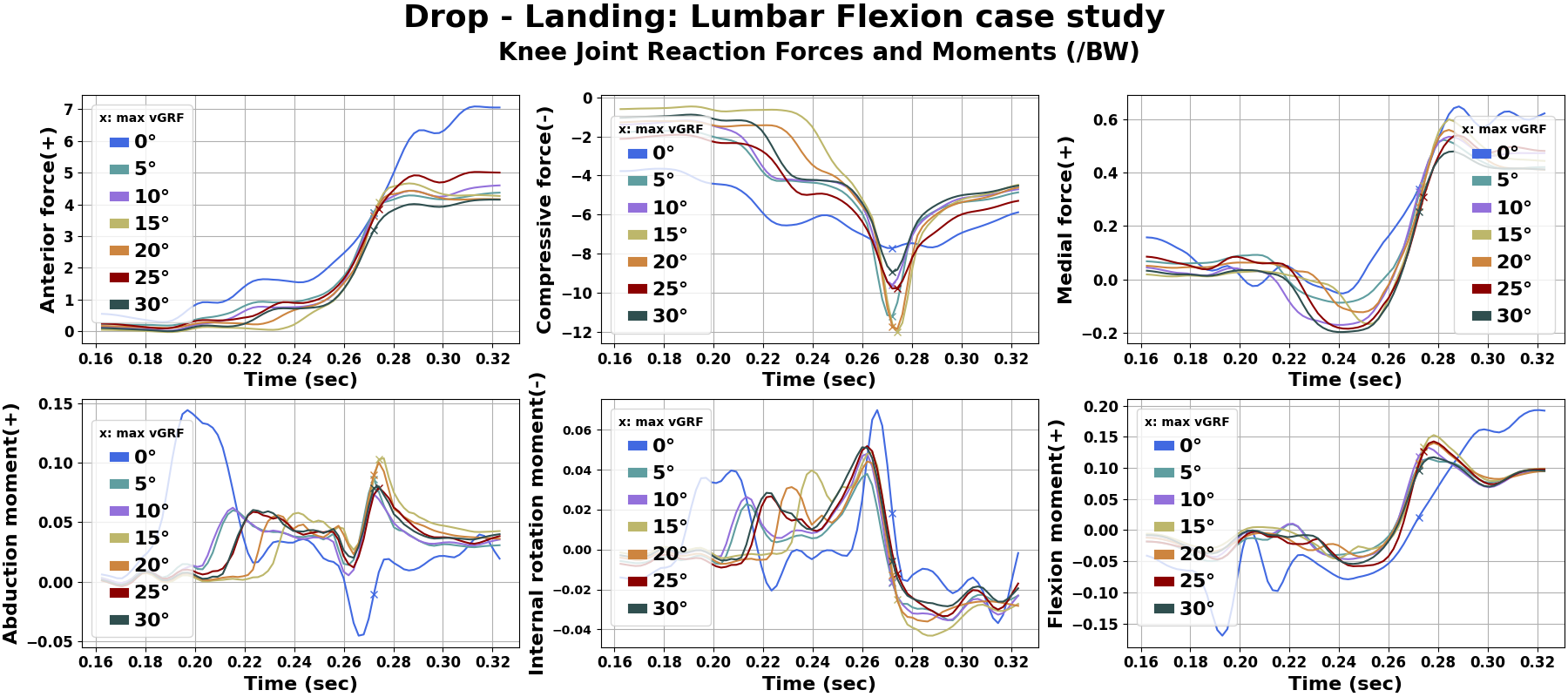}
    \caption{Knee joint \gls{jrf} and \gls{jrm} for multiple lumbar flexion angles.}
    \label{fig:predict_trunk_flexion_JRA}
\end{figure}

Subsequently, in \autoref{fig:predict_trunk_flexion_JRA} we demonstrate  \gls{kjrf} and \gls{kjrm} for the respective scenarios. It is noticeable that before and after peak \gls{vgrf}, \gls{af} is greater for the upright position. For \gls{cf}, the upright position shows a different behavior compared with the remaining cases with greater values except of the time around peak \gls{vgrf} where it takes the lowest value. Again, we notice a deviation in the behavior of \gls{abdm} and \gls{erm} for the upright case. Comparisons between the cases cannot be easily made.

\begin{figure}[H]
    \centering
    \includegraphics[width=\textwidth, height = 8cm]{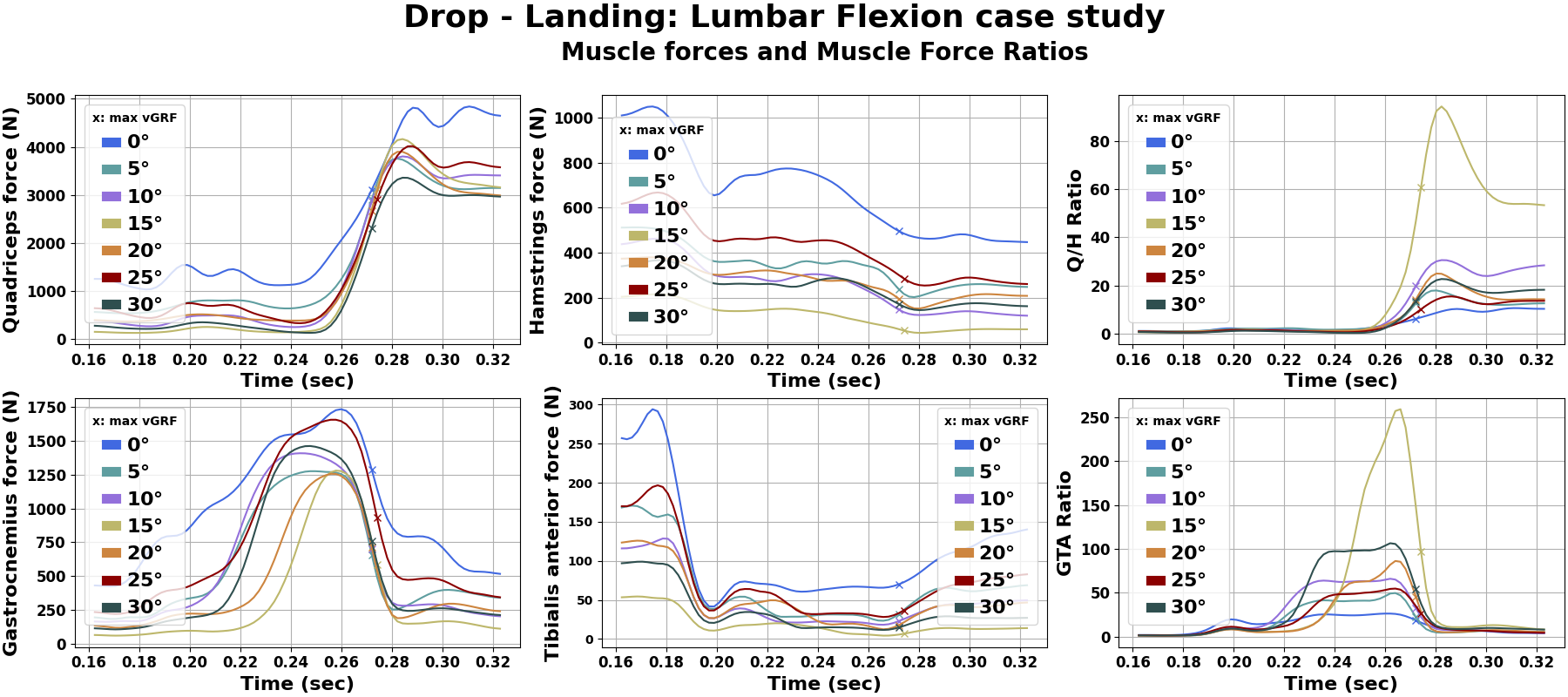}
    \caption{Muscle forces and muscle force ratios for multiple lumbar flexion angle s.}
    \label{fig:predict_trunk_flexion_muscle_forces}
\end{figure}

In \autoref{fig:predict_trunk_flexion_muscle_forces} we demonstrate  muscle forces and muscle force ratios for scenarios with multiple trunk flexion - extension angles. We can see that for the upright position quadriceps, hamstrings, gastrocnemius and tibialis anterior forces obtain higher values compared with other cases with trunk flexion. As a result, at maximum \gls{vgrf} both \gls{q/h} and \gls{gta} obtain lower values for the upright case. The greatest ratios are observed for the case with \ang{15} trunk flexion.

Moreover, in \autoref{tab:trunk_flexion_1} and \autoref{tab:trunk_flexion_2}, values for parameters at the peak \gls{vgrf} for all cases are presented. If we take a closer look at \gls{vgrf} we can notice that the lowest \gls{vgrf} and \gls{q/h} values are detected for \ang{0}, and then \ang{25} trunk flexion.

\begin{table}[H]
    \begin{center}
        \begin{tabular}{c c c c c c}
             \toprule
             Case & pGRF & vGRF & mGRF & Q/H\_ratio & GTA\_ratio \\   
             \midrule
             0 &  0.167 & 1.383 & -0.176 & 6.275 & 18.351 \\ 
             5 & 1.380 & 6.571 & 0.016 & 12.713 & 20.565 \\ 
             10 & 0.725 & 4.485 & -0.019 & 19.775 & 30.855  \\ 
             15 & 1.412 & 7.647 & 0.055 & 59.534 & 81.153 \\ 
             20 & 1.645 & 7.892 & 0.095 & 13.545 & 42.070 \\ 
             25 & 0.505 & 3.674 & -0.038 & 10.211 & 25.128 \\ 
             30 & 0.673 & 4.139 & 0.022 & 13.440 & 52.721\\ 
              \bottomrule
        \end{tabular}
        \caption{\gls{grf}, \gls{q/h} and \gls{gta} at the time instant of peak \gls{vgrf} for multiple lumbar flexion angles.}
        \label{tab:trunk_flexion_1}
    \end{center}
\end{table}

Also, in \autoref{tab:trunk_flexion_2} we notice that for the upright case \gls{abdm} is lower at max \gls{vgrf} compared with the remaining scenarios. Nonetheless, \gls{cf} obtains lower value for the upright case, while \gls{af} obtains the lowest value for the \ang{30} case following by \ang{10}, \ang{0} and \ang{5} scenarios respectively.

\begin{table}[H]
    \begin{center}
        \begin{tabular}{c c c c c c c} 
             \toprule
            Case & AF(+) & CF(-) & MF(+) & AbdM(+) & IRM(-) & FM(+) \\  
             \midrule
             0 &  3.756 & -7.761 & 0.334 & 0.006 & 0.000 & 0.023 \\ 
             5 & 3.862 & -12.160 & 0.290 & 0.094 & -0.021 & 0.114 \\ 
             10 & 3.731 & -10.090 & 0.323 & 0.077 & -0.014 & 0.126 \\  
             15 & 4.245 & -13.329 & 0.373 & 0.105 & -0.032 & 0.140 \\  
             20 & 3.829 & -13.372 & 0.268 & 0.132 & -0.023 & 0.107 \\  
             25 & 3.899 & -10.024 & 0.308 & 0.072 & -0.013 & 0.129 \\  
             30 &  3.241 & -9.421 & 0.256 & 0.093 & -0.008 & 0.101 \\ 
            \bottomrule
        \end{tabular}
        \caption{Knee joint \gls{jrf} and \gls{jrm} at the time instant of peak \gls{vgrf} for multiple lumbar flexion angles.}
        \label{tab:trunk_flexion_2}
    \end{center}
\end{table}

\begin{figure}[H]
    \centering
    \includegraphics[width=1\textwidth, height = 4cm]{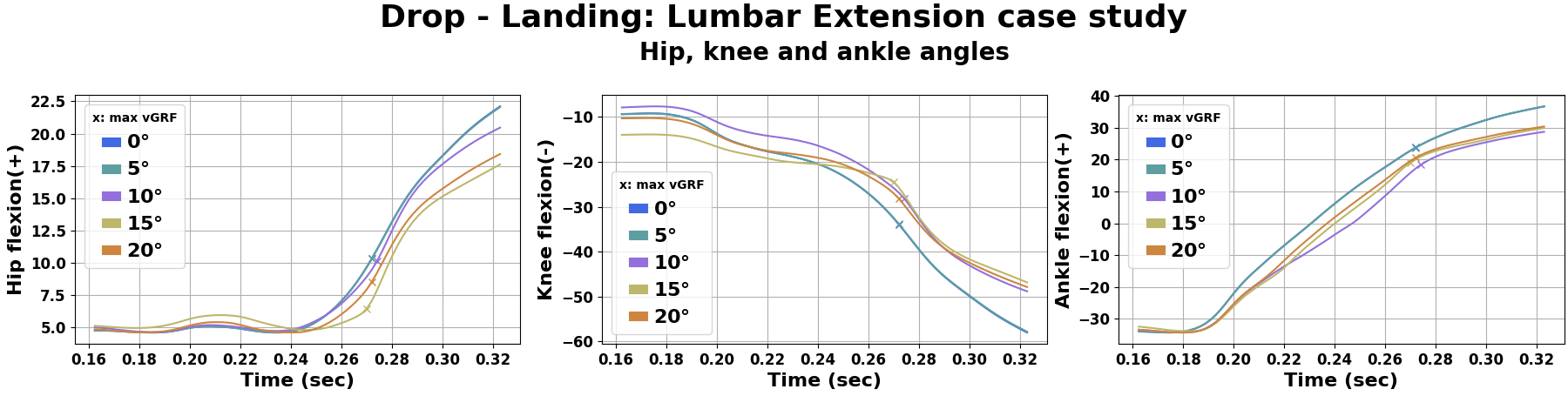}
    \caption{Hip, knee and ankle kinematics for multiple lumbar extension angles.}
    \label{fig:predict_trunk_extension_sol}
\end{figure}

Next, we present the resulted plots for the scenarios of single - leg landings with the trunk in extension. First, in \autoref{fig:predict_trunk_extension_sol} hip, knee and ankle flexion angles are displayed for multiple trunk extension angles. We observe that small deviations between the different cases exist for all three joint angles.

\begin{figure}[H]
    \centering
    \includegraphics[width=1\textwidth, height = 8cm]{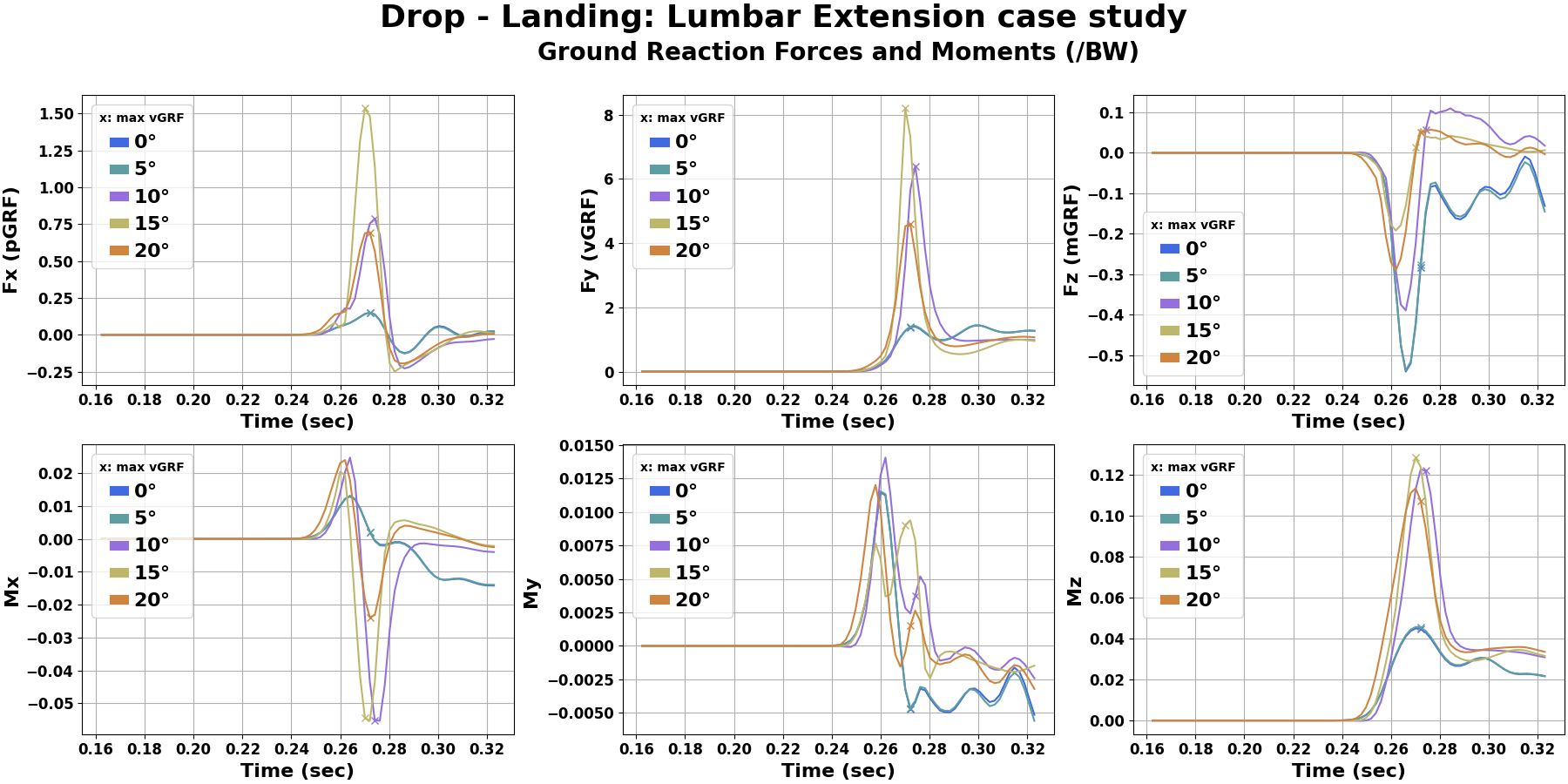}
    \caption{\gls{grf} and \gls{grm} for multiple lumbar extension angles.}
    \label{fig:predict_trunk_extension_GRF}
\end{figure}

Afterwards, in \autoref{fig:predict_trunk_extension_GRF} we demonstrate \gls{grf} and \gls{grm} for cases with different trunk extension angles. The lowest \gls{vgrf} peaks are observed for the cases with \ang{0} and \ang{5} following by \ang{20}, \ang{10} and \ang{15} trunk extension. The same order is noticed for \gls{pgrf} at peak \gls{vgrf}.

\begin{figure}[H]
    \centering
    \includegraphics[width=\textwidth, height = 8cm]{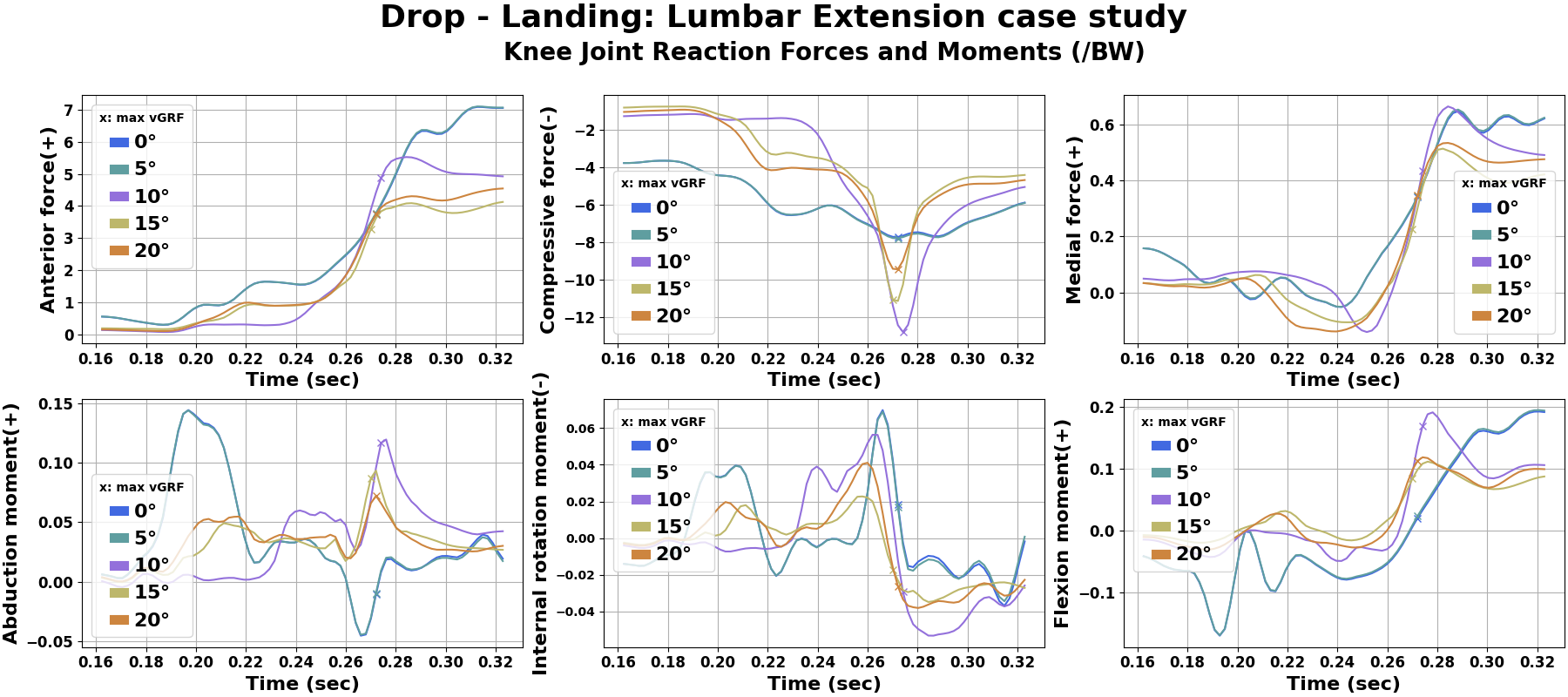}
    \caption{Knee joint \gls{jrf} and \gls{jrm} for multiple lumbar extension angles.}
    \label{fig:predict_trunk_extension_JRA}
\end{figure}

Following, in \autoref{fig:predict_trunk_extension_JRA} we demonstrate \gls{kjrf} and \gls{kjrm} for the examined scenarios. \gls{af} after max \gls{vgrf} time instant obtains greater peaks for the cases of lower trunk extension values. Though, \gls{cf} obtains lower peaks for cases of \ang{0} and \ang{5}, while the highest peak is for the case of \ang{10} of trunk extension.

Regarding \gls{abdm} and \gls{addm} on the knee joint we can detect different behavior for the different cases. Around \gls{igc}, drop - landings with \ang{0} and \ang{5} of trunk extension presents higher  \gls{abdm} values. For \gls{irm} and \gls{erm} things are more compatible between the scenarios. A first \gls{erm} peak is observed at \gls{igc} following by a greater one before  max \gls{vgrf}. 
of trunk extension.

\begin{figure}[H]
    \centering
    \includegraphics[width=\textwidth, height = 8cm]{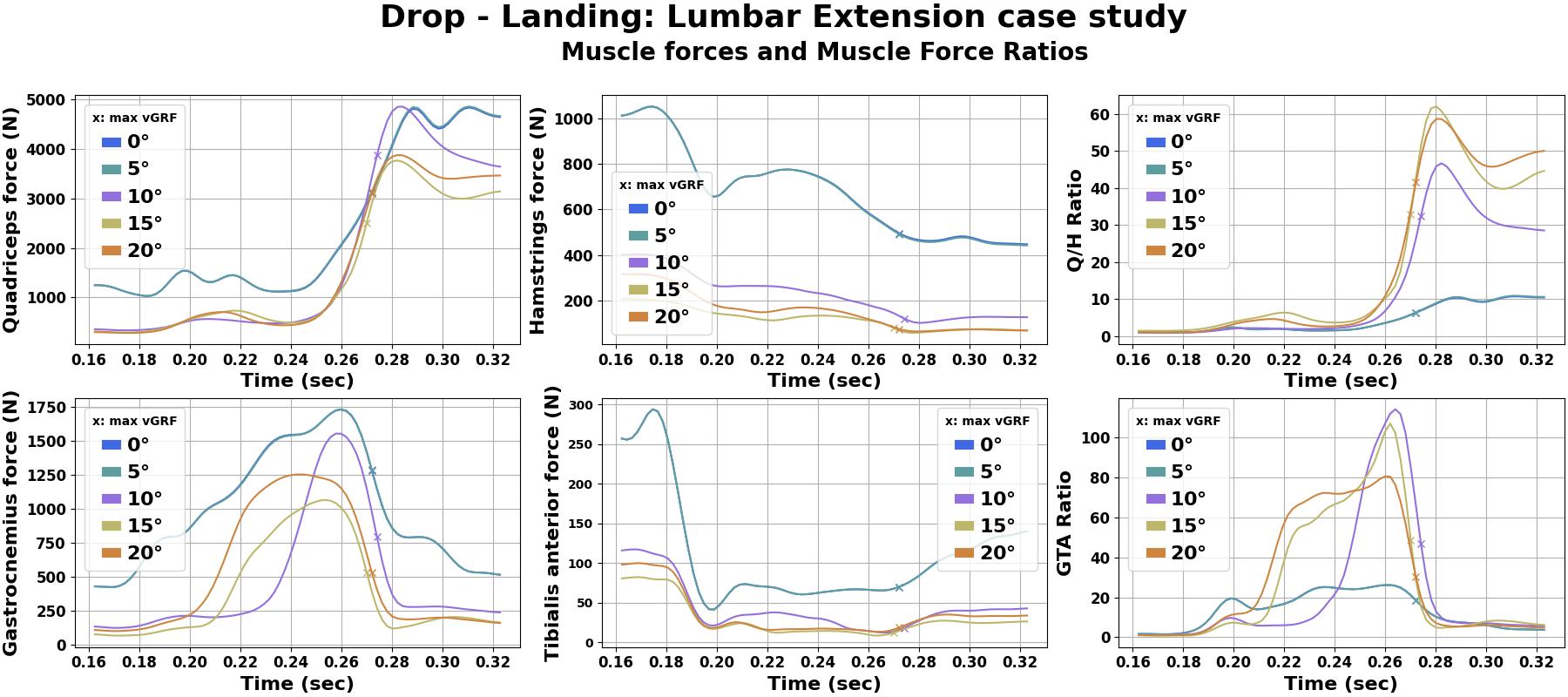}
    \caption{Muscle forces and muscle force ratios for multiple lumbar extension angles.}
    \label{fig:predict_trunk_extension_muscle_forces}
\end{figure}

In \autoref{fig:predict_trunk_extension_muscle_forces} we demonstrate  muscle forces and muscle force ratios. Hamstrings, gastrocnemius and tibialis anterior forces are greater for 0 and 5 degrees of trunk extension following by the rest of the scenarios. As a result, \gls{q/h} is lower for the previous cases.

In  \autoref{tab:trunk_extension_1} and \autoref{tab:trunk_extension_2}, values for parameters at the  maximum \gls{vgrf} for all cases are presented. We can verify the observation made previously regarding \gls{vgrf} and \gls{pgrf}, highest values for an extended trunk.  

\begin{table}[H]
    \begin{center}
        \begin{tabular}{c c c c c c}
             \toprule
             Case & pGRF & vGRF & mGRF & Q/H\_ratio & GTA\_ratio \\   
             \midrule
             0 &  0.167 & 1.383 & -0.176 & 6.275 & 18.351 \\ 
             5 &  0.165 & 1.401 & -0.171 & 6.391 & 18.483 \\ 
             10 &  1.083 & 6.403 & 0.074 & 32.214 & 42.112  \\ 
             15 & 2.211 & 8.198 & 0.006 & 30.940 & 40.500 \\ 
             20 & 0.857 & 4.593 & 0.031 & 41.636 & 28.596 \\ 
             \bottomrule
        \end{tabular}
        \caption{\gls{grf}, \gls{q/h} and \gls{gta} at the time instant of peak \gls{vgrf} for multiple lumbar extension angles.}
        \label{tab:trunk_extension_1}
    \end{center}
\end{table}

In \autoref{tab:trunk_extension_2} we can investigate in details the deviations between the cases at  maximum \gls{vgrf} time. \gls{af} takes the lower value for the case with \ang{15} trunk extension and the greater one for \ang{10} trunk extension. Moreover, \gls{cf} is lower for the scenario with no extension and higher for the case with \ang{10} trunk extension. Concerning \gls{abdm}, the greatest value is observed for \ang{10} trunk extension and the lowest for the cases of \ang{0} and \ang{5} lumbar extension.

\begin{table}[H]
    \begin{center}
        \begin{tabular}{c c c c c c c} 
             \toprule
             Case & AF(+) & CF(-) & MF(+) & AbdM(+) & IRM(-) & FM(+) \\  
             \midrule
             0 & 3.756 & -7.761 & 0.334 & 0.006 & 0.000 & 0.023 \\ 
             5 & 3.793 & -7.835 & 0.338 & 0.007 & -0.001 & 0.026 \\ 
             10 & 4.988 & -13.681 & 0.454 & 0.118 & -0.036 & 0.176 \\  
             15 & 3.507 & -12.850 & 0.254 & 0.082 & -0.029 & 0.091 \\  
             20 & 3.800 & -9.908 & 0.338 & 0.074 & -0.028 & 0.122 \\  
            \bottomrule
        \end{tabular}
        \caption{Knee joint \gls{jrf} and \gls{jrm} at the time instant of peak \gls{vgrf} for multiple trunk extension angles.}
        \label{tab:trunk_extension_2}
    \end{center}
\end{table}

\subsection{Trunk medial - lateral bending}

\begin{figure}[H]
    \centering
    \includegraphics[width=1\textwidth, height = 4cm]{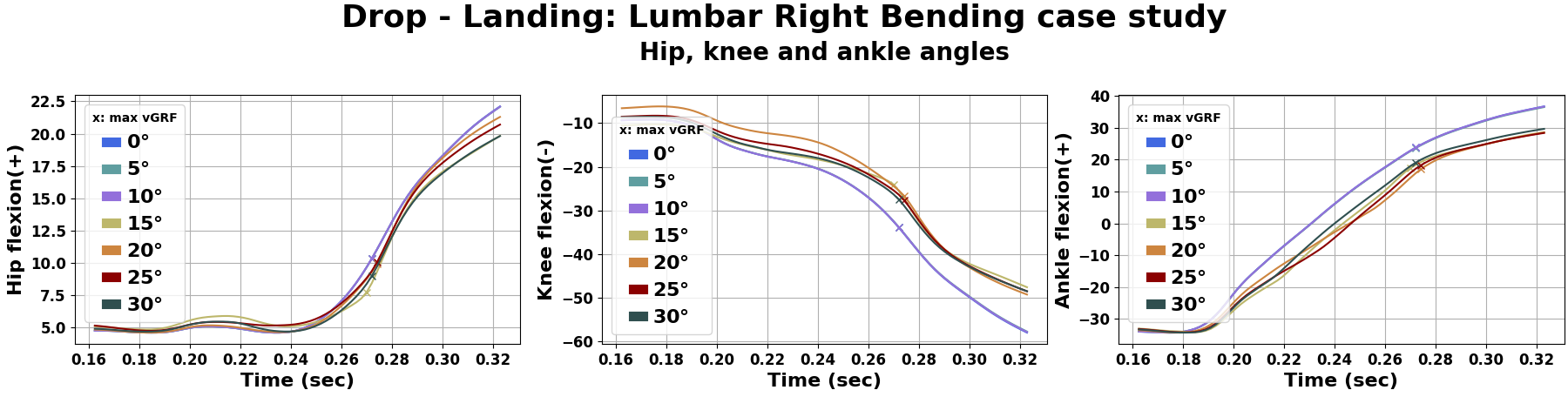}
    \caption{Hip, knee and ankle kinematics for multiple trunk right bending angles.}
    \label{fig:predict_trunk_ben_right_sol}
\end{figure}

In this section we demonstrate results for cases where the model lands with different trunk bending angles as described in \autoref{sec:trunk_info}. First, we present results about the scenarios of trunk right bending studies. Then, resulted plots about scenarios with left trunk bending are displayed. The investigated angles were \ang{0}, \ang{5}, \ang{10}, \ang{15}, \ang{20}, \ang{25} and \ang{30} for both right and left bending. Initially, in \autoref{fig:predict_trunk_ben_right_sol} we display angles for the hip, knee and ankle joints for scenarios of trunk right bending. 

\begin{figure}[H]
    \centering
    \includegraphics[width=1\textwidth, height = 8cm]{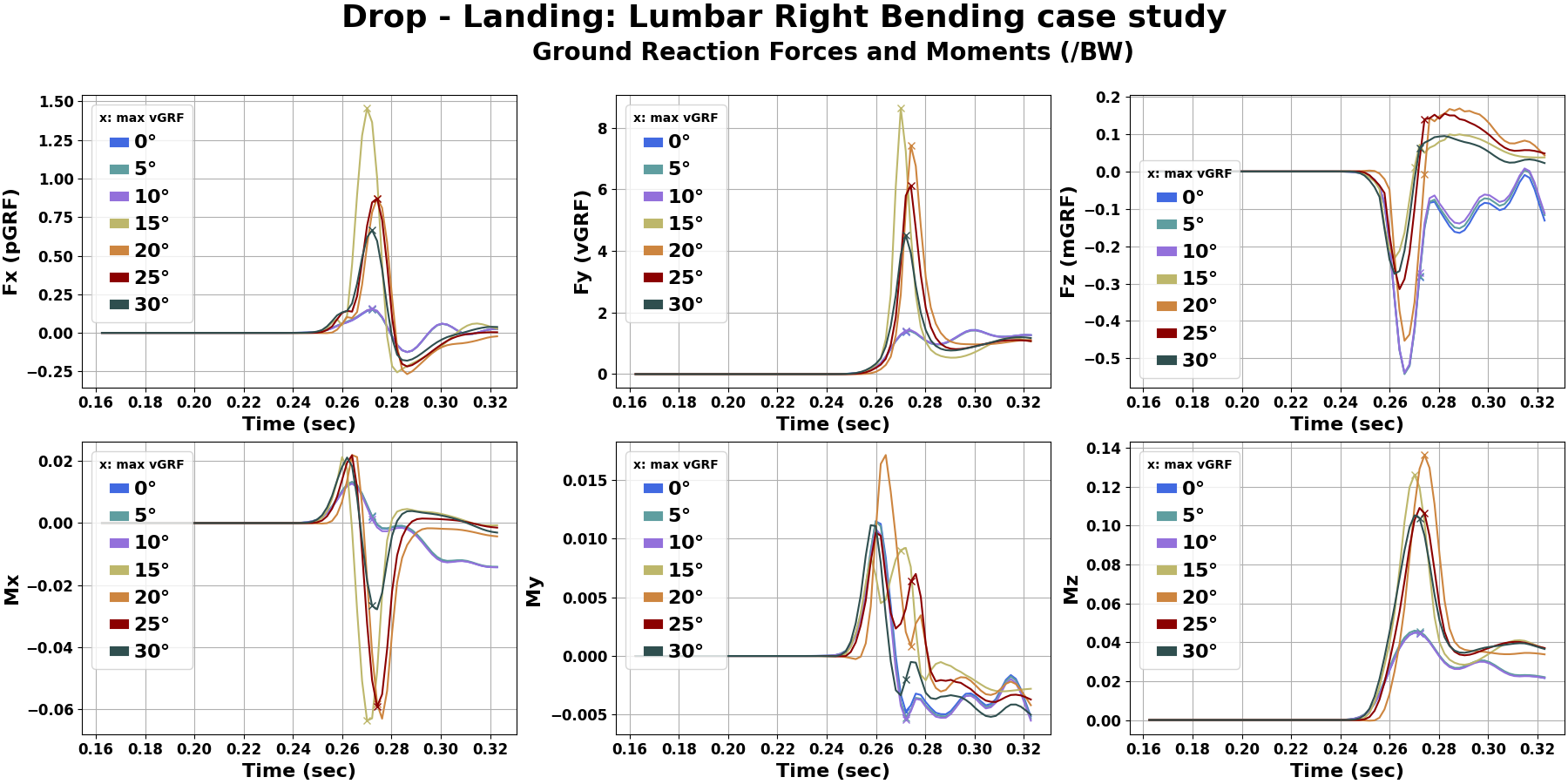}
    \caption{\gls{grf} and \gls{grm} for multiple trunk right bending angles.}
    \label{fig:predict_trunk_ben_right_GRF}
\end{figure}

In \autoref{fig:predict_trunk_ben_right_GRF} we demonstrate \gls{grf} and \gls{grm} for the respective cases. \gls{pgrf} and \gls{vgrf} peak values are lower for the cases with \ang{0}, \ang{5} and \ang{10} right bending while much greater peak values are observed for \ang{15}, \ang{20}, \ang{25} and \ang{30} trunk right bending.

\begin{figure}[H]
    \centering
    \includegraphics[width=\textwidth, height = 8cm]{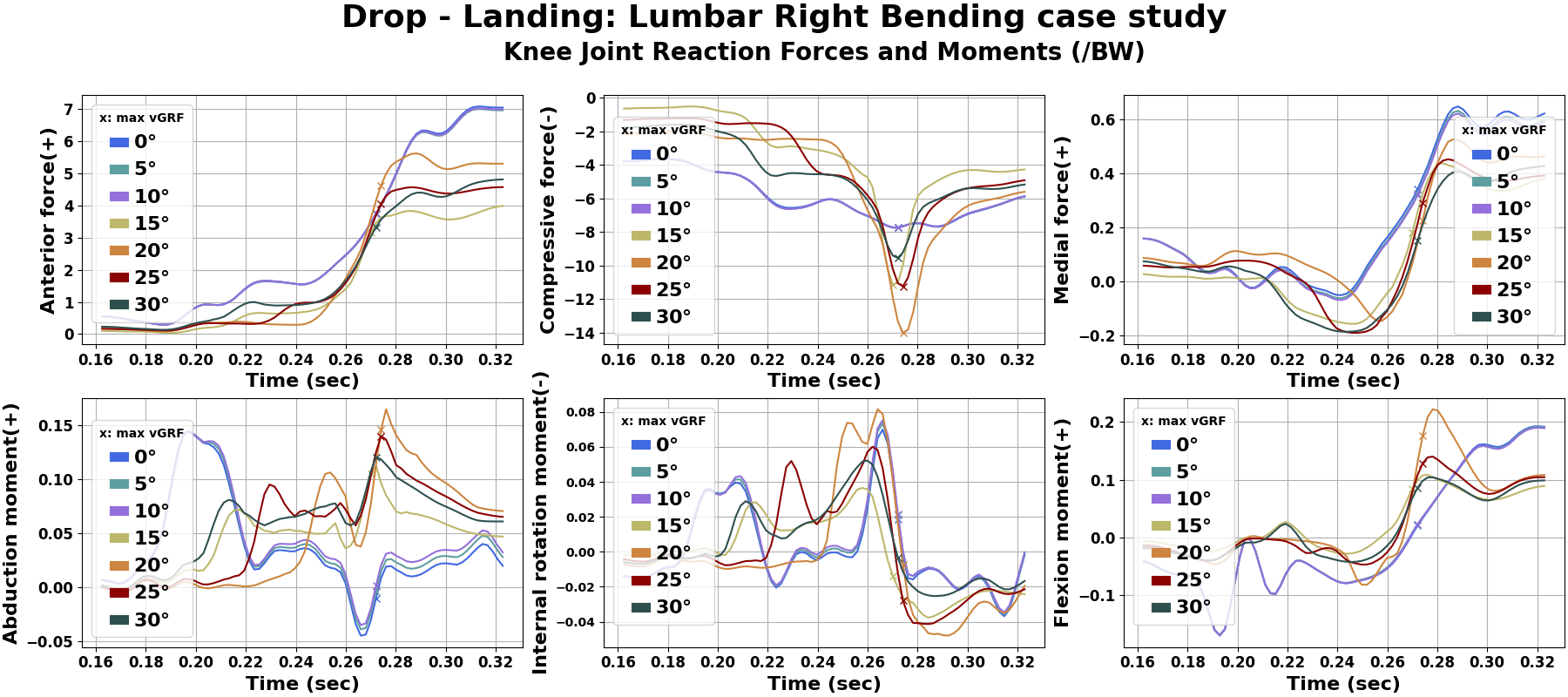}
    \caption{Knee joint \gls{jrf} and \gls{jrm} for multiple trunk right bending angles.}
    \label{fig:predict_trunk_ben_right_JRA}
\end{figure}

In \autoref{fig:predict_trunk_ben_right_JRA} we demonstrate  \gls{kjrf} and \gls{kjrm} for the studied scenarios.\gls{af} raise a peak value right after max \gls{vgrf} time. On the contrary,  \gls{cf} displays lower peak values for the scenarios with lower \gls{vgrf} peaks at max \gls{vgrf} time. Results for \gls{abdm}, \gls{addm}  and \gls{irm}, \gls{erm} are more complicated. The main observation is  for the different behavior of the \gls{abdm} at \gls{igc} time, where for increased right bending (towards the medial aspect) the \gls{abdm} rapidly increases.

\begin{figure}[H]
    \centering
    \includegraphics[width=\textwidth, height = 8cm]{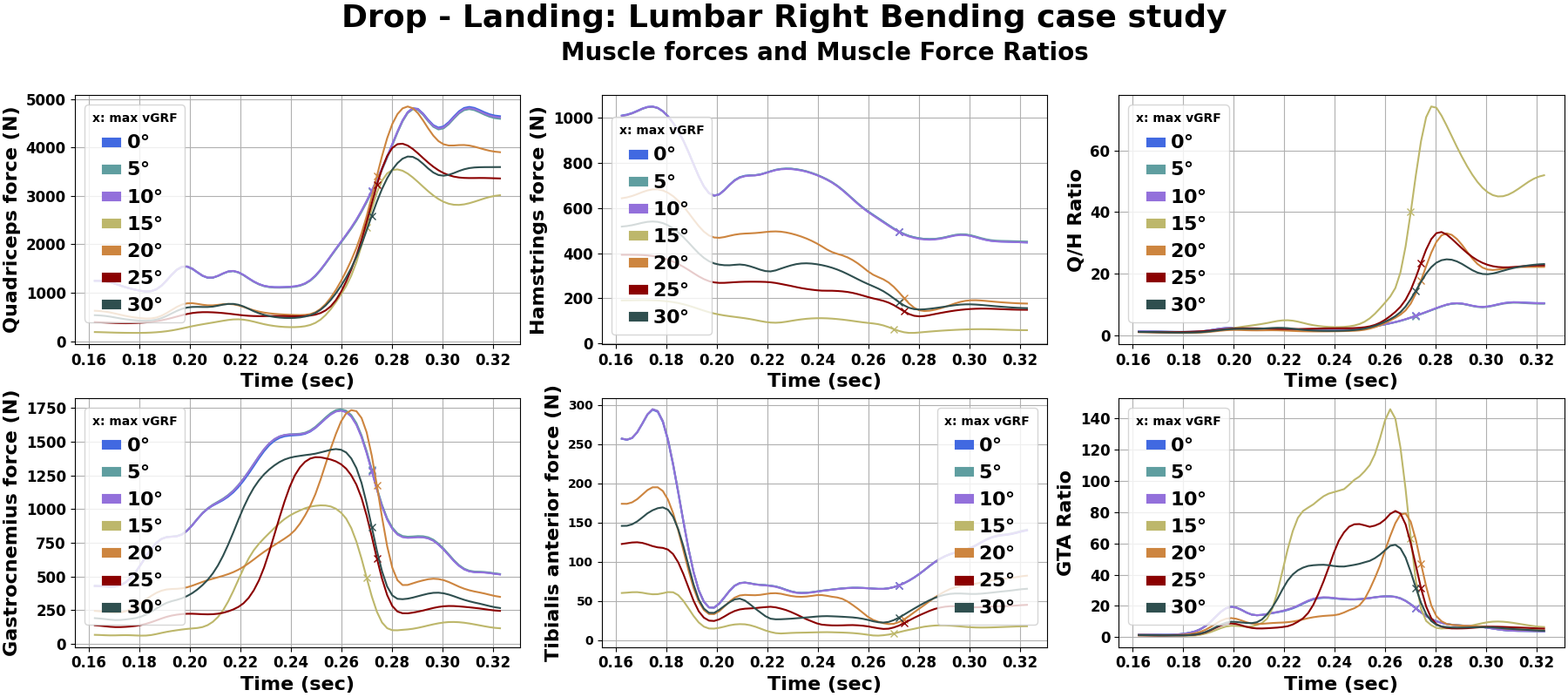}
    \caption{Muscle forces and muscle force ratios for multiple trunk right bending angles.}
    \label{fig:predict_trunk_ben_right_muscle_forces}
\end{figure}

Subsequently, in \autoref{fig:predict_trunk_ben_right_muscle_forces} we demonstrate  muscle forces and muscle force ratios for the examined cases.  We can note that hamstrings force is lower for the scenario with \ang{15} trunk right bending followed by \ang{25}, \ang{30} and \ang{20}, with the greatest hamstring forces for the cases of \ang{0}, \ang{5} and \ang{10}. Correspondingly, \gls{q/h} is lower for the cases of \ang{0}, \ang{5} and \ang{10} and further increases as the bending angle increases.

In  \autoref{tab:trunk_ben_right_1} and \autoref{tab:trunk_ben_right_2}, values for parameters at the  maximum \gls{vgrf} time for all cases are presented. The lowest values for \gls{vgrf} and \gls{q/h} are detected for \ang{0}, \ang{5} and \ang{10} of right trunk bending, while the highest values are detected for the case of \ang{15}.

\begin{table}[H]
    \begin{center}
        \begin{tabular}{c c c c c c}
             \toprule
             Case & pGRF & vGRF & mGRF & Q/H\_ratio & GTA\_ratio \\ 
             \midrule
             0 & 0.167 & 1.383 & -0.176 & 6.275 & 18.351 \\ 
             5 & 0.167 & 1.387 & -0.173 & 6.232 & 18.470 \\ 
             10 & 0.173 & 1.413 & -0.162 & 6.322 & 18.312 \\ 
             15 & 2.126 & 8.632 & 0.033 & 37.554 & 50.881 \\ 
             20 & 1.248 & 7.432 & 0.080 & 17.024 & 43.116 \\ 
             25 & 1.174 & 6.131 & 0.074 & 23.210 & 27.825 \\ 
             30 & 0.866 & 4.515 & 0.056 & 14.284 & 30.678\\ 
              \bottomrule
        \end{tabular}
        \caption{\gls{grf}, \gls{q/h} and \gls{gta} at the time instant of peak \gls{vgrf} for multiple trunk right bending angles.}
        \label{tab:trunk_ben_right_1}
    \end{center}
\end{table}

Based on \autoref{tab:trunk_ben_right_2}, we can spot that at peak \gls{vgrf} time, \gls{af} is greater for the case of \ang{20} and \ang{25} right bending scenarios. Also, \gls{abdm} is lower for the the case with \ang{0} trunk right bending.

\begin{table}[H]
    \begin{center}
        \begin{tabular}{c c c c c c c} 
             \toprule
             Case & AF(+) & CF(-) & MF(+) & AbdM(+) & IRM(-) & FM(+) \\   
             \midrule
             0 & 3.756 & -7.761 & 0.334 & 0.006 & 0.000 & 0.023 \\ 
             5 & 3.729 & -7.769 & 0.318 & 0.012 & 0.002 & 0.025 \\ 
             10 & 3.754 & -7.784 & 0.313 & 0.018 & 0.003 & 0.025 \\  
             15 & 3.397 & -13.052 & 0.218 & 0.101 & -0.026 & 0.093 \\  
             20 & 4.794 & -15.326 & 0.261 & 0.153 & -0.021 & 0.185 \\  
             25 & 4.176 & -12.075 & 0.295 & 0.131 & -0.032 & 0.134 \\  
             30 & 3.396 & -10.125 & 0.144 & 0.127 & -0.004 & 0.092 \\ 
            \bottomrule
        \end{tabular}
        \caption{Knee joint \gls{jrf} and \gls{jrm} at the time instant of peak \gls{vgrf} for multiple trunk right bending angles.}
        \label{tab:trunk_ben_right_2}
    \end{center}
\end{table}

\begin{figure}[H]
    \centering
    \includegraphics[width=1\textwidth, height = 4cm]{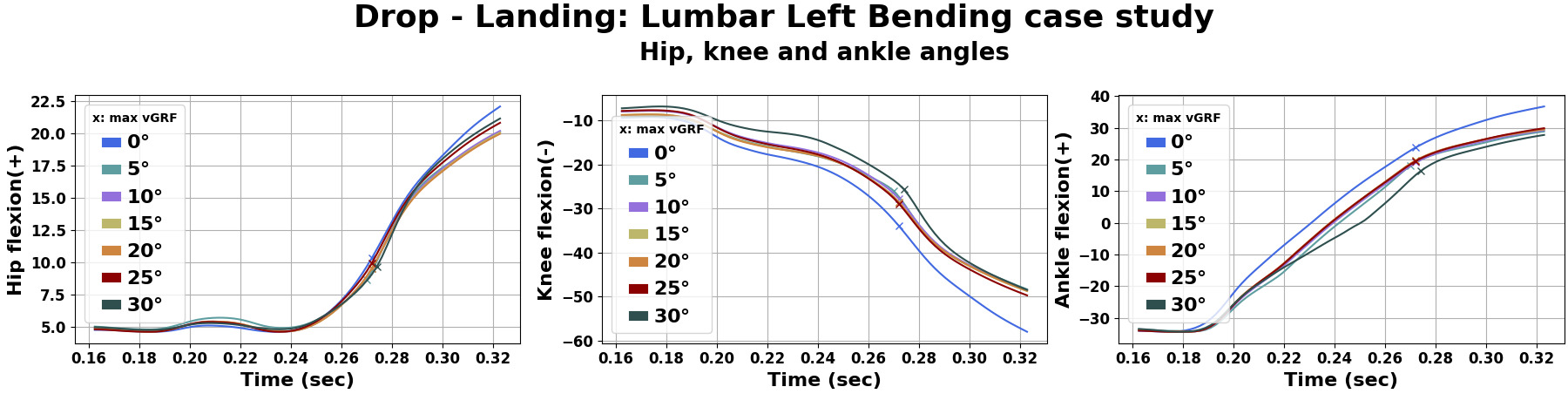}
    \caption{Hip, knee and ankle kinematics for multiple trunk left bending angles.}
    \label{fig:predict_trunk_ben_left_sol}
\end{figure}

Next, we exhibit the resulted plots regarding the cases of single - leg landings with left bending angles. When landing with the left leg, left bending corresponds to a bending towards the lateral left side. In \autoref{fig:predict_trunk_ben_left_sol} we display angles for the hip, knee and ankle joints. 

\begin{figure}[H]
    \centering
    \includegraphics[width=1\textwidth, height = 8cm]{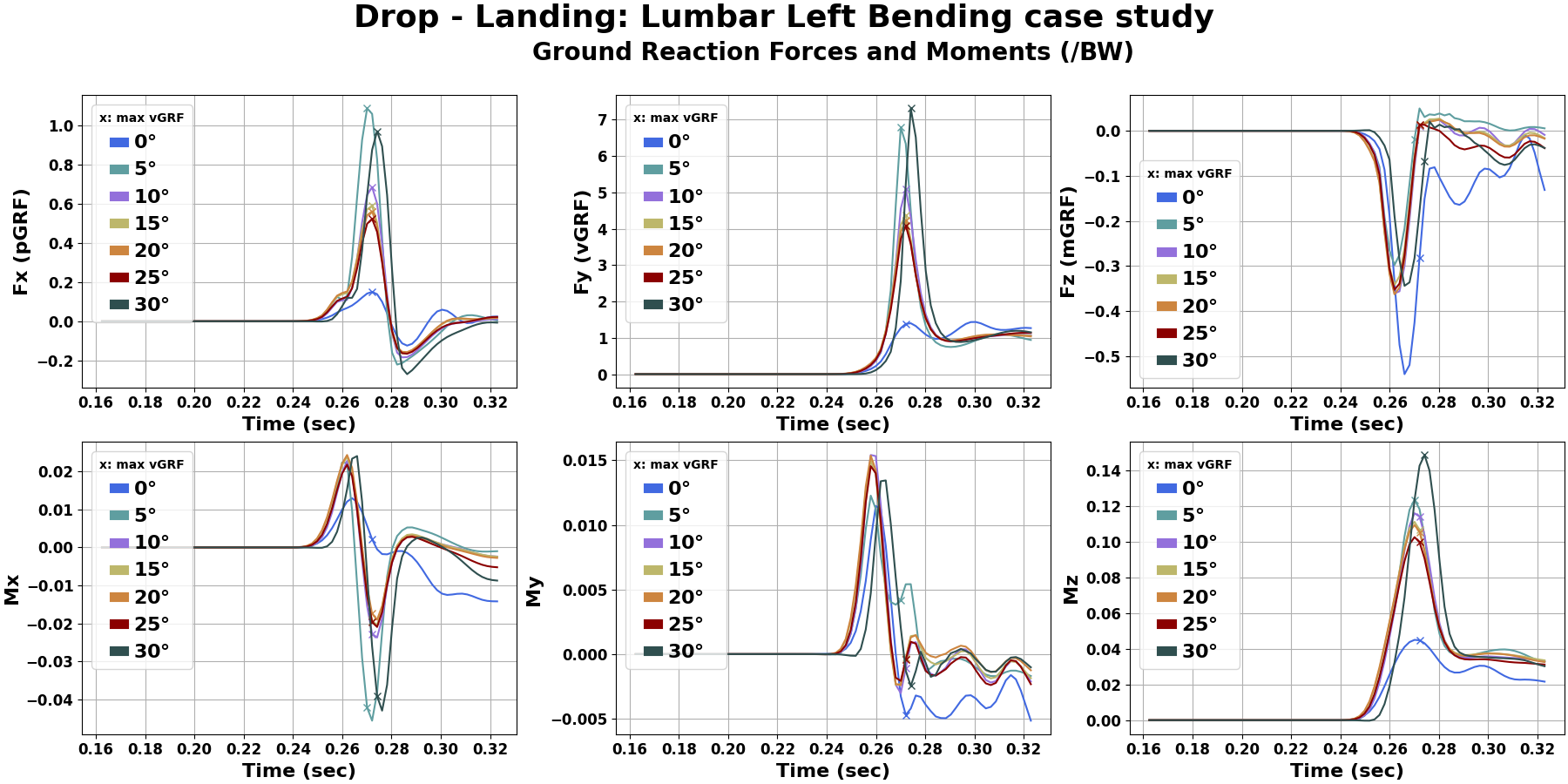}
    \caption{\gls{grf} and \gls{grm} for multiple trunk left bending angles.}
    \label{fig:predict_trunk_ben_left_GRF}
\end{figure}

Next, in \autoref{fig:predict_trunk_ben_left_GRF} we demonstrate \gls{grf} and \gls{grm} for the already mentioned cases. We notice that \gls{pgrf} and \gls{vgrf} peak values obtain the lower value for no right bending while much greater peak values are observed for the remaining examined scenarios. The highest value is spotted for the case with \ang{30} trunk right bending.

\begin{figure}[H]
    \centering
    \includegraphics[width=\textwidth, height = 8cm]{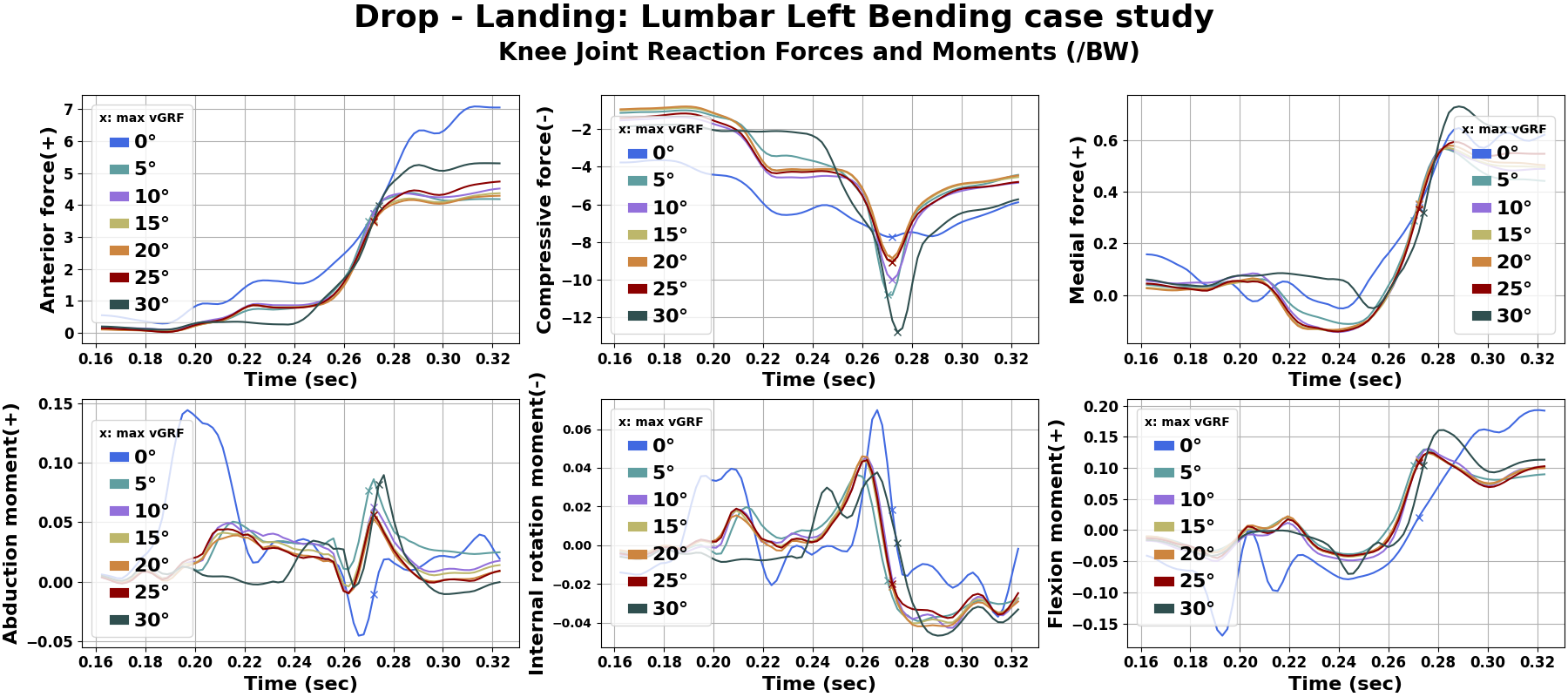}
    \caption{Knee joint \gls{jrf} and \gls{jrm} for multiple trunk left bending angles.}
    \label{fig:predict_trunk_ben_left_JRA}
\end{figure}

Also, in \autoref{fig:predict_trunk_ben_left_JRA} we demonstrate \gls{kjrf} and \gls{kjrm} for the studied scenarios. \gls{af} reaches a peak value right after max \gls{vgrf} time with greater value for the case with no bending, following by the scenario with \ang{30}. Regarding \gls{cf}, a peak value is noticed at max \gls{vgrf} time instant. Furthermore, the greatest peak is observed for the case with \ang{30} of trunk left bending, while the lowest one for the case with no bending. Concerning \gls{abdm} and \gls{irm} we can detect a deviation on behavior between the case with no trunk bending and the cases with left trunk bending.

\begin{figure}[H]
    \centering
    \includegraphics[width=\textwidth, height = 8cm]{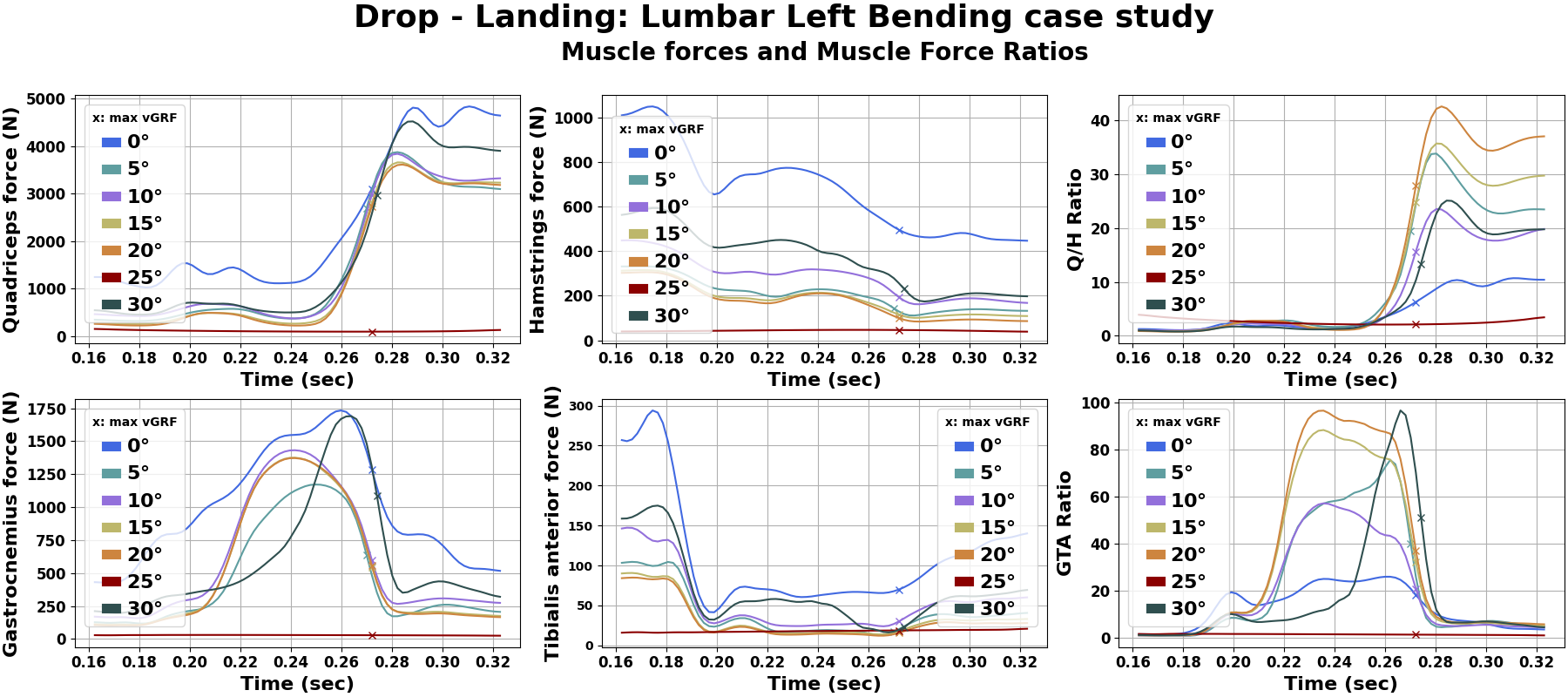}
    \caption{Muscle forces and muscle force ratios for multiple trunk left bending angles.}
    \label{fig:predict_trunk_ben_left_muscle_forces}
\end{figure}

Following, in \autoref{fig:predict_trunk_ben_left_muscle_forces} we demonstrate  muscle forces and muscle force ratios for the examined scenarios. It is obvious that \gls{q/h} is lower for the cases with \ang{0} and \ang{25} left trunk bending. The greatest values are detected for the case with \ang{30} trunk left bending. 

Finally, in \autoref{tab:trunk_ben_left_1} and \autoref{tab:trunk_ben_left_2}, values for parameters at the time instant of maximum \gls{vgrf} for all cases are presented. \gls{cf} at peak \gls{vgrf} is lower for the cases with lower \gls{vgrf} peaks. 

\begin{table}[H]
    \begin{center}
        \begin{tabular}{c c c c c c}
             \toprule
             Case & pGRF & vGRF & mGRF & Q/H\_ratio & GTA\_ratio \\   
             \midrule
             0 &  0.167 & 1.383 & -0.176 & 6.275 & 18.351 \\ 
             5 & 1.529 & 6.787 & 0.016 & 18.631 & 36.024 \\ 
             10 & 0.897 & 5.091 & -0.020 & 15.568 & 20.236  \\ 
             15 & 0.748 & 4.378 & -0.018 & 24.773 & 30.945 \\ 
             20 & 0.712 & 4.224 & -0.021 & 27.775 & 35.635 \\ 
             25 & 0.665 & 4.088 & -0.013 & 2.095 & 1.496 \\ 
             30 & 1.377 & 7.309 & -0.011 & 12.901 & 46.030 \\ 
              \bottomrule
        \end{tabular}
        \caption{\gls{grf}, \gls{q/h} and \gls{gta} at the time instant of peak \gls{vgrf} for multiple trunk left bending angles.}
        \label{tab:trunk_ben_left_1}
    \end{center}
\end{table}

\begin{table}[H]
    \begin{center}
        \begin{tabular}{c c c c c c c} 
             \toprule
            Case & AF(+) & CF(-) & MF(+) & AbdM(+) & IRM(-) & FM(+) \\    
             \midrule
             0 & 3.756 & -7.761 & 0.334 & 0.006 & 0.000 & 0.023 \\ 
             5 & 3.645 & -11.960 & 0.315 & 0.078 & -0.028 & 0.109 \\ 
             10 & 3.810 & -10.618 & 0.333 & 0.067 & -0.017 & 0.123 \\  
             15 & 3.594 & -9.494 & 0.341 & 0.058 & -0.022 & 0.117 \\  
             20 & 3.512 & -9.259 & 0.343 & 0.055 & -0.020 & 0.116 \\  
             25 & 3.552 & -9.446 & 0.325 & 0.056 & -0.021 & 0.115 \\  
             30 & 4.135 & -14.095 & 0.342 & 0.085 & -0.008 & 0.111 \\ 
            \bottomrule
        \end{tabular}
        \caption{Knee joint \gls{jrf} and \gls{jrm} at the time instant of peak \gls{vgrf} for multiple trunk left bending angles.}
        \label{tab:trunk_ben_left_2}
    \end{center}
\end{table}

\subsection{Trunk internal - external rotation}
\begin{figure}[H]
    \centering
    \includegraphics[width=1\textwidth, height = 4cm]{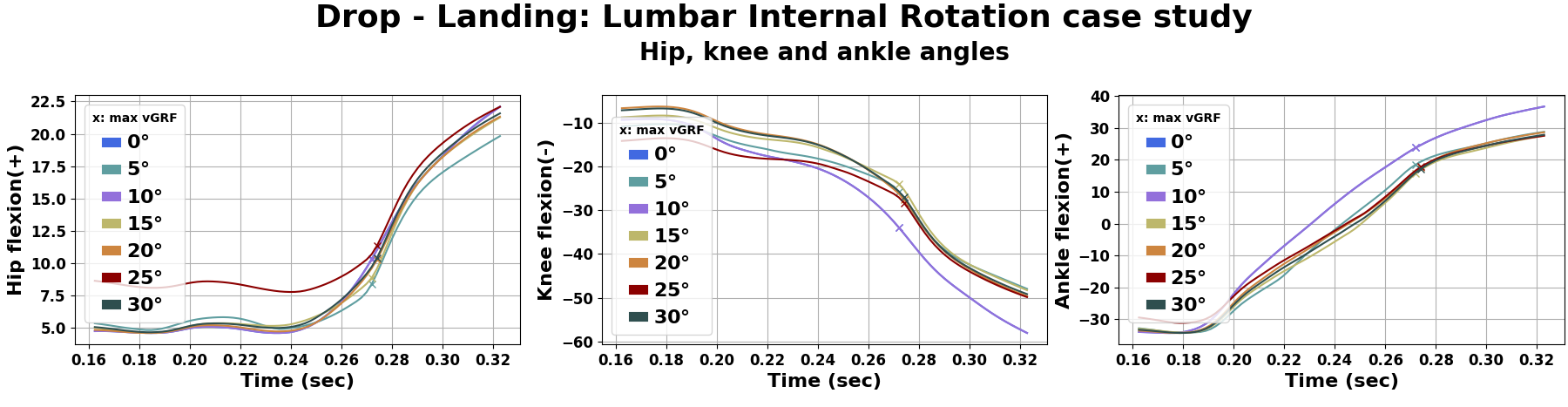}
    \caption{Hip, knee and ankle kinematics for multiple trunk internal rotation angles.}
    \label{fig:predict_trunk_rot_int_sol}
\end{figure}

In this part, we demonstrate results regarding drop landings with different trunk rotation angles as described in \autoref{sec:trunk_info}. In \autoref{fig:predict_trunk_rot_int_sol} we display angles of the hip, knee and ankle joints corresponding to internal rotation angles. 

\begin{figure}[H]
    \centering
    \includegraphics[width=1\textwidth, height = 8cm]{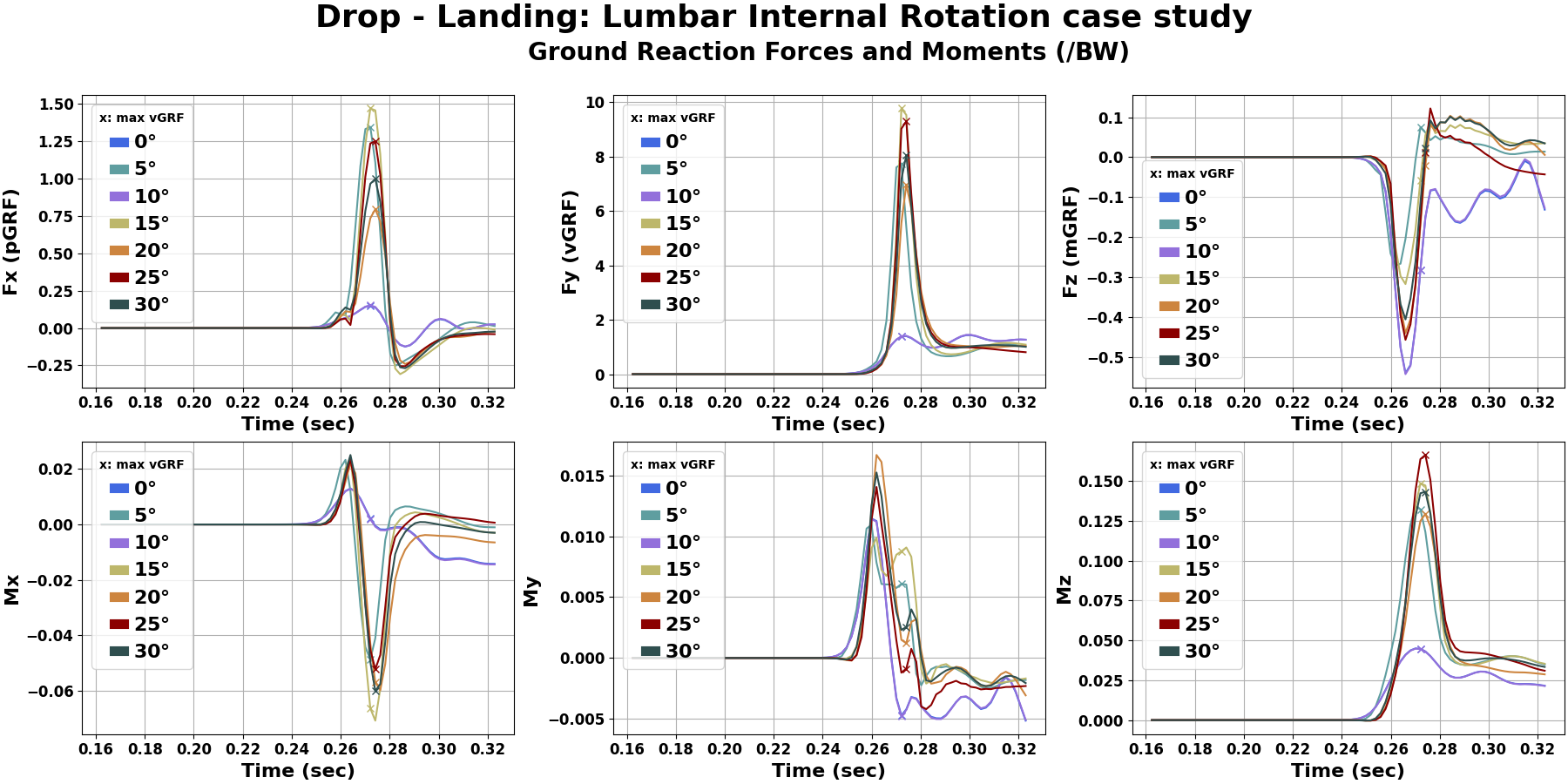}
    \caption{\gls{grf} and \gls{grm} for multiple trunk internal rotation angles.}
    \label{fig:predict_trunk_rot_int_GRF}
\end{figure}

In \autoref{fig:predict_trunk_rot_int_GRF} we demonstrate \gls{grf} and \gls{grm} for the previously mentioned scenarios. Minimum peak \gls{vgrf} values are detected for the cases with \ang{0} and \ang{10} internal rotation, while maximum peak values are spotted for the \ang{15} and \ang{25} cases. The same observation is valid for the \gls{pgrf}.

\begin{figure}[H]
    \centering
    \includegraphics[width=\textwidth, height = 8cm]{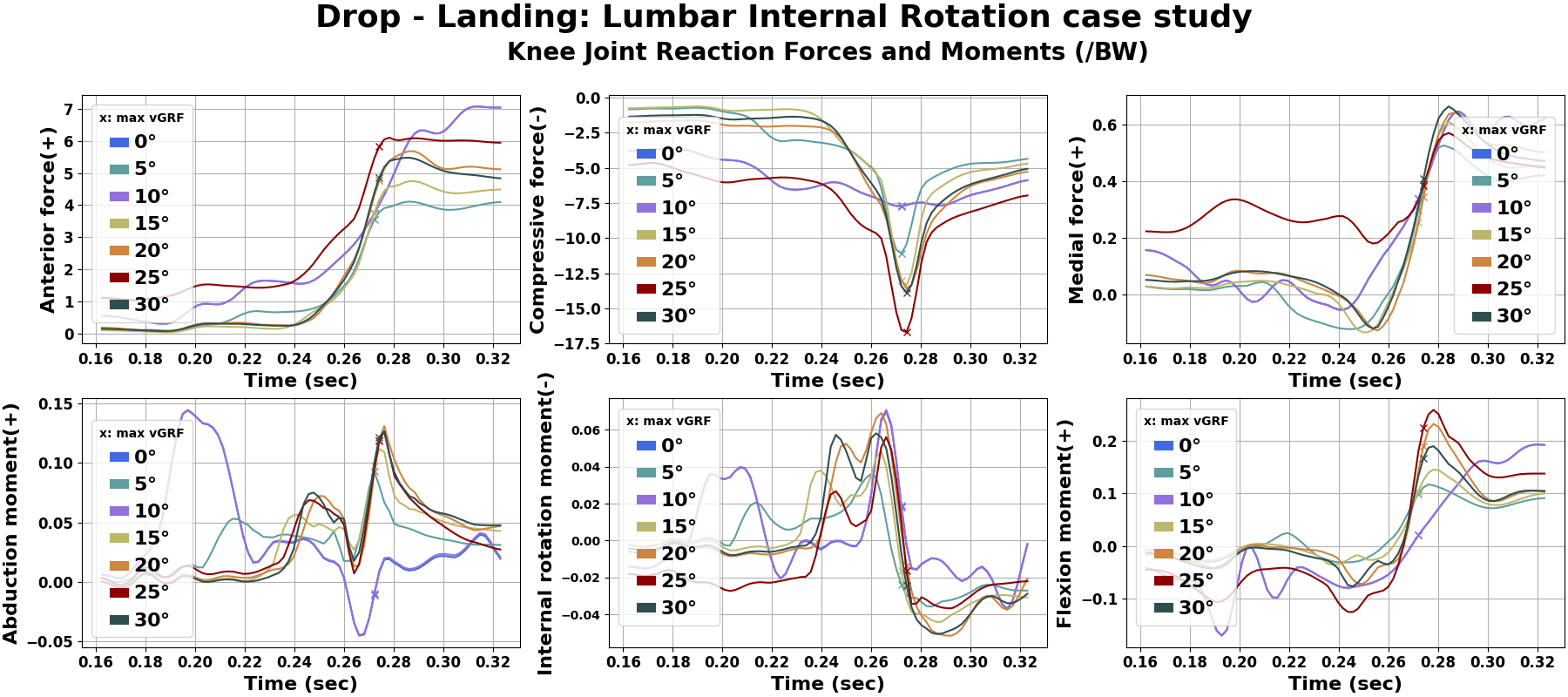}
    \caption{Knee joint \gls{jrf} and \gls{jrm} for multiple trunk internal rotation angles.}
    \label{fig:predict_trunk_rot_int_JRA}
\end{figure}

In addition, in \autoref{fig:predict_trunk_rot_int_JRA} we demonstrate  \gls{kjrf} and \gls{kjrm} for the described cases. \gls{cf} displays lower values for the cases with \ang{0} and \ang{10} of trunk internal rotations and increases for the cases with \ang{5}, \ang{15}, \ang{20}, \ang{30}, getting the highest peak for \ang{25} at max \gls{vgrf} time instant.  

\begin{figure}[H]
    \centering
    \includegraphics[width=\textwidth, height = 8cm]{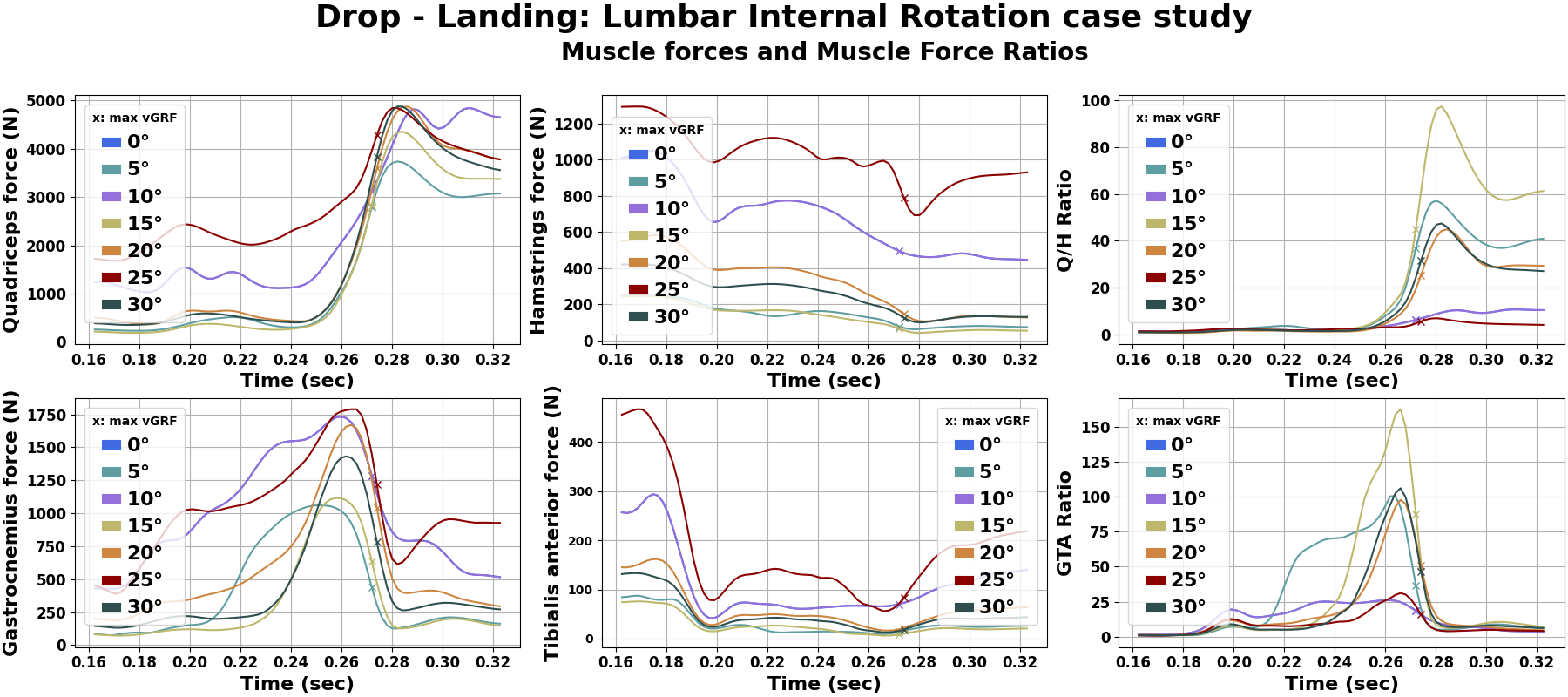}
    \caption{Muscle forces and muscle force ratios for multiple trunk internal rotation angles.}
    \label{fig:predict_trunk_rot_int_muscle_forces}
\end{figure}

In \autoref{fig:predict_trunk_rot_int_muscle_forces} we demonstrate  muscle forces and muscle force ratios. In general, we can observe that maximum muscle forces are detected for the case of \ang{25} of trunk internal rotation and the minimum muscle forces are detected for the cases of \ang{15}, \ang{5} and \ang{30}.

Also, in \autoref{tab:trunk_rot_int_1} and \autoref{tab:trunk_rot_int_2}, values of parameters at the  maximum \gls{vgrf} time instant for all cases are presented. As mentioned previously, lower \gls{vgrf} and \gls{pgrf} peaks are spotted for the \ang{0} and \ang{10} trunk internal rotation cases, while maximum values are detected for the \ang{15} and \ang{25} cases. 

\begin{table}[H]
    \begin{center}
        \begin{tabular}{c c c c c c}
             \toprule
             Case & pGRF & vGRF & mGRF & Q/H\_ratio & GTA\_ratio \\  
             \midrule
             0 &  0.167 & 1.383 & -0.176 & 6.275 & 18.351 \\ 
             5 & 1.781 & 7.732 & 0.087 & 36.534 & 31.173 \\ 
             10 & 0.169 & 1.393 & -0.175 & 6.289 & 18.345  \\ 
             15 & 2.254 & 9.776 & 0.007 & 41.032 & 79.928 \\ 
             20 & 1.126 & 6.975 & 0.066 & 24.696 & 46.359 \\ 
             25 & 1.742 & 9.305 & 0.088 & 5.596 & 13.551 \\ 
             30 & 1.432 & 8.061 & 0.080 & 30.997 & 40.003\\ 
              \bottomrule
        \end{tabular}
        \caption{\gls{grf}, \gls{q/h} and \gls{gta} at the time instant of peak \gls{vgrf} for multiple trunk internal rotation angles.}
        \label{tab:trunk_rot_int_1}
    \end{center}
\end{table}

In \autoref{tab:trunk_rot_int_2} we can observe \gls{kjrf} and \gls{kjrm} at maximum \gls{vgrf} for all cases. \gls{af} obtains the lower value for the case of \ang{5} and the highest for \ang{25}. In general, greater \gls{af} values at max \gls{vgrf} are observed for increased trunk internal rotation angles. Also, \gls{cf} has lower values for \ang{0} and \ang{10} and increases for the remaining cases. \gls{abdm} and \gls{irm} receives the lowest values for the case with no rotation, following by \ang{10} of trunk internal rotation.  

\begin{table}[H]
    \begin{center}
        \begin{tabular}{c c c c c c c} 
             \toprule
             Case & AF(+) & CF(-) & MF(+) & AbdM(+) & IRM(-) & FM(+) \\   
             \midrule
             0 & 3.756 & -7.761 & 0.334 & 0.006 & 0.000 & 0.023 \\ 
             5 & 3.687 & -12.384 & 0.271 & 0.104 & -0.028 & 0.108 \\ 
             10 & 3.757 & -7.768 & 0.332 & 0.007 & 0.001 & 0.023 \\  
             15 & 4.006 & -15.332 & 0.237 & 0.137 & -0.029 & 0.102 \\  
             20 & 4.921 & -14.675 & 0.377 & 0.127 & -0.023 & 0.198 \\  
             25 & 6.076 & -18.200 & 0.399 & 0.126 & -0.027 & 0.238 \\  
             30 & 5.049 & -15.227 & 0.436 & 0.128 & -0.032 & 0.177 \\ 
            \bottomrule
        \end{tabular}
        \caption{Knee joint \gls{jrf} and \gls{jrm} at the time instant of peak \gls{vgrf} for multiple trunk internal rotation angles.}
        \label{tab:trunk_rot_int_2}
    \end{center}
\end{table}

\begin{figure}[H]
    \centering
    \includegraphics[width=1\textwidth, height = 4cm]{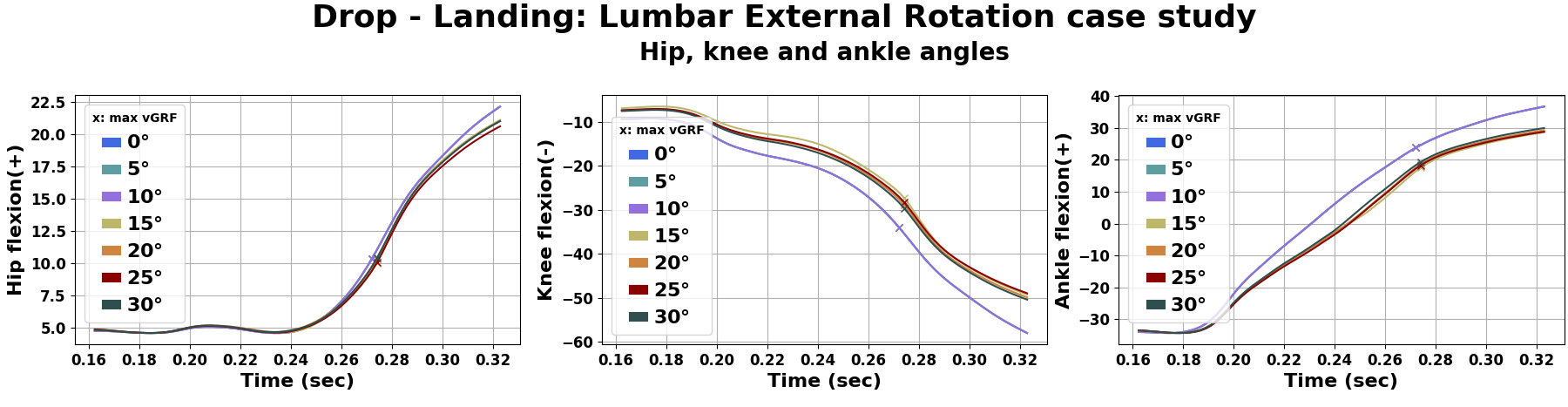}
    \caption{Hip, knee and ankle kinematics for multiple trunk external rotation angles.}
    \label{fig:predict_trunk_rot_ext_sol}
\end{figure}

In \autoref{fig:predict_trunk_rot_ext_sol} we display angles of the hip, knee and ankle joints for single - leg landings with externally rotated trunk scenarios. 

\begin{figure}[H]
    \centering
    \includegraphics[width=1\textwidth, height = 8cm]{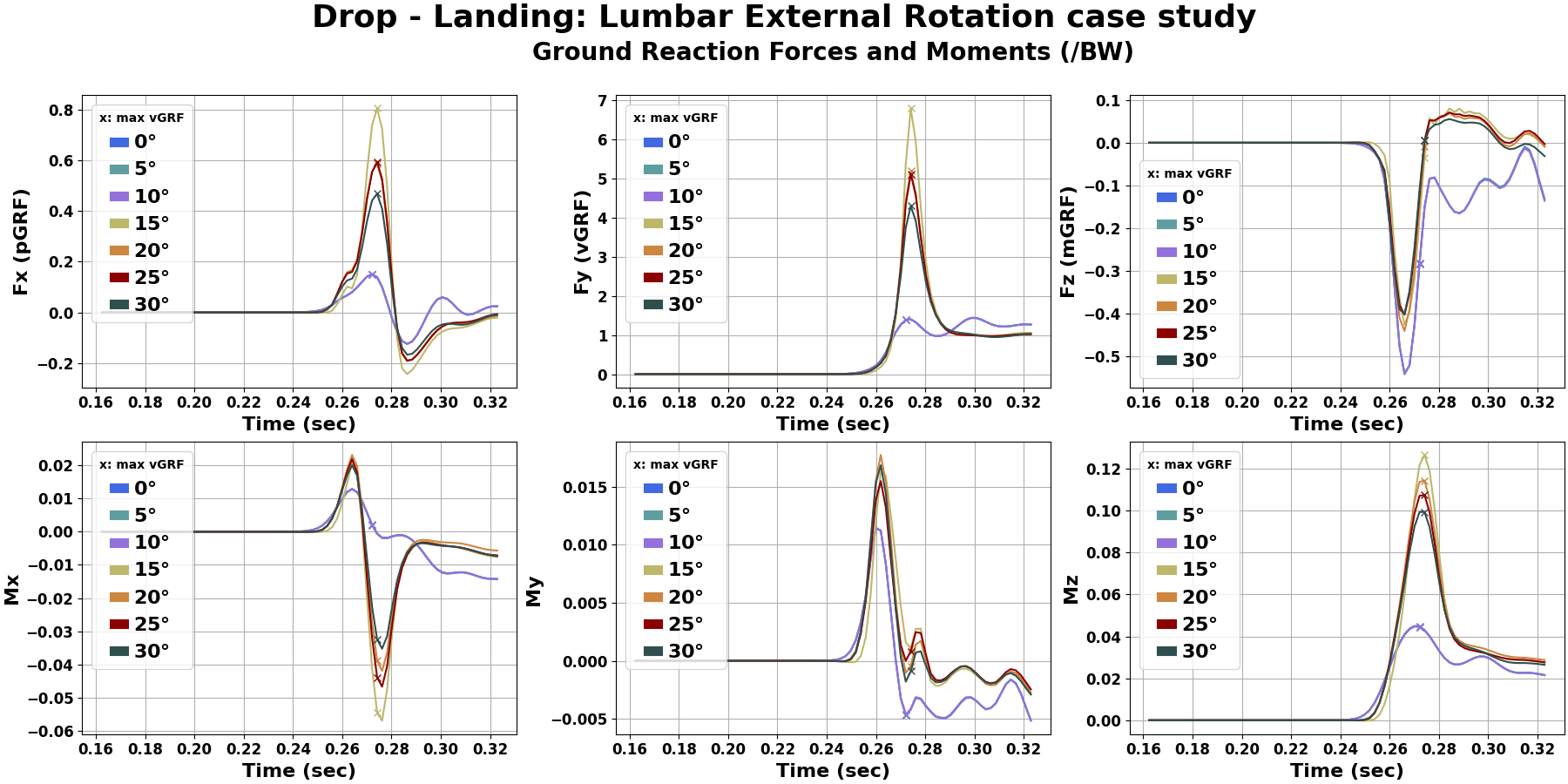}
    \caption{\gls{grf} and \gls{grm} for multiple trunk external rotation angles.}
    \label{fig:predict_trunk_rot_ext_GRF}
\end{figure}

In \autoref{fig:predict_trunk_rot_ext_GRF} we demonstrate \gls{grf} and \gls{grm} for the studied scenarios. We can detect greater \gls{vgrf} peaks for the cases with trunk external rotation greater than \ang{10}, while lower values are observed for \ang{0}, \ang{5} and \ang{10} of trunk external rotation. 

\begin{figure}[H]
    \centering
    \includegraphics[width=\textwidth, height = 8cm]{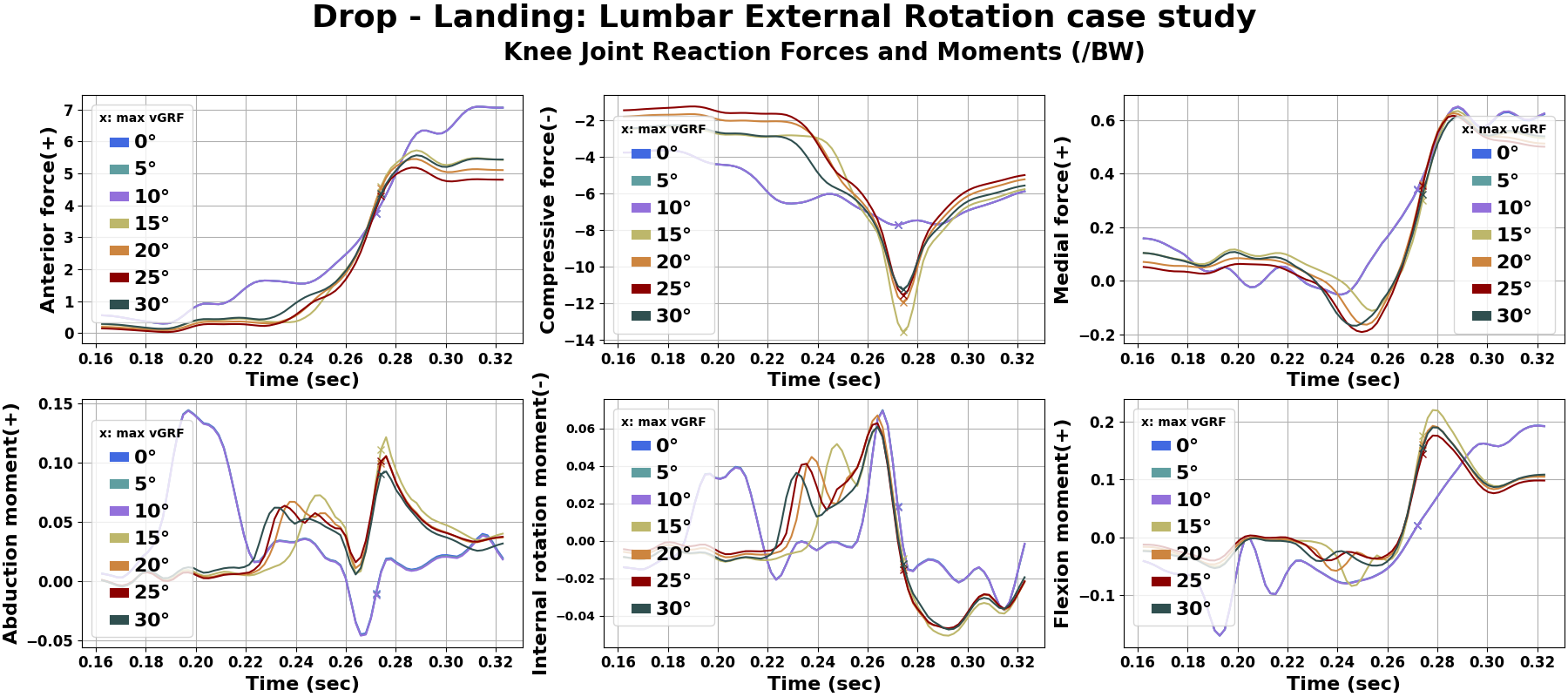}
    \caption{Knee joint \gls{jrf} and \gls{jrm} for multiple trunk external rotation angles.}
    \label{fig:predict_trunk_rot_ext_JRA}
\end{figure}

In \autoref{fig:predict_trunk_rot_ext_JRA} we demonstrate  \gls{kjrf} and \gls{kjrm}. Greater peaks of \gls{cf} are noticed for \ang{15} trunk external rotation following by \ang{20}, \ang{25}, \ang{30}, \ang{5}, \ang{10}, \ang{0} degrees respectively. 

\begin{figure}[H]
    \centering
    \includegraphics[width=\textwidth, height = 8cm]{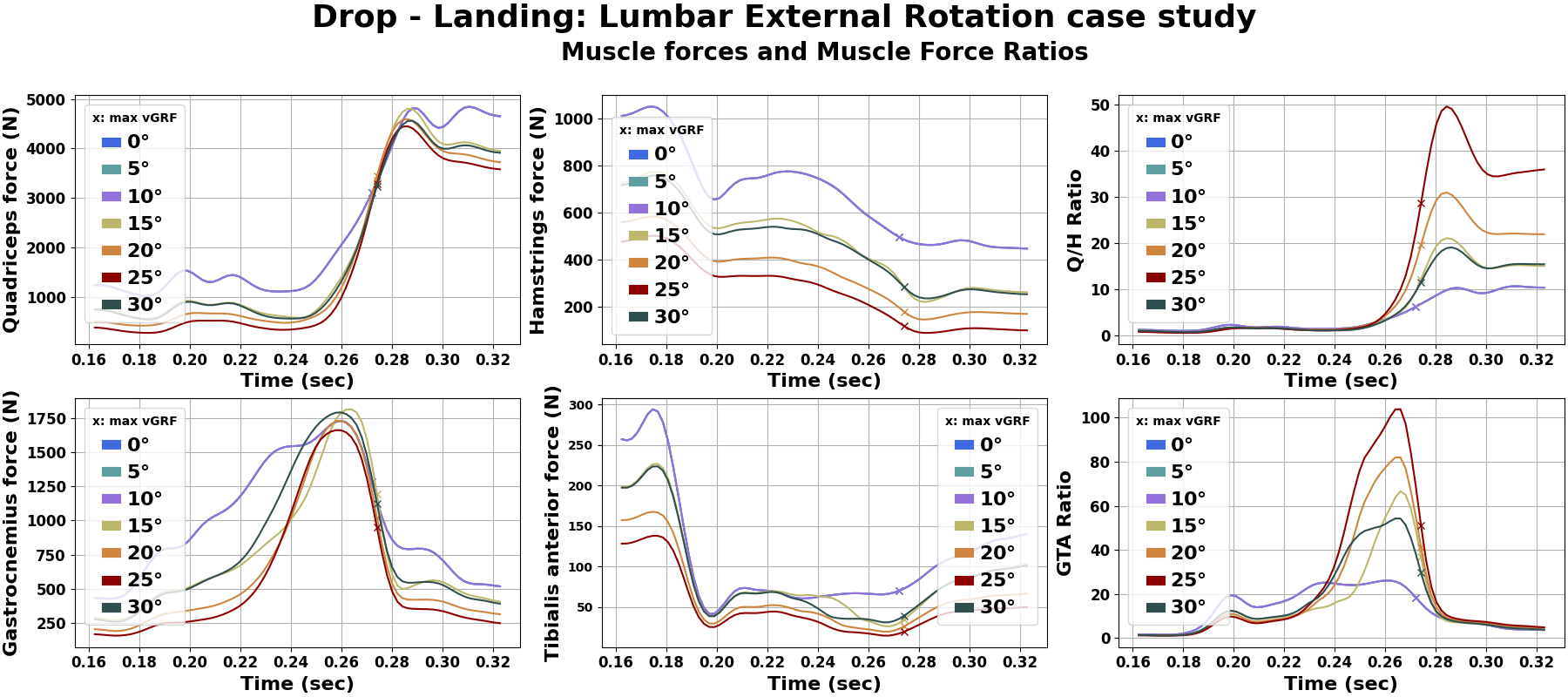}
    \caption{Muscle forces and muscle force ratios for multiple trunk external rotation angles.}
    \label{fig:predict_trunk_rot_ext_muscle_forces}
\end{figure}

In \autoref{fig:predict_trunk_rot_ext_muscle_forces} we demonstrate  muscle forces and muscle force ratios. Greater hamstrings force and lower \gls{q/h} are spotted for the cases of 
\ang{0}, \ang{5} and \ang{10} of trunk external rotation.

In  \autoref{tab:trunk_rot_ext_1} and \autoref{tab:trunk_rot_ext_2}, values of parameters at the time instant of maximum \gls{vgrf} for all cases are presented. 

\begin{table}[H]
    \begin{center}
        \begin{tabular}{c c c c c c}
             \toprule
             Case & pGRF & vGRF & mGRF & Q/H\_ratio & GTA\_ratio \\ 
             \midrule
             0 & 0.167 & 1.383 & -0.176 & 6.275 & 18.351 \\ 
             5 & 0.167 & 1.387 & -0.177 & 6.273 & 18.315 \\ 
             10 & 0.168 & 1.387 & -0.178 & 6.257 & 18.332  \\ 
             15 & 1.138 & 6.799 & 0.042 & 11.877 & 34.097 \\ 
             20 & 0.782 & 5.194 & 0.022 & 19.366 & 38.563 \\ 
             25 & 0.783 & 5.118 & 0.021 & 28.187 & 47.747 \\ 
             30 & 0.604 & 4.304 & -0.007 & 11.482 & 28.361 \\ 
              \bottomrule
        \end{tabular}
        \caption{\gls{grf}, \gls{q/h} and \gls{gta} at the time instant of peak \gls{vgrf} for multiple trunk external rotation angles.}
        \label{tab:trunk_rot_ext_1}
    \end{center}
\end{table}

In \autoref{tab:trunk_rot_ext_2} we can compare easily the values of the desired parameters at maximum \gls{vgrf} instant. \gls{af} and \gls{cf} obtain greater values for cases with greater \gls{vgrf} peaks compared with cases with lower \gls{vgrf} peaks.

\begin{table}[H]
    \begin{center}
        \begin{tabular}{c c c c c c c} 
             \toprule
             Case & AF(+) & CF(-) & MF(+) & AbdM(+) & IRM(-) & FM(+) \\  
             \midrule
             0 & 3.756 & -7.761 & 0.334 & 0.006 & 0.000 & 0.023 \\ 
             5 & 3.757 & -7.762 & 0.336 & 0.006 & 0.000 & 0.023 \\ 
             10 & 3.754 & -7.762 & 0.337 & 0.005 & 0.000 & 0.022\\  
             15 & 4.730 & -14.651 & 0.330 & 0.119 & -0.019 & 0.182\\  
             20 & 4.623 & -12.552 & 0.377 & 0.102 & -0.020 & 0.165 \\  
             25 & 4.373 & -12.141 & 0.370 & 0.102 & -0.021 & 0.149 \\  
             30 & 4.421 & -11.667 & 0.329 & 0.088 & -0.015 & 0.158 \\ 
            \bottomrule
        \end{tabular}
        \caption{Knee joint \gls{jrf} and \gls{jrm} at the time instant of peak \gls{vgrf} for multiple trunk external rotation angles.}
        \label{tab:trunk_rot_ext_2}
    \end{center}
\end{table}

\section{Muscle force of knee joint agonists and antagonists case}\label{sec:muslces_results}

In this section we demonstrate the results regarding cases where specific muscle groups were strengthened and weakened as presented in \autoref{sec:muslces_info}. In the figures following, \textit{nq} refers to normal quadriceps, \textit{sq} refers to strong quadriceps,  \textit{wq} refers to weak quadriceps,  \textit{nh} refers to normal hamstrings,  \textit{sh} refers to strong hamstrings and  \textit{wh} refers to weak hamstrings.  

\begin{figure}[H]
    \centering
    \includegraphics[width=1\textwidth, height = 4cm]{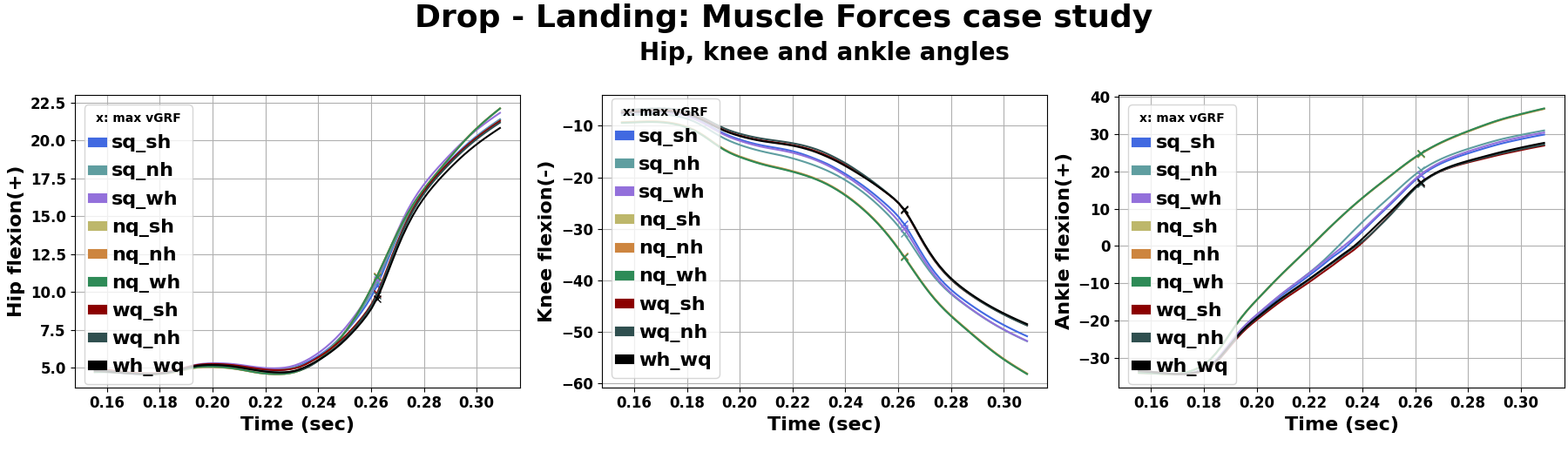}
    \caption{Hip, knee and ankle kinematics for scenarios of normal, weak and strong knee joint agonist antagonist muscles.}
    \label{fig:predict_muscles_sol}
\end{figure}

Next, in \autoref{fig:predict_muscles_sol}, angles for the hip, knee and ankle joints are displayed. If we take a close look at knee flexion, we can observe that for the scenarios where quadriceps are normal, the knee and ankle are more flexed.

\begin{figure}[H]
    \centering
    \includegraphics[width=1\textwidth, height = 8cm]{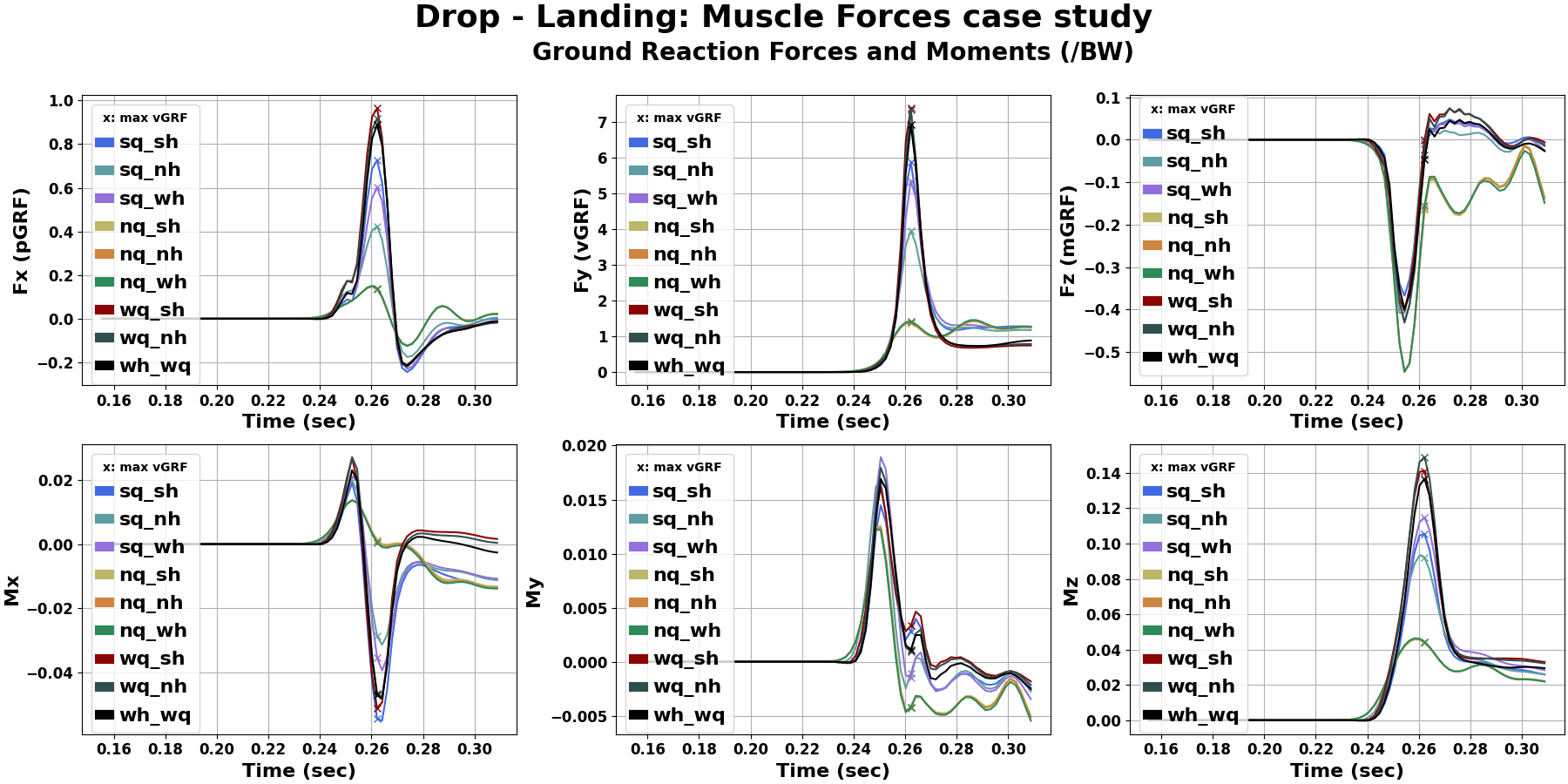}
    \caption{\gls{grf} and \gls{grm} for scenarios of normal, weak and strong knee joint agonist antagonist muscles.}
    \label{fig:predict_muscles_GRF}
\end{figure}

Moreover, in \autoref{fig:predict_muscles_GRF} we demonstrate \gls{grf} and \gls{grm} for the examined scenarios. We detect the lowest peak \gls{vgrf} values for the three cases with normal quadriceps. Also, the highest peak is noticed for the scenario with weak quadriceps and strong hamstrings.

\begin{figure}[H]
    \centering
    \includegraphics[width=\textwidth, height = 8cm]{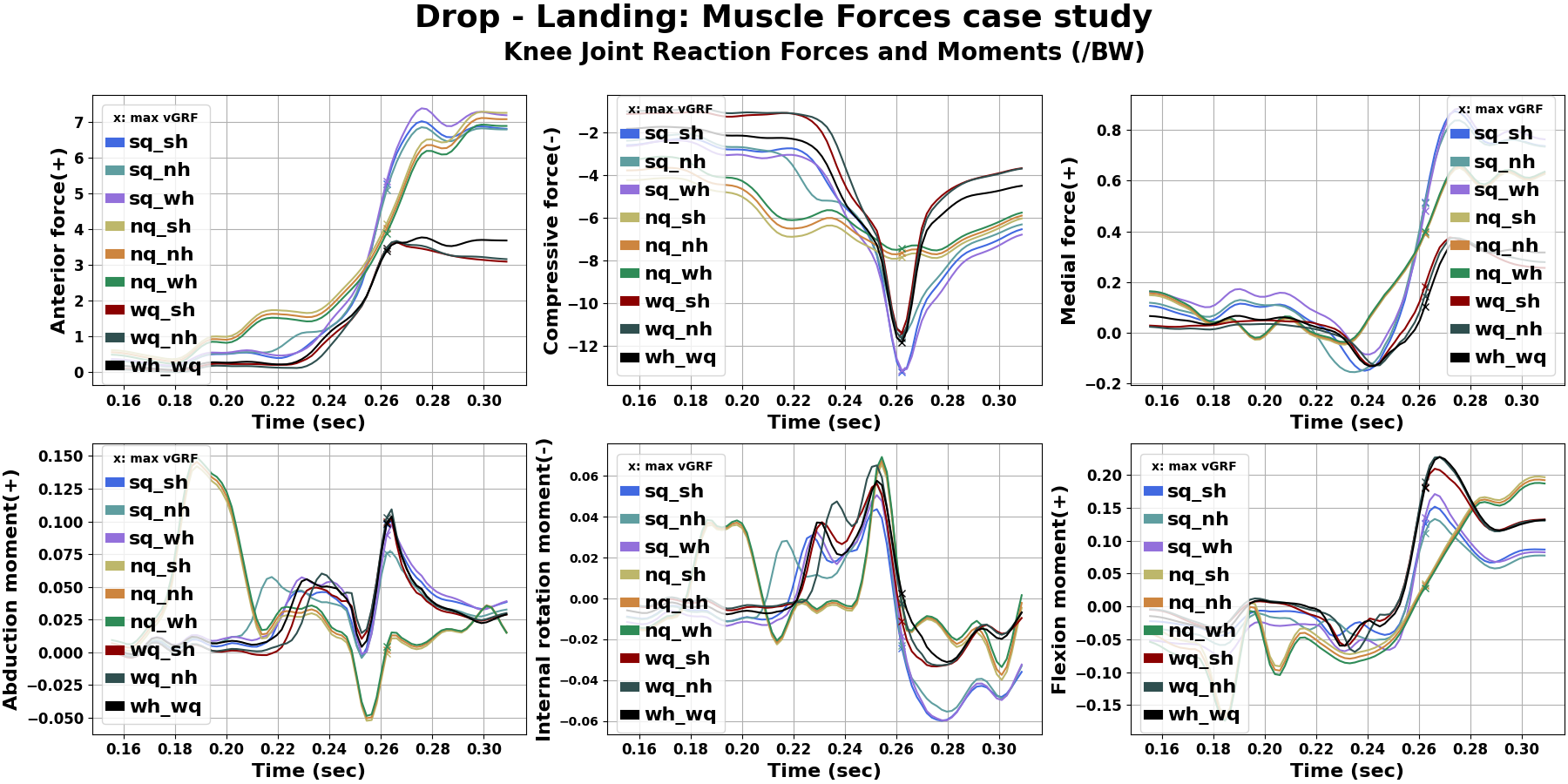}
    \caption{Knee joint \gls{jrf} and \gls{jrm} for scenarios of normal, weak and strong knee joint agonist antagonist muscles.}
    \label{fig:predict_muscles_JRA}
\end{figure}

Furthermore, in \autoref{fig:predict_muscles_JRA} we demonstrate \gls{kjrf} and \gls{kjrm} for all studied scenarios. If we take a quick look at all knee joint \gls{jrf} and \gls{jrm} we can detect similar behavior for the three cases with normal quadriceps. For these three scenarios we notice lower \gls{cf} peak around maximum \gls{vgrf} and greater \gls{abdm} and \gls{irm} peaks at \gls{igc}.  

\begin{figure}[H]
    \centering
    \includegraphics[width=\textwidth, height = 8cm]{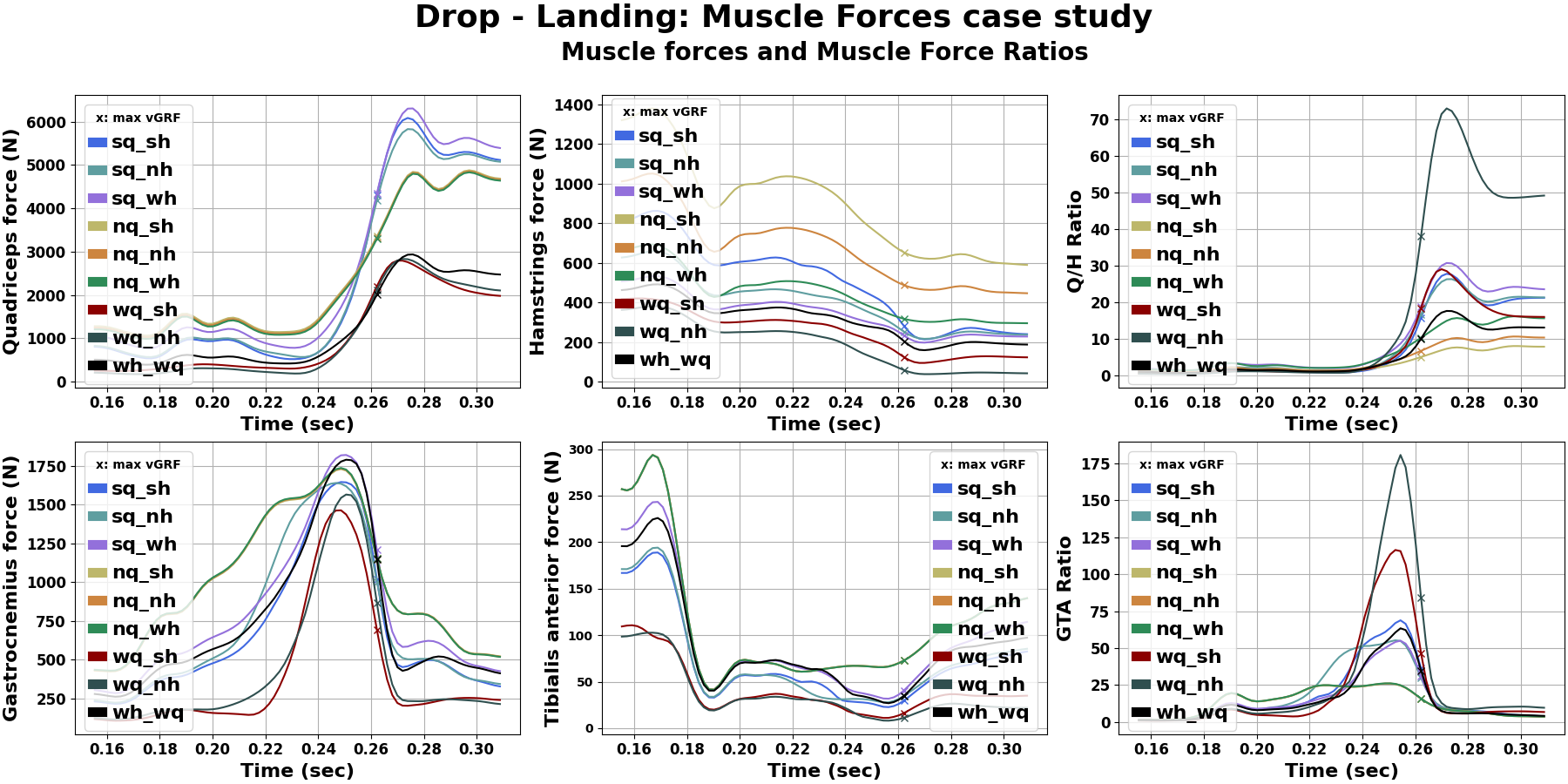}
    \caption{Muscle forces and muscle force ratios for scenarios of normal, weak and strong knee joint agonist antagonist muscles.}
    \label{fig:predict_muscles_muscle_forces}
\end{figure}

Afterwards, we demonstrate muscle forces and muscle force ratios for the previously mentioned cases (\autoref{fig:predict_muscles_muscle_forces}). Hamstrings group force obtains the highest value for the scenario with strong hamstrings and normal quadriceps. Also, the lowest values for hamstrings forces are observed for the scenarios with weak quadriceps.

Finally, in  \autoref{tab:muscles_1} and \autoref{tab:muscles_2} we present values for the studied parameters at the maximum \gls{vgrf}. We observe that the highest \gls{vgrf} peak occurs for the weak quadriceps and strong hamstrings case. Moreover, the weak quadriceps and normal hamstrings scenario obtains the highest \gls{q/h} value at peak \gls{vgrf}. On the contrary, the normal quadriceps and strong hamstrings case obtains the lowest \gls{q/h} value at peak \gls{vgrf}. 

\begin{table}[H]
    \begin{center}
        \begin{tabular}{c c c c c c}
             \toprule
             Case & pGRF & vGRF & mGRF & Q/H\_ratio & GTA\_ratio \\   
             \midrule
             sq\_sh & 1.009 & 5.889 & -0.001 & 15.529 & 33.423 \\ 
             sq\_nh &  0.542 & 3.964 & -0.037 & 16.627 & 28.490  \\ 
             sq\_wh & 0.827 & 5.355 & 0.005 & 18.631 & 29.356 \\
             nq\_sh & 0.146 & 1.362 & -0.119 & 5.149 & 15.552 \\
             nq\_nh & 0.149 & 1.388 & -0.114 & 6.836 & 15.597 \\
             nq\_wh & 0.152 & 1.417 & -0.109 & 10.406 & 15.649 \\ 
             wq\_sh & 1.362 & 7.387 & 0.051 & 18.101 & 40.877 \\
             wq\_nh & 1.305 & 7.344 & 0.040 & 36.620 & 74.210 \\
             wq\_wh & 1.259 & 6.930 & 0.014 & 9.886 & 32.315 \\ 
             \bottomrule
        \end{tabular}
        \caption{\gls{grf}, \gls{q/h} and \gls{gta} at the time instant of peak \gls{vgrf} for scenarios of normal, weak and strong knee joint agonist antagonist muscles.}
        \label{tab:muscles_1}
    \end{center}
\end{table}

Furthermore, in \autoref{tab:muscles_2} we observe that the scenarios with greater peak \gls{vgrf} acquire lower \gls{af} and \gls{mf} values. On the contrary, \gls{abdm} values are greater for these scenarios at peak \gls{vgrf}.

\begin{table}[H]
    \begin{center}
        \begin{tabular}{c c c c c c c c} 
             \toprule
             Case & AF(+) & CF(-) & MF(+) & AbdM(+) & IRM(-) & FM(+) \\   
             \midrule
             sq\_sh & 5.379 & -14.038 & 0.532 & 0.098 & -0.030 & 0.132 \\ 
             sq\_nh & 5.142 & -11.786 & 0.513 & 0.072 & -0.025 & 0.114  \\
             sq\_wh & 5.450 & -13.827 & 0.502 & 0.094 & -0.025 & 0.140 \\ 
             nq\_sh & 4.168 & -7.852 & 0.391 & 0.006 & -0.010 & 0.035 \\
             nq\_nh & 4.035 & -7.657 & 0.395 & 0.008 & -0.009 & 0.032 \\ 
             nq\_wh & 3.901 & -7.460 & 0.401 & 0.011 & -0.009 & 0.029 \\  
             wq\_sh & 3.639 & -12.622 & 0.207 & 0.105 & -0.019 & 0.189 \\
             wq\_nh & 3.642 & -12.906 & 0.174 & 0.111 & -0.012 & 0.197 \\
             wq\_wh & 3.553 & -13.030 & 0.129 & 0.104 & -0.006 & 0.186 \\ 
            \bottomrule
        \end{tabular}
        \caption{Knee joint \gls{jrf} and \gls{jrm} at the time instant of peak \gls{vgrf} for scenarios of normal, weak and strong knee joint agonist antagonist muscles.}
        \label{tab:muscles_2}
    \end{center}
\end{table}

\section{Moco Control goal weight case}\label{sec:effort_results}

As described in \autoref{sec:effort_info}, we assigned different weights to the Moco control objective functions that can be interpreted as the effort minimization cost. The examined weights for this goal were 0, 0.1, 0.2, 0.5, 1, 2, 5 and 10, and the results are presented comparatively in this section. 

\begin{figure}[H]
    \centering
    \includegraphics[width=1\textwidth, height = 4cm]{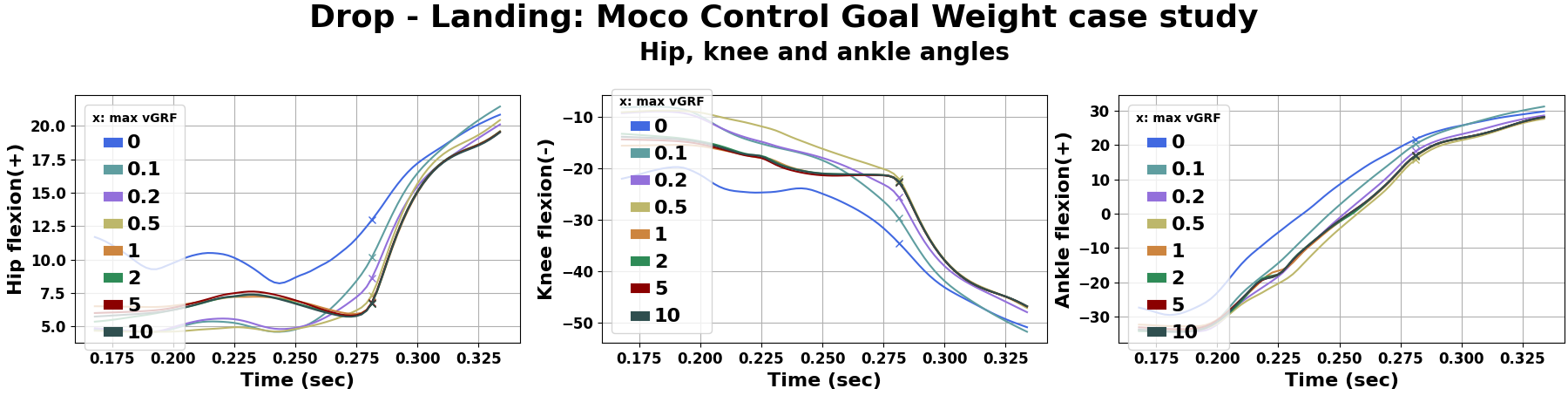}
    \caption{Hip, knee and ankle kinematics for different weights of Moco Control goal.}
    \label{fig:predict_effort_sol}
\end{figure}

Initially, in \autoref{fig:predict_effort_sol} we display the angle trajectories for the hip, knee and ankle joints for all case studies. We notice that for zero weight, ankle and hip flexion angles are greater for the entire motion. On the contrary, knee flexion angle obtain lower values compared with higher weight cases.

\begin{figure}[H]
    \centering
    \includegraphics[width=1\textwidth, height = 8cm]{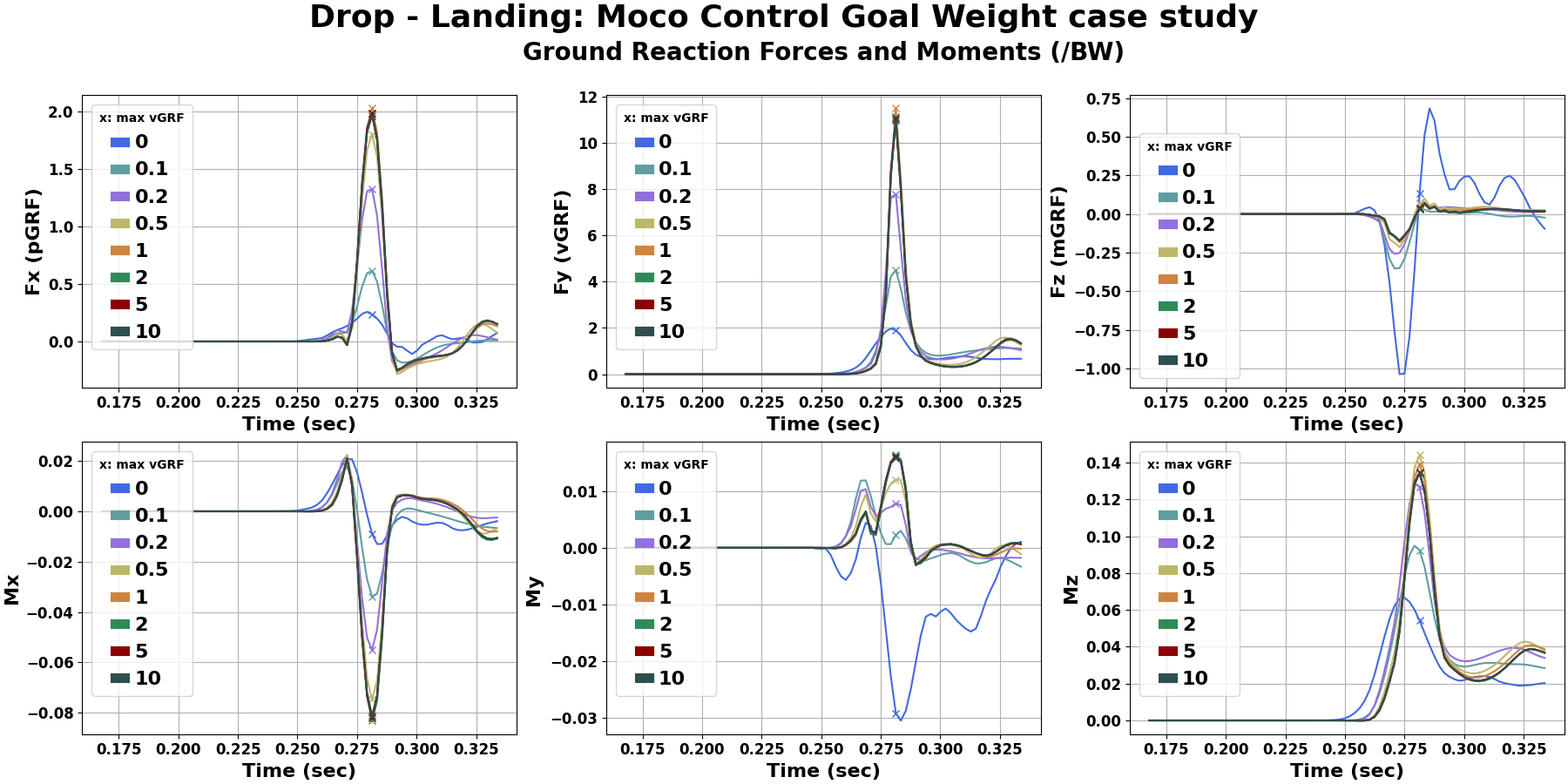}
    \caption{\gls{grf} and \gls{grm} for different weights of Moco Control goal.}
    \label{fig:predict_effort_GRF}
\end{figure}

Next, in \autoref{fig:predict_effort_GRF} we demonstrate \gls{grf} and \gls{grm} for the examined cases. We observe that the peak \gls{vgrf} is lower for zero weight and it is increased for 0.1 and 0.2, reaching even greater values for 1, 2, 5 and 10 weights. The same observation is also valid for the maximum \gls{pgrf} values.

\begin{figure}[H]
    \centering
    \includegraphics[width=\textwidth, height = 8cm]{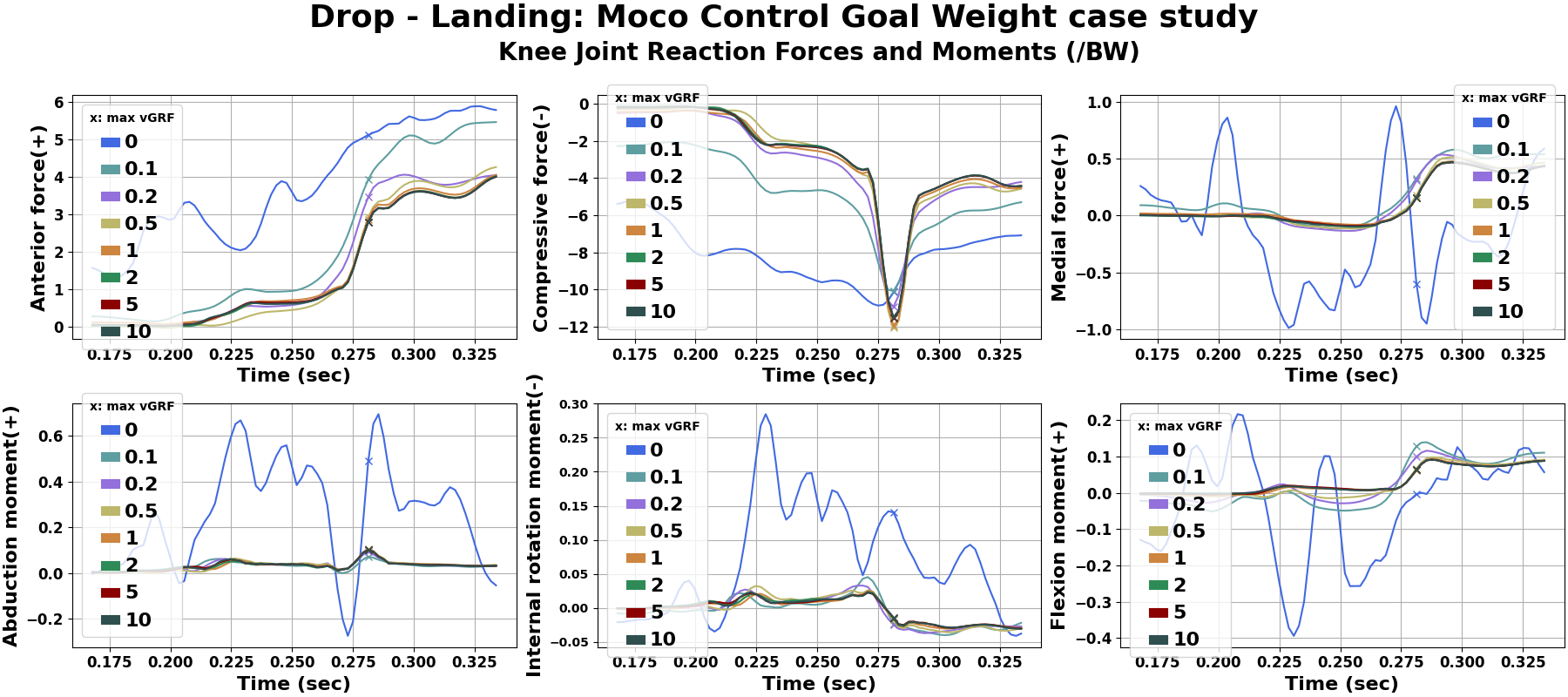}
    \caption{Knee joint \gls{jrf} and \gls{jrm} for different weights of Moco Control goal.}
    \label{fig:predict_effort_JRA}
\end{figure}

Moreover, \autoref{fig:predict_effort_JRA} demonstrates \gls{kjrf} and \gls{kjrm} for the respective scenarios. We notice that for control goal with 0 weight the results are quite different compared to the remaining cases. More specifically, \gls{af} is higher for zero weight and as the weight is increased, the force is decreased at max \gls{vgrf} time. Also, for the cases with 1,2,4 and 10 weight we notice that the results are close. Moreover, we detect greater \gls{abdm} and rotation moment for the case with zero weight compared with higher values of effort goal weight. It is obvious that the lack of the Control Goal (zero weight) highly affects the forces and moments on the knee joint resulting in much greater values for the entire motion despite the lower \gls{vgrf} peak.

\begin{figure}[H]
    \centering
    \includegraphics[width=\textwidth, height = 8cm]{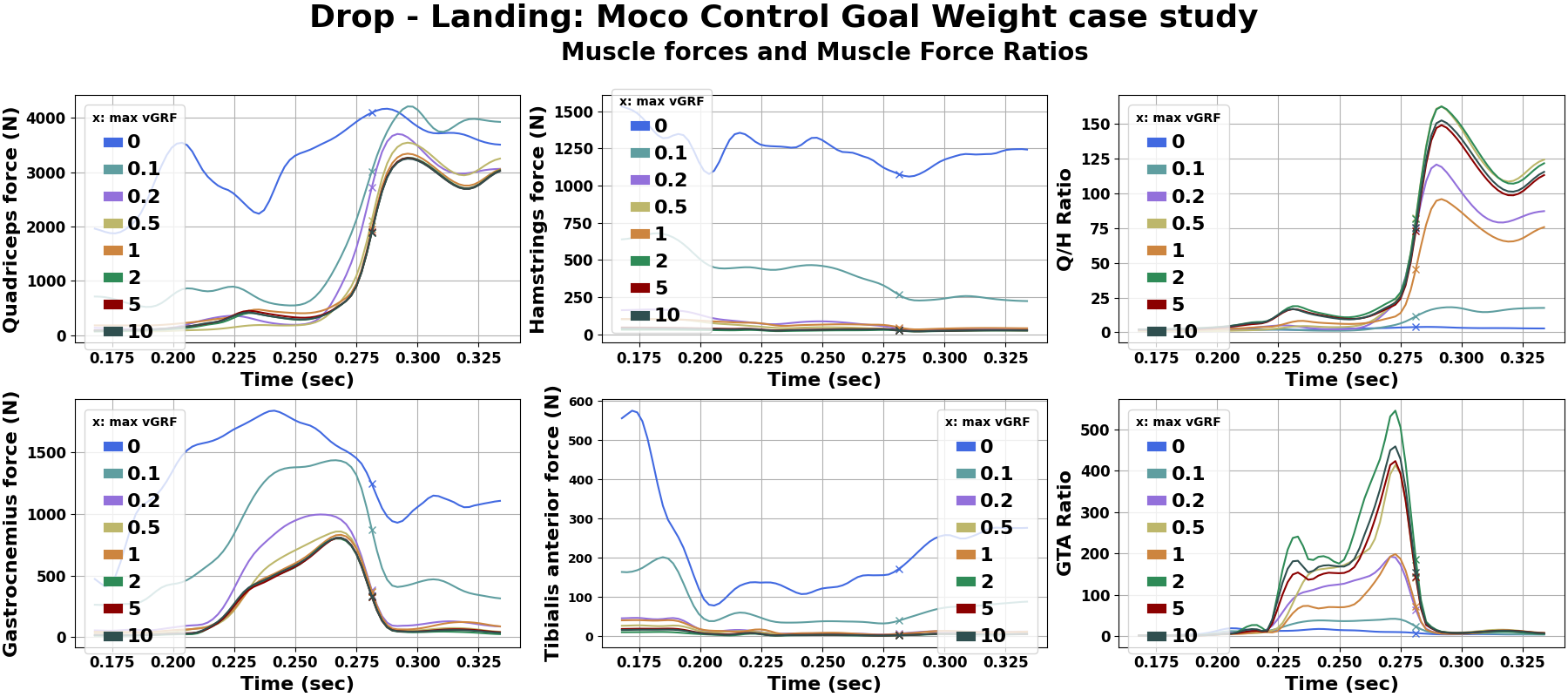}
    \caption{Muscle forces and muscle force ratios for different weights of Moco Control goal.}
    \label{fig:predict_effort_muscle_forces}
\end{figure}

Furthermore, in \autoref{fig:predict_effort_muscle_forces} muscle forces and muscle force ratios for all cases examined with different control goal weights are presented. As expected, in general muscle forces are greater for lower effort goal values. Furthermore, for weight greater than 0.5 no great differences are observed. Again we notice a great deviation for the case with no effort goal with much higher values of muscle forces compared with the other cases. As a result, \gls{q/h} and \gls{gta} at max \gls{vgrf} are greater for larger effort weights and very low in the absence of effort goal. 

In addition, in \autoref{tab:effort_1} and \autoref{tab:effort_2}, values for the previously examined parameters at maximum \gls{vgrf} for all cases are presented. We notice that as the effort goal weight is increased \gls{vgrf} is also increased until weight of 1, then for weights 2, 5 and 10 values greater than 10 times \gls{bw} are detected. \gls{q/h} is also very low for the case with no Control Goal.

\begin{table}[H]
    \begin{center}
        \begin{tabular}{c c c c c c}
             \toprule
             Case & pGRF & vGRF & mGRF & Q/H\_ratio & GTA\_ratio \\  
             \midrule
             0 & 0.284 & 1.967 & -0.469 & 3.724 & 7.918 \\ 
             0.1 & 0.792 & 4.499 & -0.001 & 11.544 & 22.167 \\ 
             0.2 &  1.772 & 7.779 & 0.081 & 74.678 & 56.125  \\ 
             0.5 & 2.658 & 11.116 & 0.206 & 78.301 & 113.327 \\ 
             1 & 3.013 & 11.479 & 0.163 & 42.970 & 55.000 \\ 
             2 & 2.923 & 11.113 & 0.146 & 77.616 & 137.173 \\ 
             5 & 2.885 & 11.043 & 0.148 & 69.774 & 104.952 \\ 
             10 & 2.876 & 10.964 & 0.140 & 72.103 & 117.775 \\ 
             \bottomrule
        \end{tabular}
        \caption{\gls{grf}, \gls{q/h} and \gls{gta} at the time instant of peak \gls{vgrf} for different weights of Moco Control goal.}
        \label{tab:effort_1}
    \end{center}
\end{table}

Moreover, based on \autoref{tab:effort_2} we observe that \gls{af} and \gls{mf} obtain lower values as the Control Goal weight is increased at peak \gls{vgrf}. On the contrary, \gls{cf} is augmented for higher control goal weights at the same time instant. In addition, for 0 weight we notice greater value of  \gls{abdm} compared with other weights. Also, at max \gls{vgrf} \gls{erm} is detected while for the other cases low values of \gls{irm} are detected.

\begin{table}[H]
    \begin{center}
        \begin{tabular}{c c c c c c c} 
             \toprule
             Case & AF(+) & CF(-) & MF(+) & AbdM(+) & IRM(-) & FM(+) \\  
             \midrule
             0 & 4.956 & -10.666 & -0.080 & 0.203 & 0.137 & -0.047 \\ 
             0.1 & 4.002 & -10.602 & 0.320 & 0.076 & -0.020 & 0.138 \\ 
             0.2 & 3.604 & -12.223 & 0.277 & 0.103 & -0.032 & 0.107 \\  
             0.5 & 3.225 & -14.880 & 0.149 & 0.144 & -0.035 & 0.071 \\  
             1 &  3.235 & -15.044 & 0.127 & 0.145 & -0.031 & 0.065 \\  
             2 & 3.143 & -14.505 & 0.132 & 0.139 & -0.028 & 0.067 \\  
             5 & 3.141 & -14.430 & 0.131 & 0.139 & -0.027 & 0.066 \\  
             10 & 3.122 & -14.344 & 0.133 & 0.135 & -0.030 & 0.067 \\  
            \bottomrule
        \end{tabular}
        \caption{Knee joint \gls{jrf} and \gls{jrm} at the time instant of peak \gls{vgrf} for different weights of Moco Control goal.}
        \label{tab:effort_2}
    \end{center}
\end{table}

%% file: Chapters/Chapter5.tex
\chapter{Discussion} 

\label{Chapter5} 

In this chapter, we present our results in comparison with other studies in literature in an attempt to identify common conclusions and discuss any divergences that may arise.

\section{Initial height case}\label{sec:height_disc}

In \autoref{sec:height_results}, we presented the simulation results for drop - landings from different landing heights. We examined single - leg landings from 30, 35, 40, 45, 50 and 55cm height. The results clearly indicated the association of the  landing height with the assessed \gls{acl} injury risk factors. 

First, we noticed that as the  landing height was increased the landing forces (\gls{grf}) were also incremented. This was in agreement with other research studies \cite{Mokhtarzadeh2010,Mokhtarzadeh2013}.  Mokhtarzadeh et al. examined single - leg landings from 30 and 60cm of height and observed greater \gls{grf} peaks for landings from 60cm \cite{Mokhtarzadeh2012}. More specifically, peak \gls{vgrf} was 3.24 times \gls{bw} when landing from 30cm height and 4.38 when landing from 60cm height. Nevertheless, in our study we observed lower peak values of \gls{vgrf} ranging from 1,43 to 1,79 times \gls{bw} for all examined landing heights. However, the \gls{grf} should be evaluated taking into account the muscle forces as there is a correlation between these two parameters as it will be described below and in \autoref{sec:effort_disc}.

Furthermore, in \autoref{tab:height_2} we observed that at max \gls{vgrf} the \gls{af} was increased for greater heights. On the contrary, \gls{cf} and \gls{mf} were decreased for increased landing heights. In a research study conducted by Verniba et al., peak \gls{cf} was greater for greater heights \cite{Verniba2017} at peak knee flexion moment. In our case, this peak coincides with the peak \gls{vgrf}. However, in this instant our results demonstrate that \gls{cf} does not increase with height. Nonetheless, we observe that \gls{cf} was greatly increased for greater heights at a short time instant after \gls{igc} (at about 0.22 sec in \autoref{fig:predict_height_effort0.001_JRA}). The same holds true for the \gls{abdm}.

This discrepancy could be related to the modified  Hunt – Crossley ground - foot contact model adopted in our study, as presented in \autoref{susec:opensim_models}. The values for parameters like stiffness, dissipation and friction were assigned based on other research studies \cite{Serrancolí2019}. Moreover, the position and total number of the contact spheres were adopted by \cite{confluence_skyhigh} a study for simulating a standing vertical jump. In general, insight into foot - ground contact during landing motion is limited and the proposed model might not accurately capture contact geometries affecting the predicted motion. This holds true for all the simulation scenarios we evaluated.


Furthermore, in our study quadriceps and hamstring forces did not present great differences among the different scenarios. On the contrary,  Mokhtarzadeh et al. observed greater quadriceps and hamstring forces for higher initial height. In our study, the peak of quadriceps force was detected around maximum \gls{vgrf} (with values  4500-5000N). These values were greater than those reported by other research studies (\cite{Mokhtarzadeh2013}). Also, hamstrings force obtained lower values in our simulations compared to the results reported by Mokhtarzadeh et al. At max \gls{vgrf} hamstring force was around 500N in our case, while Mokhtarzadeh reported values around 1000N for landing from 60cm height. However, we should mention that muscle forces are greatly dependent on the effort goal that is defined in Moco, as will be demonstrated in \autoref{sec:effort_disc}. This also affects \gls{grf} as stated previously.

Also, Mokhtarzadeh et al. emphasized on \gls{q/h} and stated that it was decreased when the \gls{vgrf} was at its peak value  which is also valid in our study. Regarding \gls{q/h}, we detected that it was decreased as the initial landing height was increased at max \gls{vgrf}. In agreement with our results, Mokhtarzadeh et al. stated that \gls{q/h} was lower for the landing from 60cm compared with a landing from 30cm. It should be mentioned that in our study we observed much higher values of \gls{q/h} at time of max \gls{vgrf} for all examined scenarios. Again, differences between the studies could be due to the selected weight of the Moco control cost. 

In a recent study \gls{acl} loading was correlated to landing height. For drop - landings from 30 and 60cm, \gls{acl} loading was predicted and the authors concluded that for greater heights the loading was increased indicating highest risk of injury \cite{Mokhtarzadeh2017}. In our study, we did not estimate \gls{acl} forces. Nonetheless, \gls{af} is inherently related with \gls{acl} loads and in our study was greater for increased heights. 

\section{Hip rotation case} \label{sec:hip_disc}

Several studies have examined the influence of static lower extremity alignment in \gls{acl} injuries \cite{Nguyen2015,Peel2020,Koga2017}. More specifically, the hip rotational alignment has been associated with knee valgus posture and \gls{acl} injuries. Yasada et al. and Koga et al. studied landings from handball players where \gls{acl} injuries took place and observed that in the majority of landings with injury, the hip was internally rotated \cite{Yasuda2016,Koga2017}. Peel et al. at their study compared landings with toe - in, toe - out and neutral position of the lower limb and concluded that the toe - out position could protect from injury while the toe - in position is associated with increased risk \cite{Peel2020}.

In this thesis we also examined the influence of hip rotation in parameters associated with \gls{acl} injuries. In \autoref{sec:hip_results} we presented the results concerning cases with different hip rotation angles. First, we presented comparative figures for the cases with internally rotated hip and then for the externally rotated cases. 

Initially, regarding the hip internal rotation scenarios we observed that the minimum \gls{vgrf} peak was for the \ang{20} case (3.602 times \gls{bw}). Although, for that scenario we detected greater \gls{abdm} and \gls{irm} at peak \gls{vgrf} ({0.044 and -0.043 times \gls{bw} respectively}). Nevertheless, for the \ang{0} hip internal scenario scenario the peak \gls{vgrf} was 3.602 times \gls{bw} while \gls{abdm}, \gls{irm} were 0.020 and -0.034 respectively. As the hip internal rotation was increased further, peak \gls{vgrf} was also increased (reaching the value of 4.658 times \gls{bw} for the \ang{30} scenario).

Nguyen et al. claimed that when the hip was internally rotated and the knee was in a valgus posture, then the average knee \gls{erm} was greater compared with a neutral position. On the other hand, Peel et al. found an association between the toe - in landing position and knee \gls{irm}. In our study we found that at peak \gls{vgrf} we have \gls{irm} with very small differences between the values for each case (\autoref{fig:predict_hip_effort0.001_internal_JRA}). However, we can also notice that we have a peak \gls{erm} right before peak \gls{vgrf} for all cases. Also, we observed the lowest \gls{abdm} at peak \gls{vgrf} for the \ang{0} and \ang{10} scenarios, while the remaining scenarios presented greater values.

Furthermore, we examined scenarios with externally rotated hip. In these scenarios we spotted the lowest \gls{vgrf} peaks for the \ang{10} and \ang{15} hip external rotation scenarios. Also, at peak \gls{vgrf} time instant the \ang{0} hip rotation scenario had greater \gls{af}, \gls{mf}, \gls{abdm} and \gls{irm} compared with the scenarios featuring externally rotated hip. On a similar note, Peel et al. claimed that the toe - out position was associated with decreased knee \gls{addm} and therefore lower \gls{acl} injury risk. In our study, we see that for increased but not excessive external hip rotation angles, the \gls{af} is noticeably smaller converging to the same conclusion. Last but not least, we noticed that as the hip external rotation angle  was increased, the \gls{q/h} was also decreased. 

Summarizing, we could claim that the internally rotated hip position could possibly be associated with \gls{acl} injuries due to increased values in risk factors. On the contrary, the externally rotated hip position may be considered safer due to lower values in risk factors. Landing with \ang{10} or \ang{15} externally rotated hip can possibly be safer due to lower \gls{grf}.
 
\section{Trunk orientation case}\label{sec:trunk_disc}

Multiple studies have examined the influence of trunk orientation in the observed kinematics, \gls{grf} and \gls{kjrf} during drop - landings. A leaning forward trunk has been associated with more flexion of the hip and knee joints and lower values of \gls{pgrf}, \gls{vgrf} and lower mean amplitude of quadriceps \gls{emg} compared with an upright position during drop landings \cite{Blackburn2008,Blackburn2009}. In another study, results were quite similar. Upright position during landings was linked with greater peak of \gls{vgrf}, lower gastrocnemius and quadriceps activation and greater knee extensor moments \cite{Shimokochi2012}. In our study, we demonstrated that for all examined cases, landings that favored a forward leaning position resulted in lower risk of \gls{acl} injury. Specifically, we observed that when the trunk was leaning forward with \ang{25} the peak \gls{vgrf} was lower, following by \ang{30}, and \ang{10}. However, the lowest \gls{vgrf} value was observed for the case of no flexion. This can be due to the  initial guess given to the study. Also, for the case of \ang{25} of trunk flexion \gls{abdm} and \gls{irm} were lower compared with other angles of trunk flexion. 

Regarding landings with trunk leaning backwards, as a general remark, we observed greater \gls{vgrf} compared with the upright position as the lumbar extension angle was increased. Additionally, \gls{q/h}, \gls{cf}, \gls{abdm} and \gls{irm} were greater at time of maximum \gls{vgrf} for the cases of landings with trunk in backwards leaning. Also, Saito et al. observed greater knee valgus angles when landing in a trunk extension position compared with landings with trunk neutral position \cite{Saito2020}, which is in agreement with the \gls{abdm} values detected in our results.

Concerning the trunk bending, Saito et al. observed that landings with the trunk leaning towards the opposite side of the landing leg or toward to the landing side led to greater values of knee valgus angles \cite{Saito2020}. In our case we observed that greater \gls{abdm} values occurred as the right or left bending angle was increased compared with the neutral position, that would introduce an increased valgus angle. Also, Jones et al. noticed greater \gls{grf} and knee joint loads when leaning in the opposite direction of landing. Similarly, in our study we observed that the largest peak \gls{vgrf} occurred when bending towards the right direction that is opposite to the landing leg with a value of 8.632 per \gls{bw}. 

Regarding landings with internal or external rotation, we noticed that for the cases with \ang{5} and \ang{10} degrees of external rotation of the trunk and 10 degrees of internal rotation peaks of equal amplitude with the upright case occurred. For the rest of the trunk rotation angles, we noticed a large increase in the amplitude of the \gls{vgrf} peaks.


\section{Muscle force of knee joint agonists and antagonists case}\label{sec:muslces_disc}

In this study, we investigated how permutations of max isometric force for the knee joint agonist and antagonist muscles affect the parameters of interest, namely \gls{vgrf} and \gls{af} (\autoref{sec:muslces_results}). Again, a direct comparison with other research studies was not feasible. 

In our study, we observed that varying the maximum isometric force of these two muscle groups affected all simulation results. We discerned that hamstrings force was the greatest for the scenario with strong hamstrings and normal quadriceps, following by the scenario where both muscles were normal. An unexpected outcome was that the lowest hamstrings forces were detected for the cases of weak quadriceps. This outcome clearly indicates how muscles work in conjunction and the significance of the muscle redundancy issue. This concept is related to the uncertainty of contribution for each in a given task.

The lowest peaks of \gls{vgrf} were detected for the following scenarios in augmented order: normal quadriceps and strong hamstrings, normal quadriceps and normal hamstrings and normal quadriceps and weak hamstrings with values 1.362, 1.388 and 1.417 times \gls{bw} respectively. The greatest peak of \gls{vgrf} was observed for the scenario of weak quadriceps and strong hamstrings. For the cases with lower peaks of \gls{vgrf}, greater values of \gls{af} were noticed at time of maximum \gls{vgrf}.

It should be emphasized that the cases with normal quadriceps present similar behavior despite hamstrings strengthening or weakening. For these scenarios we observed lower \gls{vgrf}, \gls{q/h}, \gls{gta}, knee joint \gls{cf} and \gls{abdm}. Although, \gls{af} was higher for these scenarios compared with the scenarios with weak quadriceps, but lower compared with the scenarios with strong quadriceps. 

\section{Moco Control goal weight case}\label{sec:effort_disc}

 Finally, we investigated how the Moco Control Goal (or the effort) affects the simulation results. We cannot compare our results with other research studies because we couldn't find studies with similar content. We assumed that this case study would offer a more clear view for the other case studies in our work, especially regarding the connection between \gls{grf} and muscle forces.

In \autoref{sec:effort_results}, we observed that as we increased the weight of the control goal, peak \gls{vgrf} was also increased reaching a maximum value of  11.479 times \gls{bw} for weight of 1. In general, for the upper limit weight values we noticed increased \gls{vgrf} peaks. The same pattern was also spotted for the \gls{q/h}. In contrast, we observed that as the weight was increased, \gls{af} was decreased, although the differences between the peak values above the 0.1 weight are not very large. Regarding muscle forces, as it was expected for greater weights of the goal lower muscle forces were detected since the objective function introduced to the problem tries to minimize the sum of the absolute value of the controls. Again, no great differences were observed between the cases with weight greater than 0.1.

%% file: Chapters/Conclusions.tex
\chapter{Conclusions}\label{Conclusions}

In this work, we presented a pipeline that aims to predict and identify key parameters of interest that are related to \gls{acl} injury during single - leg landings. First, we provided the basic theoretical concepts of the subject under consideration. We started from a general description of the lower limb anatomy and physiology. Emphasis was given to the knee joint with its surrounding anatomical structures and especially the \gls{acl} ligament. We also described muscle physiology and a muscle modeling approach that captures the physiological properties of muscle functionality. Moreover, we provided a high - level description of multibody dynamics and a brief introduction into trajectory optimization methods, as the tools that we used throughout this work are built upon these fundamental concepts. We finally presented related studies in an effort to gather all risk factors related to \gls{acl} injury during landings. In this thesis, we focused on the following parameters: \gls{vgrf}, \gls{af}, \gls{abdm}, and \gls{q/h} ratio.

Next, we described in detail the methods we followed to perform our study, starting from the software packages we deployed. These were the OpenSim, SCONE and Moco interfaces. OpenSim was used to modify available musculoskeletal models to include a compliant ground - foot contact model and the Analysis tool to perform \gls{jra} analysis and extract information about muscle forces for all our assessed studies. As a fist step SCONE was used to acquire a single case of single - leg landing that acted as an initial guess for all subsequent Moco simulations. These tools are based on different approaches regarding trajectory optimization. Moco is based on direct collocation methods that lead to faster predictive simulations. This is one of the main reasons that we chose Moco as our preferred option. To obtain a quick initial guess from SCONE we used a simplified model. Then, for the following Moco studies we used models with a gradually increased complexity. This offered us the ability to provide an adequate initial guess to all investigated cases, that led to quick and efficient predictive simulations. 

Subsequently, we presented the results for the following conditions of single - leg landing: a) initial landing height, b) hip internal and external rotation, c) trunk forward - backward leaning, d) trunk medial - lateral bending, e) trunk internal - external rotation and f) muscle forces permutations. We also tested the effect of Moco control goal weight on the derived results. 

Starting from the landing height we concluded that \gls{vgrf}, \gls{af}, and \gls{abdm} increased as the height increased. On the other hand the opposite holds true for \gls{q/h} ratio. These findings are in agreement with similar studies. Next, regarding the internal hip rotation we observed that in general \gls{vgrf}, \gls{af}, and \gls{abdm} increase with larger hip rotations. On the contrary, a range of hip external rotation between \ang{10} - \ang{15} demonstrated lower values for \gls{vgrf}. \gls{af} had the lowest value for a hip external rotation of \ang{20}. These findings in general agree with literature, where a toe - out posture is associated with lower \gls{acl} injury risk. Regarding trunk forward leaning, \gls{vgrf} was lower for the upright posture. A similar behavior was observed for backward trunk leaning. As a conclusion we can state that an upright posture seems to reduce \gls{vgrf}. The same holds in general true for trunk lateral - medial bending and internal external rotation. Furthermore, we investigated how modifying muscle forces for quadriceps and hamstrings affected the injury risk factors. An interesting observation is that modifications of quadriceps force capacity greatly affects hamstrings force production. Moreover, when quadriceps where unchanged, the \gls{vgrf} and \gls{abdm} demonstrated the lowest values, regardless of hamstrings force modifications. On the other hand this is not the case for \gls{af}. This observation clearly exhibits the inherent complexity of the \gls{acl} injury mechanism. Finally, we investigated how the Moco control cost weight affects muscle forces and subsequently all landing and knee joint forces and moments. This objective function is closely related to effort that furthers affects muscle activation and force production. We observed that lower weights greatly reduced \gls{vgrf} but increased \gls{af}, \gls{abdm} and \gls{erm}. As a general conclusion we can say that \gls{af} and \gls{vgrf} demonstrate an opposite behavior in all cases, where an increase in \gls{vgrf} is accompanied by a decrease of \gls{af}. 

Our study clearly demonstrates that \gls{acl} injury is a rather complicated mechanism with many associated risk factors. These factors are clearly not independent. Therefore, one cannot clearly study the phenomenon without considering this correlation between these inherently different variables. Of course, our work does not come without any limitations, especially considering the adopted modeling assumptions. First, we deployed musculoskeletal models that feature one \gls{dof} knee joint without modeling the contribution of other structures such as cartilages and ligaments. Especially, we think that the absence of a suitable contact model between the femoral and tibial knee compartments greatly affects our results. The same observation can be made for the ground - foot contact model and \gls{cop} estimation. Also, we did not assess different population classes that feature varying anthropometric data. All these limitations can serve as the foundations for future work to improve our current pipeline. Nonetheless, our pipeline clearly showcased the promising potential of predictive simulations to evaluate different aspects of complicated phenomena, such as the \gls{acl} injury. Moreover, our findings are in general agreement with similar studies. Finally, we envision that further improvements can lead to a pipeline that can be used by physiotherapists and clinicians on adjusting rehabilitation and training plans based on subject - specific characteristics.